\documentclass[twocolumn]{aastex631}

\usepackage{float}
\usepackage{comment}
\usepackage{here}
\usepackage{amsmath}

\newcommand{\JL}{\lambda_{\mathrm{J, clump}}^{\mathrm{th}}}

\newcommand{\JM}{\mathrm{M}_{\mathrm{J, clump}}^{\mathrm{th}}}

\newcommand{\Add}[1]{\textbf{#1}}
\shorttitle{AASTeX v6.3.1 Sample article}
\shortauthors{Ishihara et al.}

\begin{document}

\title{Digging into the Interior of Hot Cores with ALMA (DIHCA). IV. Fragmentation in High-mass Star-Forming Clumps}

\author[0000-0001-7080-2808]{Kousuke Ishihara}
\affiliation{Department of Astronomical Science, School of Physical Sciences,
Graduate University for Advanced Studies (SOKENDAI), 
2-21-1 Osawa, Mitaka, Tokyo 181-8588, Japan}
\email{kousuke.ishihara@grad.nao.ac.jp}
\affiliation{National Astronomical Observatory of Japan,
 2-21-1 Osawa, Mitaka, Tokyo 181-8588, Japan}
   
\author[0000-0002-7125-7685]{Patricio Sanhueza}
\affiliation{National Astronomical Observatory of Japan,
 2-21-1 Osawa, Mitaka, Tokyo 181-8588, Japan}
\affiliation{Department of Astronomical Science, School of Physical Sciences,
Graduate University for Advanced Studies (SOKENDAI), 
2-21-1 Osawa, Mitaka, Tokyo 181-8588, Japan}

\author[0000-0001-5431-2294]{Fumitaka Nakamura}
\affiliation{National Astronomical Observatory of Japan,
 2-21-1 Osawa, Mitaka, Tokyo 181-8588, Japan}
\affiliation{Department of Astronomy, Graduate School of Science, The University of Tokyo, 7-3-1 Hongo, Bunkyo-ku, Tokyo 113-0033}
\affiliation{Department of Astronomical Science, School of Physical Sciences,
Graduate University for Advanced Studies (SOKENDAI), 
2-21-1 Osawa, Mitaka, Tokyo 181-8588, Japan}

\author[0000-0003-0769-8627]{Masao Saito}
\affiliation{National Astronomical Observatory of Japan,
 2-21-1 Osawa, Mitaka, Tokyo 181-8588, Japan}
\affiliation{Department of Astronomical Science, School of Physical Sciences, Graduate University for Advanced Studies (SOKENDAI), 2-21-1 Osawa, Mitaka, Tokyo 181-8588, Japan}

\author[0000-0002-9774-1846]{Huei-Ru Vivien Chen}
\affiliation{Institute of Astronomy, National Tsing Hua University, Hsinchu 30013, Taiwan}

\author[0000-0003-1275-5251]{Shanghuo Li}
\affiliation{Max Planck Institute for Astronomy, Kstuhl 17, D-69117 Heidelberg, Germany}

\author[0000-0002-8250-6827]{Fernando Olguin}
\affiliation{Institute of Astronomy and Department of Physics, National Tsing Hua University, Hsinchu 30013, Taiwan}

\author[0000-0003-4402-6475]{Kotomi Taniguchi}
\affiliation{National Astronomical Observatory of Japan, National Institutes of Natural Sciences, 2-21-1 Osawa, Mitaka, Tokyo 181-8588, Japan}

\author[0000-0002-6752-6061]{Kaho Morii}
\affiliation{Department of Astronomy, Graduate School of Science, The University of Tokyo, 7-3-1 Hongo, Bunkyo-ku, Tokyo 113-0033, Japan}
\affiliation{National Astronomical Observatory of Japan, National Institutes of Natural Sciences, 2-21-1 Osawa, Mitaka, Tokyo 181-8588, Japan}

\author[0000-0003-2619-9305]{Xing Lu}
\affiliation{Shanghai Astronomical Observatory, Chinese Academy of Sciences, 80 Nandan Road, Shanghai 200030, People’s Republic of China}

\author[0000-0003-4506-3171]{Qiu-yi Luo}
\affiliation{Shanghai Astronomical Observatory, Chinese Academy of Sciences, 80 Nandan Road, Shanghai 200030, People’s Republic of China}

\author[0000-0003-4521-7492]{Takeshi Sakai}
\affiliation{Graduate School of Informatics and Engineering, The University of Electro-Communications, Chofu, Tokyo 182-8585, Japan}

\author[0000-0003-2384-6589]{Qizhou Zhang}
\affiliation{Center for Astrophysics \textbar Harvard \& Smithsonian, 60 Garden Street, Cambridge, MA 02138, USA}

\begin{abstract}
Fragmentation contributes to the formation and evolution of stars. 
Observationally, high-mass stars are known to form multiple-star systems, preferentially in cluster environments.   
Theoretically, Jeans instability has been suggested to determine characteristic fragmentation scales, and thermal or turbulent motion in the parental gas clump mainly contributes to the instability. 
To search for such a characteristic fragmentation scale, we have analyzed ALMA 1.33 mm continuum observations toward 30 high-mass star-forming clumps taken by the Digging into the Interior of Hot Cores with ALMA (DIHCA) survey.
We have identified 573 cores using the dendrogram algorithm and measured the separation of cores by using the Minimum Spanning Tree (MST) technique. 
The core separation corrected by projection effects has a distribution peaked  around 5800 au. 
In order to remove biases produced by different distances and sensitivities, we further smooth the images to a common physical scale and perform completeness tests.  
Our careful analysis finds a characteristic fragmentation scale of $\sim$7000 au, comparable to the thermal Jeans length of the clumps. We conclude that thermal Jeans fragmentation plays a dominant role in determining the clump fragmentation in high-mass star-forming regions, without the need of invoking turbulent Jeans fragmentation. 
\end{abstract}

\section{Introduction} \label{sec:intro}

Fragmentation plays an important role in the formation of substructures in molecular clouds, leading to star formation. In particular, high-mass stars preferentially form in clustered environments \citep{Lada_Lada_2003} and, therefore, a full understanding of the fragmentation process is key to uncover how high-mass stars are formed. Turbulent hydrodynamic simulations of clustered star formation have demonstrated that gravitational (thermal Jeans) and turbulent Jeans fragmentation of dense molecular clumps creates stellar clusters \citep[e.g.,][]{Bonnell_2001,Klessen_1998,Wang_2010,Pelkonen_2021, Vazquez-Semadeni_2019}.

For the simplest case in which the thermal pressure balances with self-gravity, the thermal Jeans fragmentation is likely a dominant process of structure formation 
\citep[e.g.,][]{Jeans_1902,Larson_1985}. It has a characteristic scale, the so-called Jeans length, which is given by 
\begin{eqnarray}
\lambda_{\mathrm{J}}^{\mathrm{th}} &=& c_s \left( \frac{\pi}{G \rho} \right)^{\frac{1}{2}}  \ , \\
    &\sim& 5600 \left( \frac{T}{20~{\rm K}} \right)^{1/2} \left( \frac{n_{\rm cl}}{10^6~{\rm cm^{-3}}} \right)^{-1/2}  {\rm au} \ , 
    \label{eq:termal_JL}
\end{eqnarray}
where $c_s = (k_BT/\mu m_H)^{\frac{1}{2}}$ is the one-dimensional isothermal sound speed, $G$ is the gravitational constant, and $\rho$ is the mass density.
The parameters $k_B$, $T$, $\mu$, and $m_H$ are the Boltzmann constant, the gas temperature, the mean molecular weight per free particle (assuming $\mu=2.37$ from \citet{Kauffmann_2008}), and the mass of a hydrogen atom, respectively.
This length scale can be affected by various factors, such as magnetic fields and external pressure \citep{Larson_1985}.
For example, the external pressure can make the fragmentation scale shorter by producing surface deformation instabilities \citep[e.g.,][]{Chandrasekhar_1953}.
The magnetic fields shorten or lengthen the fragmentation scales, depending on the degree of the dynamical support \citep[e.g.,][]{Nagasawa_1987, Nakamura_1993, Das_2021}.

Turbulent pressure is dominant over the thermal pressure at the clump and cloud scales of $\sim$ 0.1 -- 10 pc.  
In such cases, turbulence supports the clouds and clumps on large scales against gravitational contraction, while it provokes collapse and fragmentation on local scales, by forming overdensities due to dynamic interaction of turbulent eddies or waves. 
\citet{Bonazzola_1987} derived the turbulent Jeans fragmentation condition by introducing the wavenumber ($k$) depending on the  effective sound speed, 
\begin{equation}
    c_s (k) ^2 = c_s^2 + \frac{1}{3} v^2 (k)  \ ,  
\end{equation}
where $v(k)$ is determined by the turbulent power spectrum. For turbulent interstellar clouds, this relation breaks down below $\sim$$10^{-2}$ pc scale, at which the thermal pressure becomes comparable to the turbulent pressure, as it has been seen in infrared dark clouds \citep[IRDCs; e.g.,][]{Sanhueza_2017,Sanhueza_2019,Sanhueza_2021,Morii_2021,Morii_2023,Morii_2024,Li_S_2020,Li_S_2022,Li_S_2023}. Therefore, the minimum fragmentation scale in a turbulent medium is likely to be set to that of thermal Jeans  fragmentation, that is the thermal Jeans length \citep[see also][]{Vazquez-Semadeni_1995}. If the dense filaments have a characteristic width of $\sim$0.1 pc, as suggested by {\it Herschel} observations of nearby molecular clouds \citep{Arzoumanian_2011}, 

the characteristic fragmentation scale is suggested to be around 0.3 pc, which corresponds to the wavelength of the most unstable mode \citep{Larson_1985}. 
For high-mass clumps, this filament fragmentation scale appears to be about 10 times larger than the thermal Jeans length (Equation (\ref{eq:termal_JL})).

Previous studies of core separations suggest the existence of the characteristic length scale. For example, \citet{Beuther_2018} observed 20 molecular clumps located in high-declination regions ($>24^\circ$) with the IRAM Northern Extended-Millimeter Array (NOEMA) with an angular resolution between $\sim$0\farcs32 and $\sim$0\farcs50 ($\sim 960-1500$ au at 3 kpc). They identified 121 cores and measured their separations in each region. They found that most of their separations are distributed under 5000 au. Their core separation takes its peak at around thermal Jeans length.

\citet{Lu_2020} studied 3 high-density clouds in Central Molecular Zone (CMZ). They used ALMA band 6 (1.33 mm wavelength) with an angular resolution of $\sim$0\farcs2, corresponding to $\sim 1600$ au at $\sim 8.2$ kpc. 
They identified about 800 cores.  
The thermal Jeans length range is estimated as $\sim 1.0 \times 10^4 - 1.9 \times 10^4$ au. They suggested that the core separations match the thermal Jeans length 
of $\sim$ 0.2 pc with gas density of $10^6$ cm$^{-3}$ and temperature of 20--50 K.
The CMZ is an extreme environment in the Galaxy, and the results in this region may not be applicable to other high-mass star-forming regions, even though their sample size is large.

\citet{Zhang_S_2021} studied 8 high-mass starless clumps including 4 clumps associated with \ion{H}{2} region.
Using ALMA with an angular resolution of $\sim$1\farcs3 ($\sim 5100$ au at $\sim 4$ kpc), they identified 51 cores in 8 regions and found two peaks in the separation distribution at $0.3$$\lambda_{\mathrm{J}}^{\mathrm{th}}$ and $0.8$$\lambda_{\mathrm{J}}^{\mathrm{th}}$. They argued that the existence of two peaks implies hierarchical fragmentation on $\lesssim 10^4$ au.
However, the shorter peak is almost equal to their beam size. At a similar angular resolution ($\sim$1\farcs2), the ALMA Survey of 70 $\mu$m Dark High-mass Clumps in Early Stages \citep[ASHES;][]{Sanhueza_2019,Morii_2021} explains the fragmentation in 12 IRDC clumps \citep[]{Sanhueza_2019} and the complete survey with 39 sources \citep{Morii_2023,Morii_2024} as dominated by thermal Jeans fragmentation rather than turbulent Jeans fragmentation. 


These previous studies add evidence that the thermal Jeans fragmentation is important in high-mass star-forming regions, with a few counter examples, as the study of the Carina Nebula by \cite{Rebolledo_2020}. However, the statistics and spatial resolution are still insufficient to conclude that the core separation is comparable to the thermal Jeans length of the parent clumps.

In this work, we present the Digging into the Interior of Hot Cores with ALMA (DIHCA) survey. We have observed 30 high-mass star-forming regions to search for dense compact cores at high-angular resolution in ALMA band 6 ($\sim$226 GHz; $\sim$1.33 mm). In the the DIHCA project, the targets were observed in two configurations providing $\sim$0\farcs06 (extended configuration) and $\sim$0\farcs3 (compact configuration) resolutions. As seen in a case study of DIHCA, cores are clearly resolved at 0\farcs3 angular resolution \citep{olguin_2021}, while at $\sim$0\farcs06 angular resolution, disk and/or core  fragmentation become important \citep{Olguin_2022,Olguin_2023,Li_S_2024}. With the aim of exploring clump fragmentation, we determine whether there is a characteristic length scale based on core separations, using the compact configuration data only. An improvement with respect to previous works on this topic is that we put particular emphasis in considering the effects of the different physical resolution and mass sensitivity among targets when deriving the characteristic core separation for each clump. The analysis of the core/disk fragmentation using the extended configuration data will be presented in a future work (P. Sanhueza et al., in prep).

This paper is organized as follows. 
In Section \ref{sec:sources}, we describe the target selection. 
In Section \ref{sec:obs}, we summarize the observational parameters.
We present the results of the 1.33 mm dust continuum observations in Section \ref{sec:results}. DIHCA data reveal that the target clumps contain a large number of compact structures, i.e., dense cores.  Applying the dendrogram technique, we identify dense cores and derive their physical properties. In Section \ref{sec:discussion} we discuss the characteristic fragmentation scale and possible biases in the analysis. Finally, we summarize the main results in Section \ref{sec:summary}.

\begin{figure*}[htb!]
\centering
\includegraphics[width=0.8\linewidth]{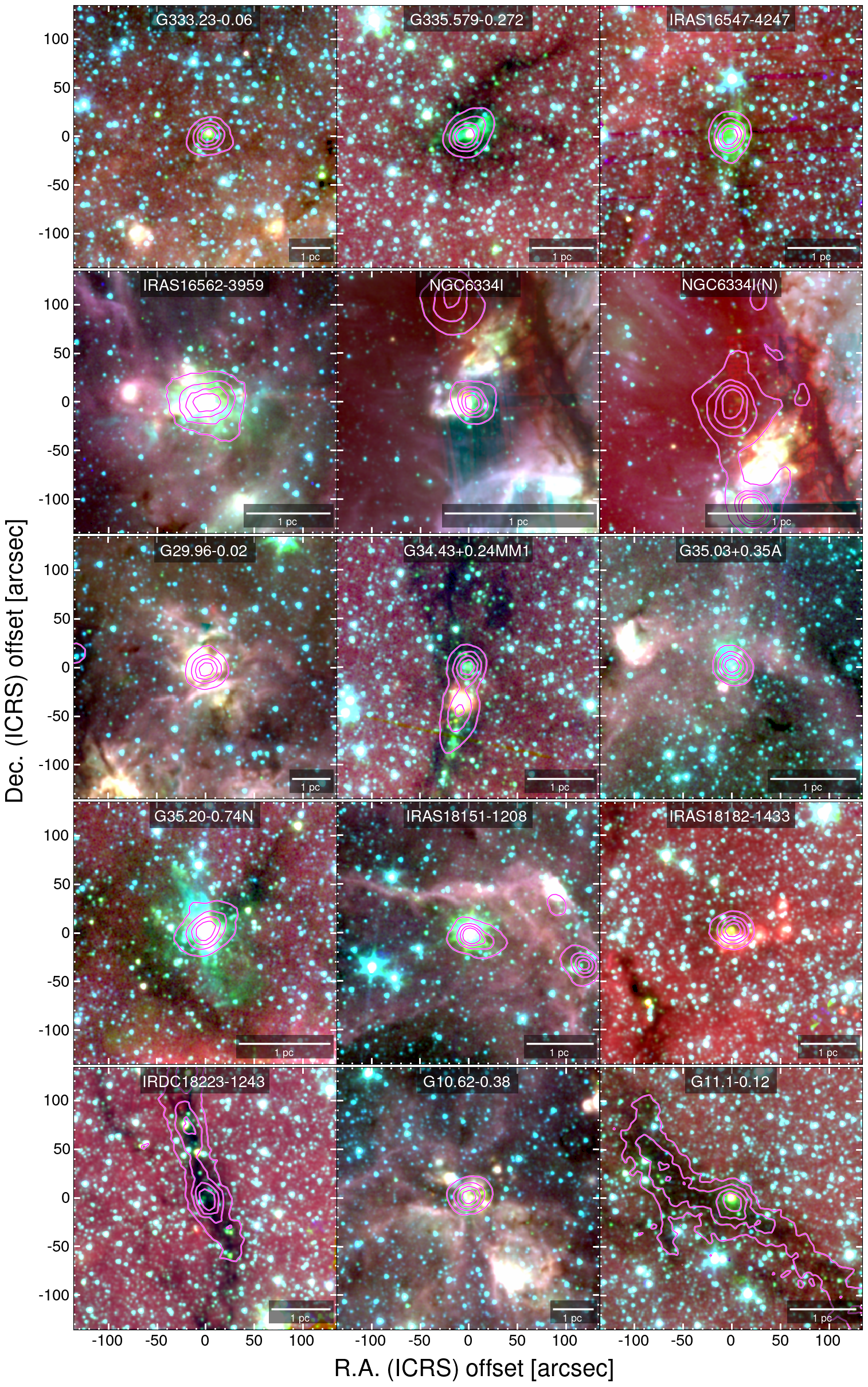}
\caption{Large-scale overview \textit{Spitzer/IRAC} three-color image as a background (3.6 $\mu$m in blue, 4.5 $\mu$m in green, and 8.0 $\mu$m in red) with overlaid ATLASGAL 870 $\mu$m dust continuum emission in contours for the entire sample, except IRAS18151-1208 and IRAS18161-2048 for which JCMT 850 $\mu$m dust continuum emission is used. The contour levels are 0.2, 0.4, 0.6 and 0.8 $\times S_{\rm{peak}}$, with $S_{\rm{peak}}$ equal to the peak flux of 870 or 850 $\mu$m emission of each clump.
\label{fig:IR&LSview1}}
\end{figure*}
%
\begin{figure*}[p!]
\centering
\includegraphics[width=0.8\linewidth]{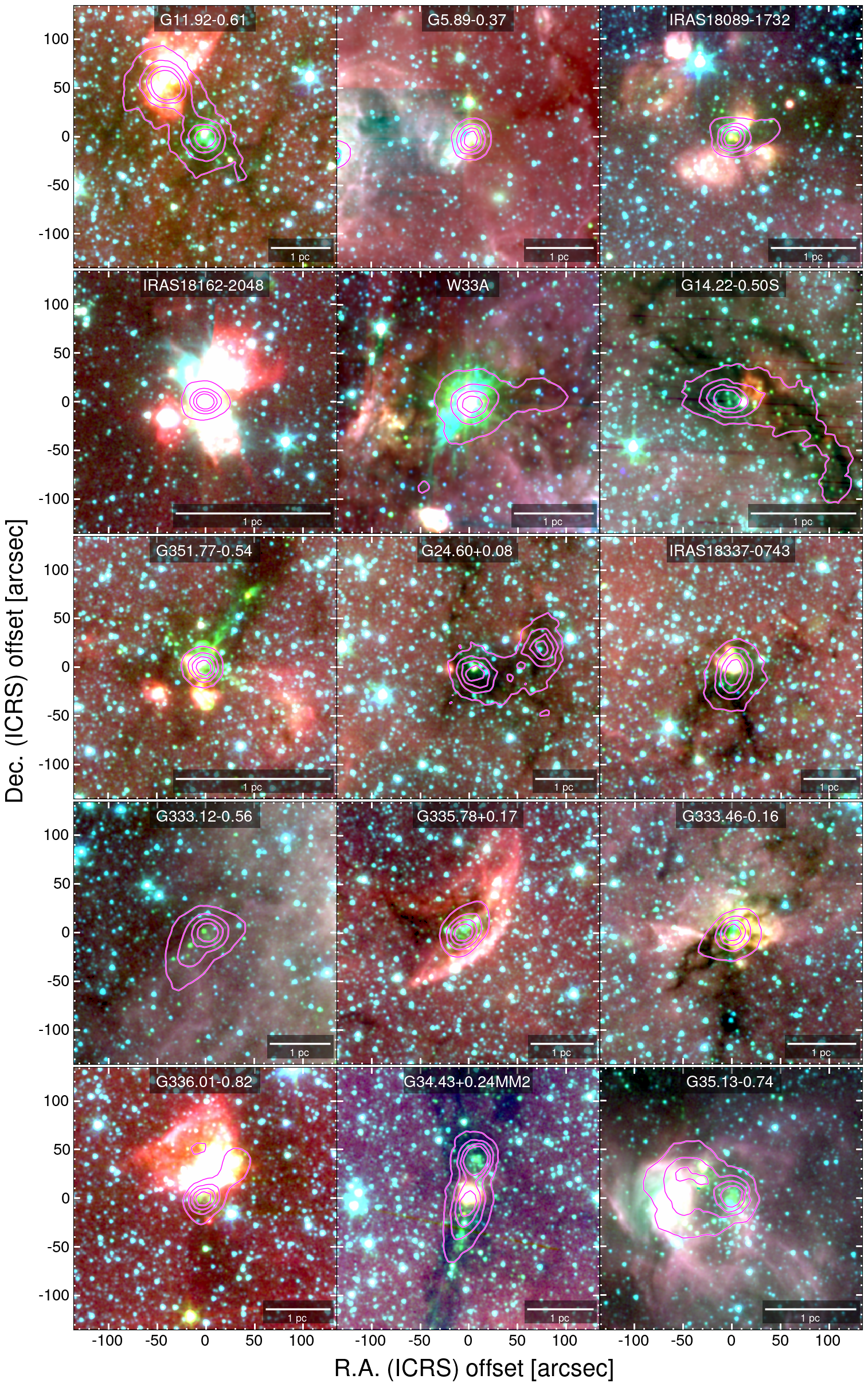}
\caption{Same as Fig. \ref{fig:IR&LSview1}.
\label{fig:IR&LSview2}}
\end{figure*}
\section{Source selection} \label{sec:sources}
Target sources are selected from the literature according to the following criteria: (i) the target must be bright, (ii) relatively close, and (iii) the clump must have the potential to form high-mass stars. 
We selected 30 regions whose parameters are summarized in Table~\ref{tab:sources}. In Figures \ref{fig:IR&LSview1} and \ref{fig:IR&LSview2}, we show the \textit{Spitzer}/IRAC three-color images (3.6 $\mu$m in blue, 4.5 $\mu$m in green, and 8.0 $\mu$m in red)\footnote{\textit{NASA/IPAC Infrared Science Archive},
\url{https://sha.ipac.caltech.edu/applications/Spitzer/SHA/}} of the targets with overlaid contours of the dust continuum emission from single-dish telescopes (APEX and JCMT).

The aforementioned  three selection criteria are described in detail below:
 
 (i) A flux limit of $>$0.1 Jy at 230 GHz was imposed to make sure that we would detect compact objects in the most extended configuration (the one providing $\sim$0\farcs06 resolution that will be presented in a future work).
 
 (ii) To assure a similar physical resolution (and mass sensitivity), we restricted the distance ($d$) to lie between 1.6 and 3.8 kpc. However, as parallax distances have been made available over time, some of the observed targets resulted to be closer (1.3 kpc) and farther (5.26 kpc), expanding the whole range of distances (see Table~\ref{tab:sources} for up to date distances). 
  
 (iii) Some target fields are evidently forming high-mass stars, based on the detection of free-free emission or the presence of hot cores with a line forest spectra. However, to have a uniform selection criteria, we also checked whether the clump masses and surface densities satisfy empirical thresholds for high-mass star formation (e.g., $M_{\mathrm{lim}} = 460\, \mathrm{M_\odot} (R/\mathrm{pc})^{1.9}$ by \citet{Larson_1981} and $M_{\mathrm{lim}} = 580\, \mathrm{M_\odot} (R/\mathrm{pc})^{1.33}$ by \citet{Kauffmann_Pillai_2010}). 

Clumps masses and radii are obtained as follows.
Using the 870 $\mu$m and 850 $\mu$m emission from the ATLASGAL survey
\footnote{\textit{The ATLASGAL Database Server}, \url{https://atlasgal.mpifr-bonn.mpg.de/cgi-bin/ATLASGAL_DATABASE.cgi}} \citep{Schuller_2009}
and SCOPE survey
\footnote{\textit{The JCMT Science Archive}, \url{https://www.cadc-ccda.hia-iha.nrc-cnrc.gc.ca/en/jcmt/}} \citep{Eden_2019}, we first fitted each target with a 2D Gaussian function to derive the integrated flux ($S_\nu$) and size of the clumps.  
The clump radius, $R_\mathrm{cl}$, is calculated as the geometrical mean of the semi-minor and semi-major full width at half maximum (FWHM) deconvolved from the beam. The clump masses are derived as,
\begin{equation}
M_{\mathrm{cl}} = \mathbb{R} \frac{S_\nu d^2}{\kappa_\nu B_\nu (T_{\rm dust})} ,
\label{eq:mass}
\end{equation}
where $\mathbb{R}$ is the gas-to-dust mass ratio (assumed to be 100), $\kappa_\nu$ is the dust opacity, and $B_\nu (T_{\rm{dust}})$ is the Planck function at the dust temperature $T_{\rm dust}$, and at the observed frequency $\nu$ (or wavelength $\lambda$).
We calculate $\kappa_\lambda$ following
\begin{equation}
\kappa_\lambda=\kappa_{1.33 \mathrm{mm}} \left( \frac{\mathrm{1.33~mm}}{\lambda} \right)^\beta .
\end{equation}
Here we adopt $\beta=1.8$ and $\kappa_{1.33 \mathrm{mm}}=0.899$ cm$^2$ g$^{-1}$ corresponding to the opacity of dust grain with thin ice mantles at a gas density of $\sim 10^6$ cm$^{^{-3}}$ \citep{Ossenkopf_Henning_1994}. 
We obtain $\kappa_{\mathrm{870\mu m}}=1.93$ cm$^2$ g$^{-1}$ 
and $\kappa_{\mathrm{850\mu m}}=2.01$ cm$^2$ g$^{-1}$.

For the dust temperatures, we adopted the values derived from  one or two components fitting to the spectral energy distribution. 
For G35.20-0.74N, we adopted the temperature derived by \citet{Konig_2017}.
For IRAS 18151-1208 and IRAS 18162-2048, the values by \citet{McCutcheon_1995}.
For the remaining 27 clumps, the values derived in  \citet{Urquhart_2018}.
Based on the above references, we adopted the temperature of cold component as the clump dust temperature for all clumps.
  
The surface density, $\Sigma_{\mathrm{cl}}$, peak column density, $N^{\mathrm{peak}}$, 
and volume density, $n_{\mathrm{cl}}$, are derived as follows,
\begin{eqnarray}
\Sigma_{\mathrm{cl}} &= \frac{M_{\mathrm{cl}}}{\pi R_\mathrm{cl}^2}~, & \label{eq:surface density}\\
N^{\mathrm{peak}} &= \mathbb{R} \frac{S^{\rm{peak}}_{\nu}}{\Omega \kappa_\nu B_\nu (T) \mu_{\rm{H}} m_{\rm{H}}}~, & \label{eq:peak column density}\\
n_{\mathrm{cl}} &= \frac{M_{\mathrm{cl}}}{\frac{4}{3} \pi R_\mathrm{cl}^3 \mu_{\mathrm{H}} m_{\mathrm{H}}}~, & \label{eq:volume density}
\end{eqnarray}
respectively, where $S^{\rm{peak}}$ is peak flux, $\Omega$ is the solid angle of the synthesized beam, $\mu_{\mathrm{H}}$ is the mean molecular weight per hydrogen molecule (assumed to be 2.8), and $m_{\mathrm{H}}$ is the atomic mass of hydrogen.
We summarize the clump physical quantities in 
Table~\ref{tab:sources}.

\begin{figure}[H]
\includegraphics[width=\linewidth]{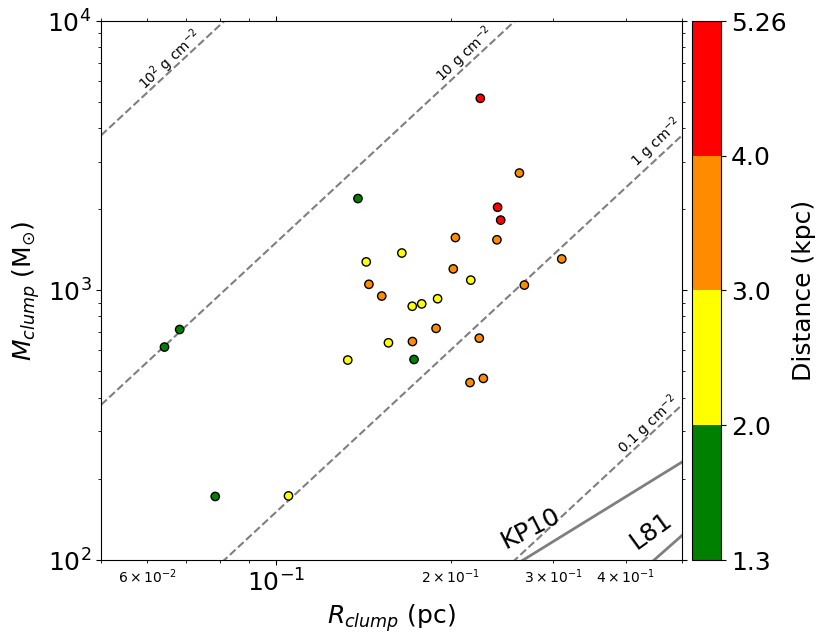}
\caption{Clump mass against radius. 
The solid lines show the relationship of \citep[L81]{Larson_1981} and \citep[KP10]{Kauffmann_Pillai_2010}. 
All clumps in the sample satisfy the empirical conditions for high-mass star formation. The color coding shows target distance.
\label{fig:clump_mass}}
\end{figure}

In Figure \ref{fig:clump_mass}, we plot the DIHCA clumps and compared their properties with several threshold relations for high-mass star formation. 
Larson's relation \citep[]{Larson_1981} is given as $M_{\mathrm{lim}} = 460\, \mathrm{M_\odot} (R/\mathrm{pc})^{1.9}$ and is approximately related to a constant surface density of $\Sigma \sim 460/\pi\ \mathrm{M_{\odot}\ pc^{-2}}$. 
\citet{Kauffmann_Pillai_2010} gives $M_{\mathrm{lim}} = 580\, \mathrm{M_\odot} (R/\mathrm{pc})^{1.33}$ for the formation of high-mass stars.  
For these empirical relations, the clumps located above the thresholds are likely to form high-mass stars. In fact, the clump volume densities obtained here are high, $> 10^5 - 10^6 $ cm$^{-3}$. 

Twelve of our target clumps were selected from \citet{Beltran_2016_0}. Our target list contains both hot cores embedded in infrared dark clouds, known to host the earliest stages of high-mass star formation \citep[e.g.,][]{Rathborne_2007,Zhang_2009,Sanhueza_2012,Sanhueza_2013,Sanhueza_2019}, as well as massive young stellar objects (MYSOs) presenting the typical hot core line forest with and without signs of compact cm (free-free) emission. 
See the Appendix \ref{asec:targets} for more details of the individual target clumps.

  
\begin{deluxetable*}{llccccccccr}[htb!]
\tabletypesize{\scriptsize}
\tablecaption{Physical Properties of High-mass Star-forming Clumps \label{tab:sources}}
\tablewidth{0pt}
\tablehead{
\colhead{Source} & \colhead{Catalogue} & \multicolumn2c{Position (ICRS)} & \colhead{Dist.} & \colhead{$T_{\rm{dust}}$} & \colhead{$M_{\rm{cl}}$} & \colhead{$R_{\rm{cl}}$} &
\colhead{$\Sigma_{\rm{cl}}(\rm{H}_2)$} & \colhead{$n_{\rm{cl}}(\rm{H}_2)$} & Ref.\\
\cline{3-4}
\colhead{Clump} & \colhead{Name} & \colhead{R.A. (h:m:s)} & \colhead{Decl. (d:m:s)} & \colhead{(kpc)} & \colhead{(K)} & \colhead{($M_\odot$)} & \colhead{(pc)} & 
\colhead{($\rm{g}~\rm{cm}^{-2}$)} & \colhead{($\times 10^6 \rm{cm}^{-3}$)} & \colhead{} \\ 
\colhead{(1)} & \colhead{(2)} & \colhead{(3)} & \colhead{(4)} & \colhead{(5)} & \colhead{(6)} & \colhead{(7)} & \colhead{(8)} & \colhead{(9)} & \colhead{(10)} & \colhead{(11)}
}
\startdata
G333.23-0.06 & AGAL333.234-00.061 & 16:19:51.20 & $-$50:15:13.00 & 5.20 & 20.9 & 2030 & 0.24 & 2.3 & 5.0 & 1 \\
G335.579-0.272 & AGAL335.586-00.291 & 16:30:58.76 & $-$48:43:54.01 & 3.25 & 23.1 & 1570 & 0.20 & 2.5 & 6.4 & 2 \\
IRAS16547-4247 & AGAL343.128-00.062 & 16:58:17.22 & $-$42:52:07.49 & 2.90 & 28.9 & 1380 & 0.16 & 3.4 & 10.6 & 1 \\
IRAS16562-3959 & AGAL345.493+01.469 & 16:59:41.63 & $-$40:03:43.62 & 2.38 & 42.3 & 1090 & 0.22 & 1.5 & 3.7 & 3 \\
NGC6334I & AGAL351.416+00.646 & 17:20:53.30 & $-$35:47:00.02 & 1.35 & 30.6 & 720 & 0.07 & 10.2 & 77.6 & 4 \\
NGC6334I(N) & AGAL351.444+00.659 & 17:20:54.90 & $-$35:45:10.02 & 1.35 & 21.4 & 2190 & 0.14 & 7.6 & 28.5 & 4 \\
G29.96-0.02 & AGAL029.954-00.016 & 18:46:03.76 & $-$02:39:22.60 & 5.26 & 35.5 & 1820 & 0.24 & 2.0 & 4.3 & 5 \\
G34.43+0.24MM1 & AGAL034.411+00.234 & 18:53:18.01 & $+$01:25:25.50 & 3.03 & 22.7 & 1050 & 0.14 & 3.4 & 12.0 & 6 \\
G35.03+0.35A & AGAL035.026+00.349 & 18:54:00.65 & $+$02:01:19.30 & 2.32 & 31.8 & 170 & 0.11 & 1.0 & 5.1 & 4 \\
G35.20-0.74N & AGAL035.197-00.742 & 18:58:13.03 & $+$01:40:36.00 & 2.19 & 29.5 & 640 & 0.16 & 1.7 & 5.8 & 4 \\
IRAS18151-1208 & SCOPEG018.34+01.77 & 18:17:58.17 & $-$12:07:25.02 & 3.00 & 35.0 & 720 & 0.19 & 1.4 & 3.7 & 7 \\
IRAS18182-1433 & AGAL016.586-00.051 & 18:21:09.13 & $-$14:31:50.58 & 3.58 & 24.7 & 650 & 0.17 & 1.5 & 4.4 & 5 \\
IRDC18223-1243 & AGAL018.606-00.074 & 18:25:08.55 & $-$12:45:23.32 & 3.40 & 13.0 & 660 & 0.22 & 0.9 & 2.0 & 1 \\
G10.62-0.38 & AGAL010.624-00.384 & 18:10:28.65 & $-$19:55:49.52 & 4.95 & 31.0 & 5160 & 0.22 & 6.8 & 15.7 & 8 \\
G11.1-0.12 & AGAL011.107-00.114 & 18:10:28.27 & $-$19:22:30.92 & 3.00 & 15.8 & 470 & 0.23 & 0.6 & 1.4 & 1 \\
G11.92-0.61 & AGAL011.917-00.612 & 18:13:58.02 & $-$18:54:19.02 & 3.37 & 22.8 & 1050 & 0.27 & 1.0 & 1.9 & 5 \\
G5.89-0.37 & AGAL005.884-00.392 & 18:00:30.43 & $-$24:04:01.64 & 2.99 & 34.5 & 1280 & 0.14 & 4.1 & 15.0 & 5 \\
IRAS18089-1732 & AGAL012.888+00.489 & 18:11:51.40 & $-$17:31:28.52 & 2.34 & 23.4 & 550 & 0.13 & 2.1 & 8.1 & 9 \\
IRAS18162-2048 & SCOPEG010.84-02.59 & 18:19:12.20 & $-$20:47:29.02 & 1.30 & 41.0 & 170 & 0.08 & 1.8 & 12.2 & 10 \\
W33A & AGAL012.908-00.259 & 18:14:39.40 & $-$17:52:01.02 & 2.53 & 23.6 & 890 & 0.18 & 1.9 & 5.4 & 11 \\
G14.22-0.50S & AGAL014.114-00.574 & 18:18:13.00 & $-$16:57:21.82 & 1.90 & 15.8 & 550 & 0.17 & 1.2 & 3.7 & 1 \\
G351.77-0.54 & AGAL351.774-00.537 & 17:26:42.53 & $-$36:09:17.40 & 1.30 & 29.9 & 620 & 0.06 & 9.9 & 80.0 & 1 \\
G24.60+0.08 & AGAL024.633+00.152 & 18:35:40.50 & $-$07:18:34.02 & 3.45 & 15.9 & 450 & 0.22 & 0.6 & 1.6 & 12 \\
IRAS18337-0743 & AGAL024.441-00.227 & 18:36:40.82 & $-$07:39:17.74 & 3.80 & 19.3 & 1310 & 0.31 & 0.9 & 1.5 & 1 \\
G333.12-0.56 & AGAL333.129-00.559 & 16:21:36.00 & $-$50:40:50.01 & 3.30 & 16.1 & 2720 & 0.26 & 2.6 & 5.2 & 13 \\
G335.78+0.17 & AGAL335.789+00.174 & 16:29:47.00 & $-$48:15:52.32 & 3.20 & 23.1 & 1200 & 0.20 & 2.0 & 5.0 & 13 \\
G333.46-0.16 & AGAL333.466-00.164 & 16:21:20.26 & $-$50:09:46.56 & 2.90 & 24.6 & 870 & 0.17 & 2.0 & 6.0 & 1 \\
G336.01-0.82 & AGAL336.018-00.827 & 16:35:09.30 & $-$48:46:48.16 & 3.10 & 21.8 & 950 & 0.15 & 2.7 & 9.3 & 1 \\
G34.43+0.24MM2 & AGAL034.401+00.226 & 18:53:18.58 & $+$01:24:45.98 & 3.03 & 20.5 & 1540 & 0.24 & 1.8 & 3.8 & 6 \\
G35.13-0.74 & AGAL035.132-00.744 & 18:58:06.30 & $+$01:37:05.98 & 2.20 & 19.4 & 930 & 0.19 & 1.7 & 4.7 & 1 \\
\enddata
\tablecomments{
(1), (2) All target clumps except for IRAS18151-1208 and IRAS18162-2048 are referred from Compact Source Catalog by \textcite{Contreras_2013} \& \textcite{Urquhart_2014} in ATLASGAL survey, while the rest two targets are by \textcite{Eden_2019} in SCOPE survey.
(3), (4) Phase center of ALMA pointing observation.
(5), (11) Target distances are referred from the following.
1: \textcite{Urquhart_2018},
2: \textcite{Peretto_2013},
3: \textcite{Moises_2011},
4: \textcite{Wu_2014},
5: \textcite{Sato_2014},
6: \textcite{Mai_2023},
7: \textcite{Brand_1993},
8: \textcite{Sanna_2014},
9: \textcite{Xu_2011},
10: \textcite{Anez-Lopez_2020},
11: \textcite{Immer_2013},
12: \textcite{Dirienzo_2015},
13: \textcite{Green_2011}.
}
\end{deluxetable*}
%


\section{Observations and data reduction} \label{sec:obs}
We observed 30 high-mass star-forming clumps with the ALMA 12 m array, including more than 40 antennas in band 6 ($\sim$226.150 GHz; $\sim$1.33 mm). Observations were scheduled in Cycle 4, 5, and 6 (Project ID: 2016.1.01036.S, 2017.1.00237.S; PI: Patricio Sanhueza). The angular resolution is $\sim$0\farcs3 ($\sim$900 au at 3 kpc) and 
the maximum recoverable scale is $\sim$8\farcs7-10\farcs7 ($\sim2.6-3.2 \times 10^4$ au at 3 kpc). Observational parameters are summarized in Table \ref{tab:Obs_info}.

The data reduction and calibration were done using CASA \citep{McMullin_2007} version 4.7.0, 4.7.2, 5.1.1-5, 5.4.0-70, and 5.6.1-8. The data were then self-calibrated in steps of decreasing solution time intervals. The continuum was extracted by averaging the line-free channels following the procedure defined in DIHCA paper I \citep{olguin_2021}. All images were made using CASA task TCLEAN. The imaging of the data cubes from the continuum-subtracted visibilities was performed with the automasking routine YCLEAN \citep{Contreras_2018}. Cubes were presented in DIHCA paper III \citep{Taniguchi_2023}. The imaging of the continuum emission was done interactively, using Briggs weighting with a robust parameter of 0.5. 
The average angular resolution of the final continuum images is  0\farcs31, with an average rms noise of $0.26$ mJy beam$^{-1}$.
Primary-beam correction was applied to measure all fluxes, but the images shown in this paper are without primary-beam correction.

\begin{deluxetable*}{lclrcccc|l}
\tabletypesize{\scriptsize}
\tablecaption{Main information about ALMA observations.\label{tab:Obs_info}}
\tablewidth{0pt}
\tablehead{
\colhead{Source} & \colhead{rms Noise\tablenotemark{a}} & \colhead{Beam\tablenotemark{a}} & \colhead{res.\tablenotemark{a}} & \colhead{MRS\tablenotemark{b}} & \colhead{Baselines} & \colhead{Configuration} & \colhead{Number of} & \colhead{Calibrators\tablenotemark{c}} \\
\colhead{Clump} & \colhead{(mJy beam$^{-1}$)} & \colhead{($''\times''$, PA)} & \colhead{(au)} & \colhead{($''$)} & \colhead{(m)} & \colhead{} & \colhead{Antennas} &
}
\startdata
G333.23-0.06&0.206&0.35 $\times$ 0.30 (-46.2$^\circ$)&1680& 3.2&18.6-1100&C40-6&41 & F: J1617-5848 \\
G335.579-0.272&0.438&0.36 $\times$ 0.30 (-57.8$^\circ$)&1070& 3.2&18.6-1100&C40-6&41 & B: J1427-4206 \\
&&&&&&&& P: J1603-4904 \\
\tableline 
IRAS16547-4247&0.167&0.23 $\times$ 0.19 (-63.5$^\circ$)&610& 2.6&15.1-2600&C40-5/C43-5&44 & F: J1617-5848 \\
IRAS16562-3959&0.182&0.23 $\times$ 0.15 (-63.1$^\circ$)&440& 2.6&16.7-2600&C40-5&44 & B: J1617-5848 \\
NGC6334I&1.040&0.24 $\times$ 0.16 (-53.9$^\circ$)&260& 2.6&16.7-2600&C40-5&44 & P: J1713-3418 \\
NGC6334I(N)&0.352&0.22 $\times$ 0.14 (-58.7$^\circ$)&240& 2.6&16.7-2600&C40-5&44 &\\
\tableline 
G29.96-0.02&0.343&0.39 $\times$ 0.25 (-65.5$^\circ$)&1640& 2.5&18.6-1100&C40-6&41 & F: J1924-2914 \\
G34.43+0.24MM1&0.412&0.38 $\times$ 0.25 (-61.9$^\circ$)&1090& 2.5&18.6-1100&C40-6&41 & B: J1924-2914 \\
G35.03+0.35A&0.161&0.39 $\times$ 0.25 (-61.0$^\circ$)&730& 2.5&18.6-1100&C40-6&41 & P: J1851+0035 \\
G35.20-0.74N&0.249&0.38 $\times$ 0.25 (-61.7$^\circ$)&680& 2.5&18.6-1100&C40-6&41 &\\
\tableline 
IRAS18151-1208&0.084&0.42 $\times$ 0.24 (-72.5$^\circ$)&960& 2.9&15.1-1100&C40-5&45 & F: J1924-2914 \\
IRAS18182-1433&0.134&0.41 $\times$ 0.24 (-75.0$^\circ$)&1130& 2.9&15.1-1100&C40-5&45 & B: J1924-2914 \\
IRDC18223-1243&0.075&0.41 $\times$ 0.25 (-74.3$^\circ$)&1100& 2.9&15.1-1100&C40-5&45 & P: J1832-2039 \\
&&&&&&&& ~~~ J1832-1035 \\
\tableline 
G10.62-0.38&0.415&0.41 $\times$ 0.26 (-76.3$^\circ$)&1600& 3.8&15.1-1100&C40-5&50 & F: J1924-2914 \\
G11.1-0.12&0.089&0.41 $\times$ 0.25 (-75.7$^\circ$)&970& 3.8&15.1-1100&C40-5&50 & B: J1924-2914 \\
G11.92-0.61&0.138&0.40 $\times$ 0.26 (-76.1$^\circ$)&1080& 3.8&15.1-1100&C40-5&50 & P: J1832-2039 \\
G5.89-0.37&0.328&0.42 $\times$ 0.26 (-77.7$^\circ$)&980& 3.8&15.1-1100&C40-5&50 &\\
IRAS18089-1732&0.181&0.41 $\times$ 0.26 (-74.4$^\circ$)&750& 3.8&15.1-1100&C40-5&50 &\\
IRAS18162-2048&0.219&0.39 $\times$ 0.26 (-76.3$^\circ$)&410& 3.8&15.1-1100&C40-5&50 &\\
W33A&0.139&0.42 $\times$ 0.26 (-73.6$^\circ$)&830& 3.8&15.1-1100&C40-5&50 &\\
\tableline 
G14.22-0.50S&0.193&0.52 $\times$ 0.34 (-61.8$^\circ$)&800& 3.9&15.1-1400&C43-5&49 & F: J1924-2914 \\
&&&&&&&& B: J1924-2914 \\
&&&&&&&& P: J1832-2039 \\
\tableline 
G351.77-0.54&0.718&0.30 $\times$ 0.27 (87.8$^\circ$)&370& 3.8&15.1-1400&C43-5&44 & F: J1517-2422 \\
&&&&&&&& B: J1517-2422 \\
&&&&&&&& P: J1711-3744 \\
\tableline 
G24.60+0.08&0.082&0.32 $\times$ 0.28 (56.6$^\circ$)&1030& 3.6&15.1-1400&C43-5&47 & F: J1751+0939 \\
IRAS18337-0743&0.087&0.33 $\times$ 0.28 (55.0$^\circ$)&1140& 3.6&15.1-1400&C43-5&47 & B: J1751+0939 \\
\tableline 
G333.12-0.56&0.149&0.34 $\times$ 0.34 (9.6$^\circ$)&1120& 3.9&15.1-1300&C43-5&44 & F: J1427-4206 \\
G335.78+0.17&0.250&0.33 $\times$ 0.31 (56.1$^\circ$)&1020& 3.9&15.1-1300&C43-5&44 & B: J1427-4206 \\
G333.46-0.16&0.188&0.35 $\times$ 0.33 (35.6$^\circ$)&980& 3.9&15.1-1300&C43-5&44 & P: J1603-4904 \\
G336.01-0.82&0.194&0.34 $\times$ 0.32 (-38.0$^\circ$)&1020& 3.9&15.1-1300&C43-5&44 &\\
\tableline 
G34.43+0.24MM2&0.147&0.34 $\times$ 0.28 (-79.0$^\circ$)&1080& 3.6&15.1-1400&C43-5&47 & F: J1751+0939 \\
G35.13-0.74&0.138&0.34 $\times$ 0.27 (-80.7$^\circ$)&670& 3.6&15.1-1400&C43-5&47 & B: J1751+0939 \\
&&&&&&&& P: J1851+0035 \\
\enddata
\tablecomments{
\tablenotetext{a}{The rms noize and beamsize in the cleaned continuum images.}
\tablenotetext{b}{Maximum Recoverable Scale is calculated using $5^{th}$ percentile shortest baseline 
(\url{https://almascience.nao.ac.jp/documents-and-tools/cycle8/alma-technical-handbook})
.
}
\tablenotetext{c}{``F", ``B" and ``P" represent flux, bandpass and phase calibrator, respectively.}
}
\end{deluxetable*}


\section{Results} \label{sec:results}
\subsection{1.33 mm dust continuum emission} \label{subsec:dust continuum}

\begin{figure*}[htb!]
\centering
\includegraphics[width=0.8\linewidth]{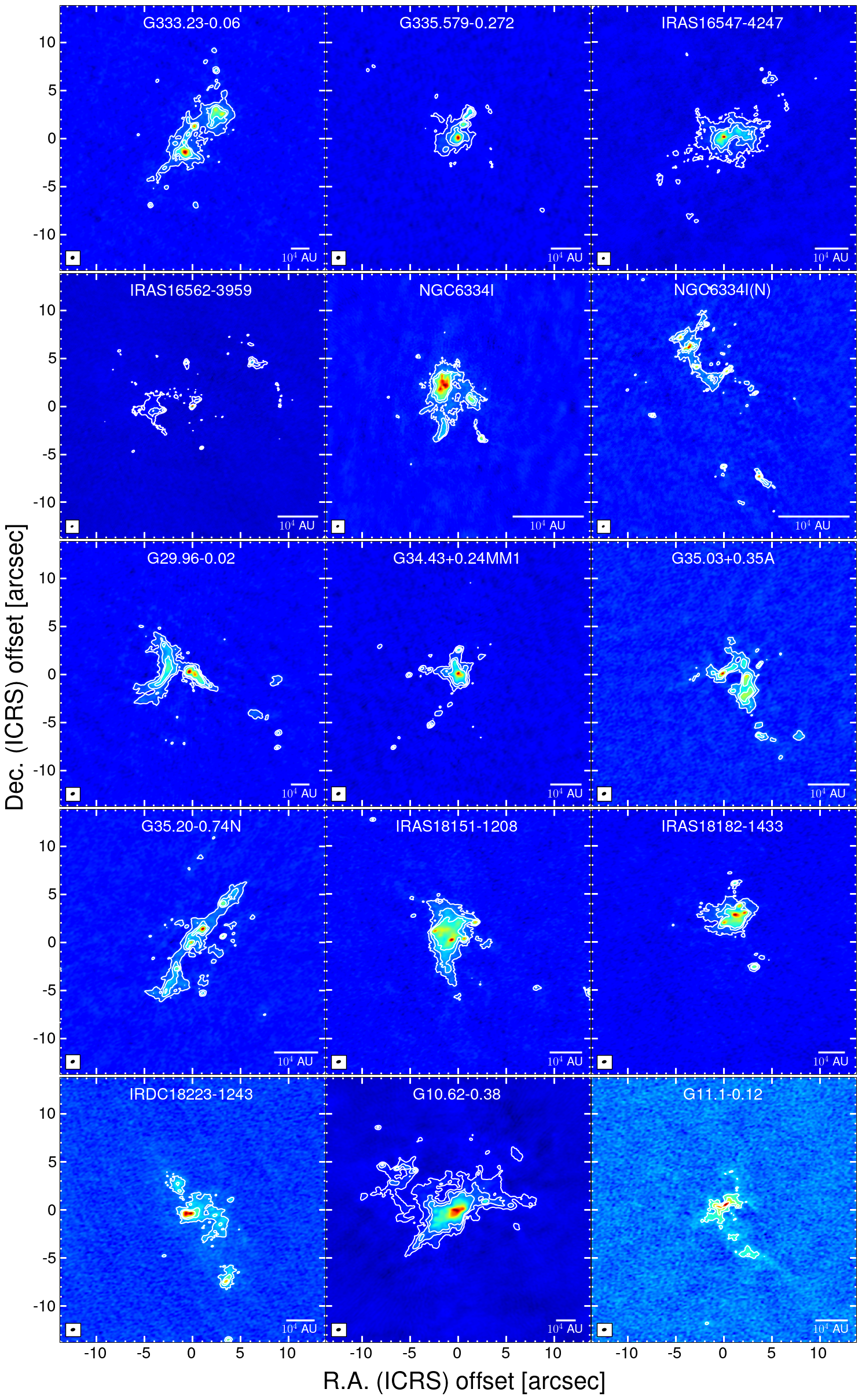}
\caption{The overview of 1.33 mm continuum images for our targets on the same angular scale.
The contour levels are [5, 15, 30] $\times \sigma$, with $\sigma$ from column 2 in Table 2.
The synthesized beam and scalebar are shown in the bottom-left and -right in each panel, respectively.
The full views of each image with colorbar are shown in Appendix \ref{asec:targets}.
\label{fig:continuums1}}
\end{figure*}

\begin{figure*}[p!]
\centering
\includegraphics[width=0.8\linewidth]{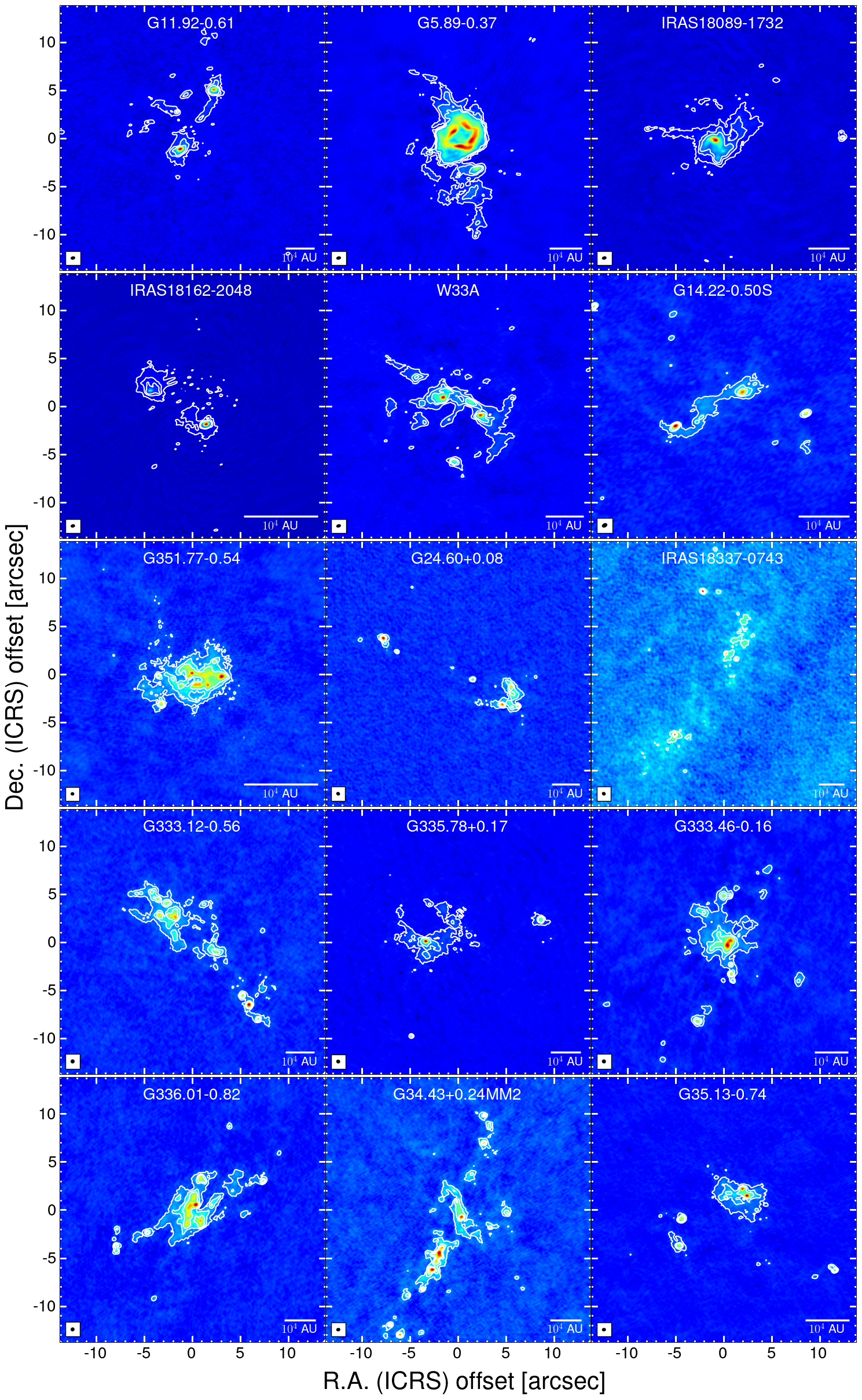}
\caption{Same as Fig. \ref{fig:continuums1}.
\label{fig:continuums2}}
\end{figure*}

Figures \ref{fig:continuums1} and \ref{fig:continuums2} show the ALMA 1.33 mm dust continuum images of all targets.
The internal structure of all the observed high-mass star-forming regions is well resolved and presents a variety of morphologies.  
Some of the DIHCA targets have emission highly concentrated near the clump center (e.g., IRAS16547-4247, NGC6334I, and G351.77-0.54), while in others the emission is relatively widespread (e.g., IRAS16562-3959, G11.92-0.61, and G333.46-0.16). 
Some clumps present linearly-aligned compact structures, resembling a fragmenting filament (e.g., G333.23-0.06, NGC6334I(N), and G35.03+0.35A). 

Individual figures on each region can be found in the Appendix \ref{asec:targets}. The ALMA observations reveal that each clump has several compact structures, known as cores. Below, we describe how we uniformly identify these cores, as well as how we derive their physical properties. 

%
%
%
%
%
%

\subsection{Core identification}
\label{subsec:core identification}

We adopted the \texttt{astrodendro} package \citep{Rosolowsky_2008}, which uses a dendrogram algorithm to compute hierarchical structures. We applied \texttt{astrodendro} to the continuum data without primary-beam correction, and identified the cores to measure their properties (integrated flux, peak flux, core size, and position). 
The \texttt{astrodendro} identifies three structures: leaf, branch, and trunk. 
We define the minimum structure, ``a leaf'', as a core. The input parameters for dendrogram are three: \texttt{min\_value} ($S_{\mathrm{min}}$), \texttt{min\_delta} ($\delta_{\mathrm{min}}$), and \texttt{min\_npix} ($\theta_{\mathrm{min}}$). 

For \texttt{min\_value}, we set $S_{\mathrm{min}}=5\sigma$ to remove non-reliable sources ($\sigma$ is the rms noise written in Table \ref{tab:Obs_info}).
For \texttt{min\_delta}, which helps to distinguish cores from neighboring structures, we adopt $\delta_{\mathrm{min}}=1\sigma$. 
The minimum pixel number, \texttt{min\_npix}, was selected to be the corresponding number of pixels contained in a synthesized beam.
After core identification, core fluxes were corrected by the primary-beam response. 
We identified 579 leaves in total and removed 6 leaves locating at the edge of the primary beam less than 30 \% response.

Table \ref{tab:dendro_out} lists the peak flux, integrated flux, position, and size of cores obtained from dendrogram. The identified cores are labeled by ascending order from 1 in descending order of the peak flux of the core.  
The final number of identified cores (hereafter $\mathcal{N}_{\mathrm{cores}}$) used for analysis is 573.  
On average, each clump contains 19 cores.
The number of identified cores $\mathcal{N}_{\mathrm{cores}}$ varies from region to region ranging from 5 to 38 (see Table~\ref{tab:core_results} for each clump). The wide range of $\mathcal{N}_{\mathrm{cores}}$ may be attributed to the differences of clump properties, the noise level (or mass sensitivity), and the linear resolutions. 
A summary of core physical properties per clump is presented in Table~\ref{tab:core_results}. 

\begin{figure*}[ht!]
\centering
\includegraphics[width=\linewidth]{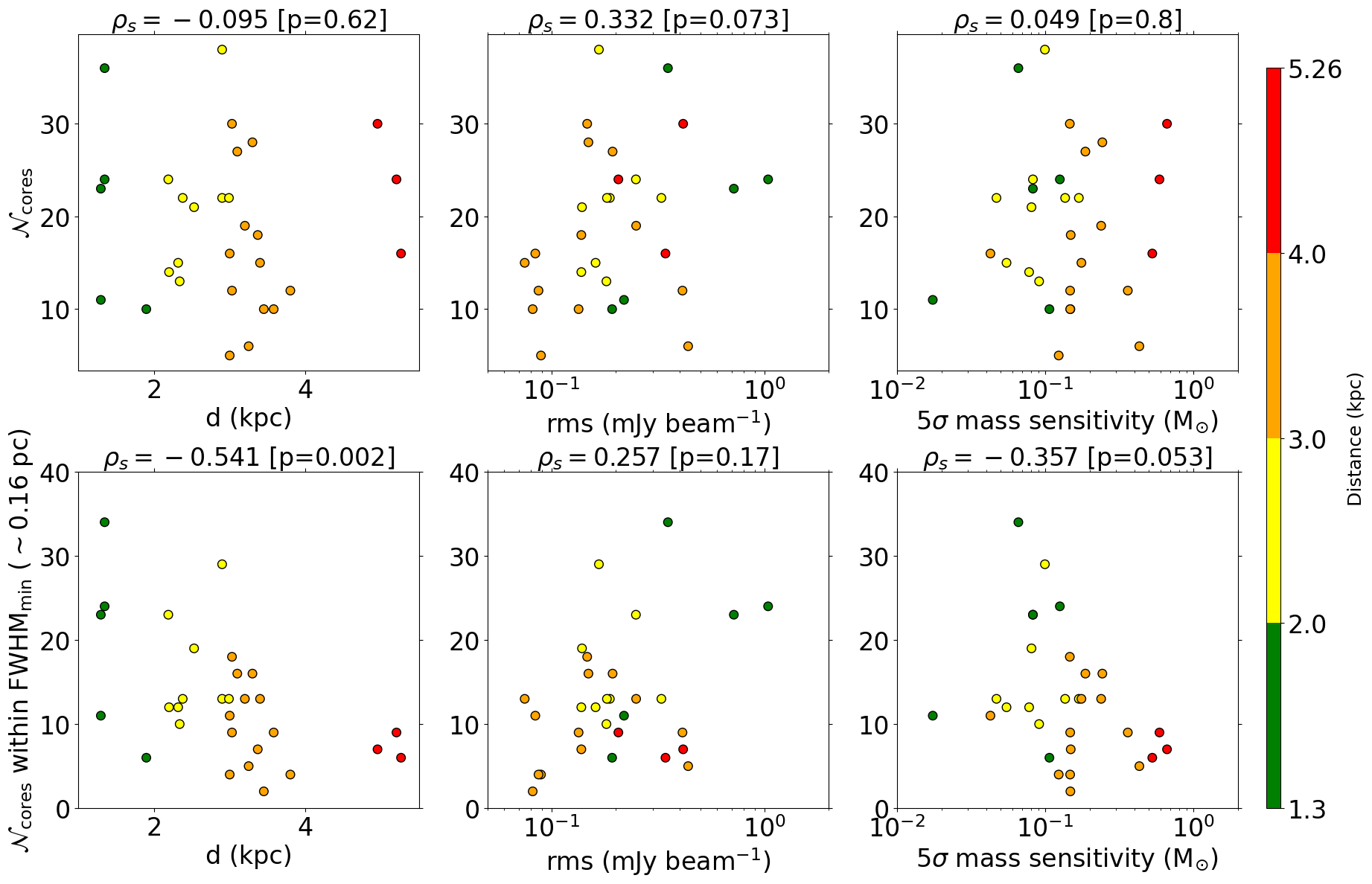}
\caption{Top panels: target distance, rms noise level and $5\sigma$ mass sensitivity against the number of identified cores.
Bottom panels: target distances, rms noise level and $5\sigma$ mass sensitivity against the number of identified cores within the same physical area whose diameter is the minimum of the FWHM ($\sim$16 pc) of the FoV in the regions.
The header of each panel shows the Spearman's rank coefficient $\rho_s$ and the p-value. 
The color coding shows the target distances.
\label{fig:correlations}}
\end{figure*}

In Figure \ref{fig:correlations}, top columns show $\mathcal{N}_{\mathrm{cores}}$ as function of distance, rms noise level, and mass sensitivity. We also present the number of identified cores, $\mathcal{N}_{\mathrm{cores}}$, within the same physical area in bottom columns. 
In this case, the total number of cores is calculated inside the circular area centered at the phase center of the observation whose diameter is the minimum FWHM among all observed targets ($\sim$25\arcsec $\sim$ 0.16 pc at 1.3 kpc). These figures indicate that the identified number of cores does not depend on the target distance, but 
the core number within same physical area strongly depends on the distance. The dependencies of $\mathcal{N}_{\mathrm{cores}}$ on rms noise level and mass sensitivity are weak.

More quantitatively, we applied the Spearman's rank correlation test.
The test provides the Spearman's rank correlation coefficient ($\rho_s$) which is a measure to estimate the strength of the correlation between two variables for the classical and non-parametric test. Non-parametric tests are valid for small sample-size and do not have to assume any distribution of the variables such as population distribution to be Gaussian. The value of $\rho_s$ is between -1 and 1. The closer $\rho_s$ is to $+1$ or $-1$, the stronger the positive or negative correlation, respectively. A value of 0 means no correlation. 
The derived values of $\rho_s$ are shown at the top of each panel of Figure \ref{fig:correlations}. 
Each correlation with $\mathcal{N}_{\mathrm{cores}}$ is weak, with $|\rho_s| \lesssim 0.3$ (Figure \ref{fig:correlations}, top panels). However, there are not statistically significant correlations because the probability value $p>0.05$ for each test, corresponding to the probability that the correlation is false relationship. On the other hand, the correlation between $\mathcal{N}_{\mathrm{cores}}$ within 0.16 pc and distance is strong, with $|\rho_s| \gtrsim 0.5$ and $p \ll 0.05$ (Figure \ref{fig:correlations}, bottom panels). This is because the closer the target clumps, the higher the spatial resolution ($\propto d$) and the mass sensitivity ($\propto d^2$). 

\subsection{Core properties}

By using the outputs from \texttt{astrodendro} analysis, we defined the physical quantities of each core. The position of each core is defined as the position of the pixel having the peak flux. The mass ($M_{\mathrm{core}}$) is calculated with Equation \ref{eq:mass}, assuming that the core temperature is the same as that of the parent clump.
The surface density, column density, and volume density are estimated, respectively, using Equations \ref{eq:surface density}, \ref{eq:peak column density}, and \ref{eq:volume density}, replacing $M_{\rm cl}$ and $R_{\rm cl}$ by $M_{\rm core}$ and $R_{\rm core}$, respectively.  The core radius ($R_{\rm core}$) was defined as half of the geometric mean of the $\rm FWHM_{major}$ and $\rm FWHM_{minor}$ provided by astrodendro\footnote{details can be seen in \url{https://dendrograms.readthedocs.io/en/stable/}}
(that is, $R_{\rm core}=1/2 \times \left( \rm FWHM_{major} \times FWHM_{minor} \right)^{0.5} = \left( 2 \ln 2 \times \sigma_{\rm major} \times \sigma_{\rm minor} \right)^{0.5}$). 
The calculated physical quantities of all the identified cores are listed in Table \ref{tab:core_catalog_sample}, and Table \ref{tab:core_results} as the average values for each clump.

%
%
\begin{deluxetable*}{rcccccccc}[htb!]
\tabletypesize{\small}
\tablecaption{Core Properties as dendrogram output \label{tab:dendro_out}}
\tablewidth{0pt}
\tablehead{
\colhead{Clump} & \colhead{Core} & \multicolumn2c{Position (ICRS)} & \colhead{Peak} & \colhead{Integrated} & 
\colhead{$\rm{FWHM_{major}}$} & 
\colhead{$\rm{FWHM_{minor}}$} & 
\colhead{Radius} \\
\cline{3-4}
\colhead{Name} & \colhead{Name} & \colhead{R.A. (J2000)} & \colhead{Decl. (J2000)} & \colhead{flux} & \colhead{flux} &&& \\
&& \colhead{(h:m:s)} & \colhead{(d:m:s)} & \colhead{(mJy beam$^{-1}$)} & \colhead{(mJy)} & \colhead{($''$)} & \colhead{($''$)} & \colhead{($''$)}
}
\startdata
G11.1-0.12 & ALMA1 & 18:10:28.25 & $-$19:22:30.37 & 6.56 & 8.50 & 0.46 & 0.20 & 0.15 \\
G11.1-0.12 & ALMA2 & 18:10:28.30 & $-$19:22:30.62 & 3.59 & 3.28 & 0.36 & 0.14 & 0.11 \\
G11.1-0.12 & ALMA3 & 18:10:28.09 & $-$19:22:35.17 & 0.93 & 2.69 & 0.75 & 0.35 & 0.26 \\
G11.1-0.12 & ALMA4 & 18:10:28.06 & $-$19:22:35.67 & 0.88 & 0.93 & 0.32 & 0.21 & 0.13 \\
G11.1-0.12 & ALMA5 & 18:10:28.19 & $-$19:22:33.77 & 0.72 & 1.90 & 0.61 & 0.37 & 0.24 \\
\enddata
\tablecomments{
The complete table is available in machine-readable form.
}
\end{deluxetable*}
%
\begin{deluxetable*}{rcccccccc}[htb!]
\tabletypesize{\small}
\tablecaption{Calculated physical Properties of identified cores \label{tab:core_catalog_sample}}
\tablewidth{0pt}
\tablehead{
\colhead{Clump} & 
\colhead{Core} & 
\multicolumn2c{Position (ICRS)} & 
\colhead{$M_{\rm{core}}$} & 
\colhead{$R_{\rm{core}}$} & 
\colhead{$n_{\rm{ave}}(\rm{H}_2)$} & 
\colhead{$\Sigma_{\rm{ave}}(\rm{H}_2)$} & 
\colhead{$N_{\rm{peak}}(\rm{H}_2)$}\\
\cline{3-4}
\colhead{Name} & 
\colhead{ID} & 
\colhead{R.A. (h:m:s)} & 
\colhead{Decl. (d:m:s)} & 
\colhead{($M_\odot$)} & 
\colhead{(au)} & 
\colhead{($\times 10^8 \rm{cm}^{-3}$)} & \colhead{($\rm{g~cm}^{-2}$)} & 
\colhead{($\times 10^{23} \rm{cm}^{-2}$)}
}
\startdata
G11.1-0.12 & ALMA1 & 18:10:28.25 & $-$19:22:30.37 & 2.36 &  450 & 7.71 & 32.59 & 32.20 \\
G11.1-0.12 & ALMA2 & 18:10:28.30 & $-$19:22:30.62 & 0.91 &  340 & 6.89 & 22.04 & 17.61 \\
G11.1-0.12 & ALMA3 & 18:10:28.09 & $-$19:22:35.17 & 0.75 &  770 & 0.50 & 3.57 & 4.55 \\
G11.1-0.12 & ALMA4 & 18:10:28.06 & $-$19:22:35.67 & 0.26 &  380 & 1.39 & 4.97 & 4.35 \\
G11.1-0.12 & ALMA5 & 18:10:28.19 & $-$19:22:33.77 & 0.53 &  710 & 0.44 & 2.94 & 3.52 \\
\enddata
\tablecomments{
The complete table is available in machine-readable form.
}
\end{deluxetable*}
\begin{deluxetable*}{lccccccccccc}[htb!]
\tabletypesize{\scriptsize}
\tablecaption{Core average properties per clump \label{tab:core_results}}
\tablewidth{0pt}
\tablehead{
\colhead{Source} & \colhead{$1\sigma$ Mass} & \colhead{$\mathcal{N}_{\mathrm{core}}$} & \multicolumn3c{$M_{\mathrm{core}}$} & \multicolumn3c{$R_{\mathrm{core}}$} & \multicolumn3c{Mean} \\
\cline{4-6} \cline{7-9} \cline{10-12} 
\colhead{Clump} & \colhead{Sensitivity} & \colhead{} & \colhead{Min} & \colhead{Max} & \colhead{\Add{Median}} & \colhead{Min} & \colhead{Max} & \colhead{\Add{Median}} & \colhead{$\Sigma_{\mathrm{ave}}(\mathrm{H_2})$} & \colhead{$N_{\mathrm{peak}}(\mathrm{H_2})$} & \colhead{$\bar{n}(\mathrm{H_2})$} \\
\colhead{$\mathrm{}$} & \colhead{($\mathrm{M_{\odot}}$)} & & \colhead{($\mathrm{M_{\odot}}$)} & \colhead{$\mathrm{(M_{\odot}}$)} & \colhead{($\mathrm{M_{\odot}}$)} & \colhead{($\mathrm{au}$)} & \colhead{($\mathrm{au}$)} & \colhead{($\mathrm{au}$)} & \colhead{($\mathrm{g~cm^{-2}}$)} & \colhead{($\times 10^{23}\,\mathrm{cm^{-2}}$)} & \colhead{($\times 10^8\,\mathrm{cm^{-3}}$)}
}
\startdata
G333.23-0.06 & 0.118 & 24 & 0.72 & 132.0 & 2.7 &  559 & 1704 &  807 & 28.7 & 35.0 & 3.3 \\
 G335.579-0.272 & 0.087 &  6 & 0.98 & 127.2 & 10.8 &  386 & 1143 &  730 & 90.5 & 155.8 & 11.8 \\
 IRAS16547-4247 & 0.020 & 38 & 0.12 & 68.8 & 0.4 &  222 & 1128 &  344 & 20.0 & 26.6 & 5.4 \\
 IRAS16562-3959 & 0.009 & 22 & 0.07 & 9.1 & 0.4 &  150 &  606 &  247 & 41.8 & 55.3 & 18.9 \\
       NGC6334I & 0.025 & 24 & 0.18 & 58.9 & 0.5 &  105 &  486 &  150 & 246.1 & 247.4 & 144.4 \\
    NGC6334I(N) & 0.013 & 36 & 0.12 & 29.3 & 0.4 &   82 &  521 &  134 & 104.0 & 168.2 & 72.4 \\
    G29.96-0.02 & 0.106 & 16 & 0.91 & 44.6 & 3.9 &  596 & 1628 &  772 & 37.5 & 36.2 & 5.0 \\
 G34.43+0.24MM1 & 0.072 & 12 & 0.75 & 107.3 & 2.1 &  346 &  953 &  486 & 60.1 & 82.8 & 10.9 \\
   G35.03+0.35A & 0.011 & 15 & 0.10 & 5.2 & 0.4 &  262 &  890 &  430 & 9.1 & 12.7 & 1.9 \\
   G35.20-0.74N & 0.017 & 24 & 0.13 & 9.5 & 0.6 &  270 &  658 &  385 & 16.1 & 23.8 & 4.0 \\
 IRAS18151-1208 & 0.009 & 16 & 0.08 & 8.3 & 0.4 &  322 & 1142 &  477 & 9.1 & 12.6 & 1.7 \\
 IRAS18182-1433 & 0.030 & 10 & 0.29 & 20.5 & 4.4 &  430 &  774 &  564 & 45.2 & 50.2 & 8.5 \\
 IRDC18223-1243 & 0.035 & 15 & 0.29 & 27.1 & 0.8 &  395 & 1539 &  581 & 10.5 & 14.6 & 1.6 \\
    G10.62-0.38 & 0.133 & 30 & 1.15 & 807.3 & 4.7 &  574 & 2452 &  864 & 35.5 & 40.3 & 3.7 \\
     G11.1-0.12 & 0.025 &  5 & 0.26 & 2.4 & 0.7 &  341 &  768 &  452 & 13.2 & 12.4 & 3.4 \\
    G11.92-0.61 & 0.030 & 18 & 0.22 & 42.2 & 0.8 &  390 & 1691 &  617 & 11.2 & 28.6 & 1.5 \\
     G5.89-0.37 & 0.034 & 22 & 0.28 & 74.7 & 2.3 &  364 & 1918 &  757 & 64.2 & 69.4 & 9.5 \\
 IRAS18089-1732 & 0.018 & 13 & 0.16 & 64.7 & 0.4 &  298 & 1174 &  430 & 21.2 & 44.4 & 3.9 \\
 IRAS18162-2048 & 0.003 & 11 & 0.02 & 7.9 & 0.0 &  144 & 1042 &  220 & 25.1 & 63.6 & 9.0 \\
           W33A & 0.016 & 21 & 0.17 & 14.6 & 1.1 &  312 & 1218 &  666 & 12.5 & 22.2 & 2.0 \\
   G14.22-0.50S & 0.021 & 10 & 0.26 & 6.3 & 0.9 &  261 &  862 &  481 & 20.0 & 30.6 & 4.7 \\
   G351.77-0.54 & 0.017 & 23 & 0.16 & 9.1 & 1.2 &  130 &  353 &  217 & 104.1 & 112.7 & 49.1 \\
    G24.60+0.08 & 0.030 & 10 & 0.37 & 11.3 & 1.8 &  377 &  852 &  496 & 24.2 & 32.3 & 5.0 \\
 IRAS18337-0743 & 0.030 & 12 & 0.19 & 4.6 & 0.6 &  410 & 1189 &  540 & 6.6 & 8.8 & 1.2 \\
   G333.12-0.56 & 0.049 & 28 & 0.33 & 28.6 & 1.7 &  419 & 1367 &  642 & 19.2 & 25.3 & 3.0 \\
   G335.78+0.17 & 0.048 & 19 & 0.39 & 44.6 & 1.5 &  407 & 1347 &  687 & 19.0 & 37.0 & 2.8 \\
   G333.46-0.16 & 0.027 & 22 & 0.20 & 35.0 & 0.8 &  322 & 1468 &  535 & 12.4 & 17.2 & 2.2 \\
   G336.01-0.82 & 0.037 & 27 & 0.22 & 15.6 & 1.6 &  350 & 1047 &  558 & 21.9 & 23.4 & 4.4 \\
 G34.43+0.24MM2 & 0.029 & 30 & 0.23 & 11.5 & 1.5 &  331 &  943 &  582 & 17.9 & 23.5 & 3.5 \\
    G35.13-0.74 & 0.016 & 14 & 0.11 & 6.2 & 0.7 &  241 &  693 &  340 & 32.7 & 42.6 & 9.6 \\
\enddata
\tablecomments{$\mathcal{N}_{\mathrm{core}}$, $M_{\mathrm{core}}$, $R_{\mathrm{core}}$, $\Sigma_{\mathrm{ave}}(\mathrm{H_2})$, $N_{\mathrm{peak}}(\mathrm{H_2})$ and $\bar{n}(\mathrm{H_2})$ correspond to the number of cores, core mass, core radius, core average values of mean surface density, peak column density, and mean volume density.}
\end{deluxetable*}
%
%
%
%

\begin{figure}[htb!]
\centering
\includegraphics[width=\linewidth]{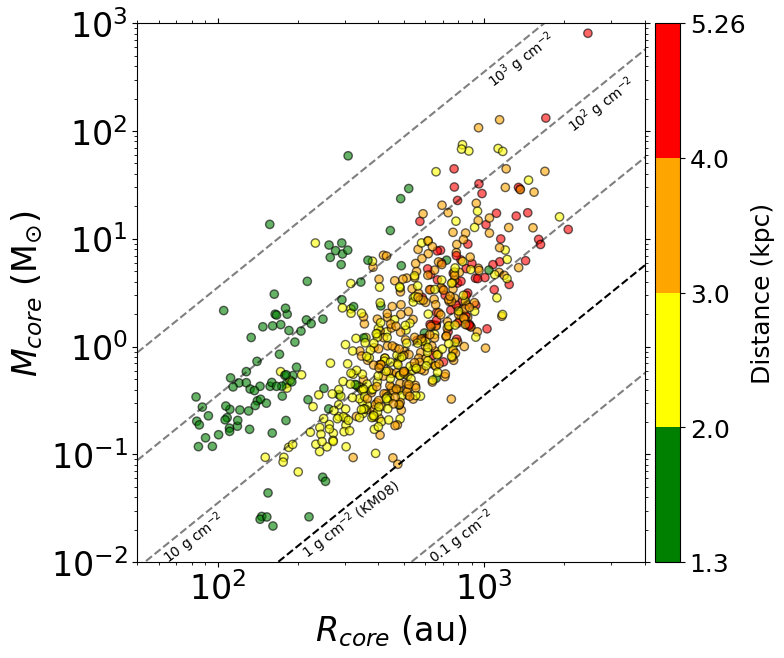}
\caption{Masses versus radius for all identified cores.
Each dotted line represents a constant column density. The dashed line labeled KM08 represents the theoretical threshold of 1 g cm$^{-1}$ suggested by \citet{Krumholz_2008}, which corresponds to the column density $\sim 2.1 \times 10^{23}$ cm$^{-2}$.
The color coding shows the target distances.
\label{fig:core_R-M}}
\end{figure}

Figure \ref{fig:core_R-M} shows the relationship between the radius and mass of all identified cores color-coded by the distance with the plotted lines of constant column densities between $10^{-1}-10^{3}$ g cm$^{-2}$, including the theoretical threshold of 1 g cm$^{-2}$ suggested by \citet{Krumholz_2008} as the minimum surface density necessary for the formation of high-mass stars.  
Almost all cores satisfy this threshold. The mean surface density range is mostly distributed between $\sim 1-100~\mathrm{g~cm}^{-2}$ and some of the most massive cores have $\geq 100~\mathrm{g~cm}^{-2}$. There is a clear trend between radius and distance due to the broad distance range.
The cores identified in clumps with the closest distances, $\lesssim$ 2 kpc, have somewhat higher column densities than those of cores in other clumps (see Table \ref{tab:sources}).

Note that several clumps contain hot cores or \ion{H}{2} region (e.g., IRAS18162-2048 and G5.89-0.37) and there is a discrepancy between clump and core average temperatures. 
This leads to an overestimation of the core mass and density, specially in the hot cores with the highest temperatures.  
Indeed, \citet{Taniguchi_2023} found 44 hot cores using the CH$_3$CN line.
The derived excitation temperatures, which can represent the gas and dust temperature of the hot cores under LTE conditions, are larger than 70 K (see also Table 3 in \citet{Taniguchi_2023}).

In the current work, we mostly focus on the core separation and leave a more detailed analysis including the core mass once temperatures at high-resolution are better constrained for the larger sample.  
For the moment we note that the distribution of core masses is comparable to the estimated thermal Jeans mass of the clumps, which is a result consistent with the subsequent discussion on thermal Jeans length.

%
%

\subsection{Core separations}
\label{sec:core_sep}

The separation of the identified cores is most likely to reflect the fragmentation process.
To measure the separations between cores, we applied the geometric algorithm, Minimum Spanning Tree \citep[MST;][]{Gower_1969} to the identified cores. This is one of the geometrical algorithms in graph theory. The connection of graph vertices with no closed loops is called a spanning tree so that given $n$ vertices, a spanning tree has $n-1$ edges. 
In the case of fragmentation analysis, vertices and edges correspond to core positions and separations, respectively. 
This analysis has been validated to detect multiple length-scales of hierarchical structures \citep[e.g.,][]{Clarke_2019}.
The separation between the nearest neighboring cores, the length of an edge, is regarded as that  produced by the fragmentation from the parental structure, following previous studies \citep[e.g.,][]{Beuther_2018, Sanhueza_2019, Lu_2020, Zhang_S_2021,Morii_2024}.
Here, we used the peak position of each core.

\begin{figure*}[htb!]
    \centering
    \includegraphics[width=\linewidth]{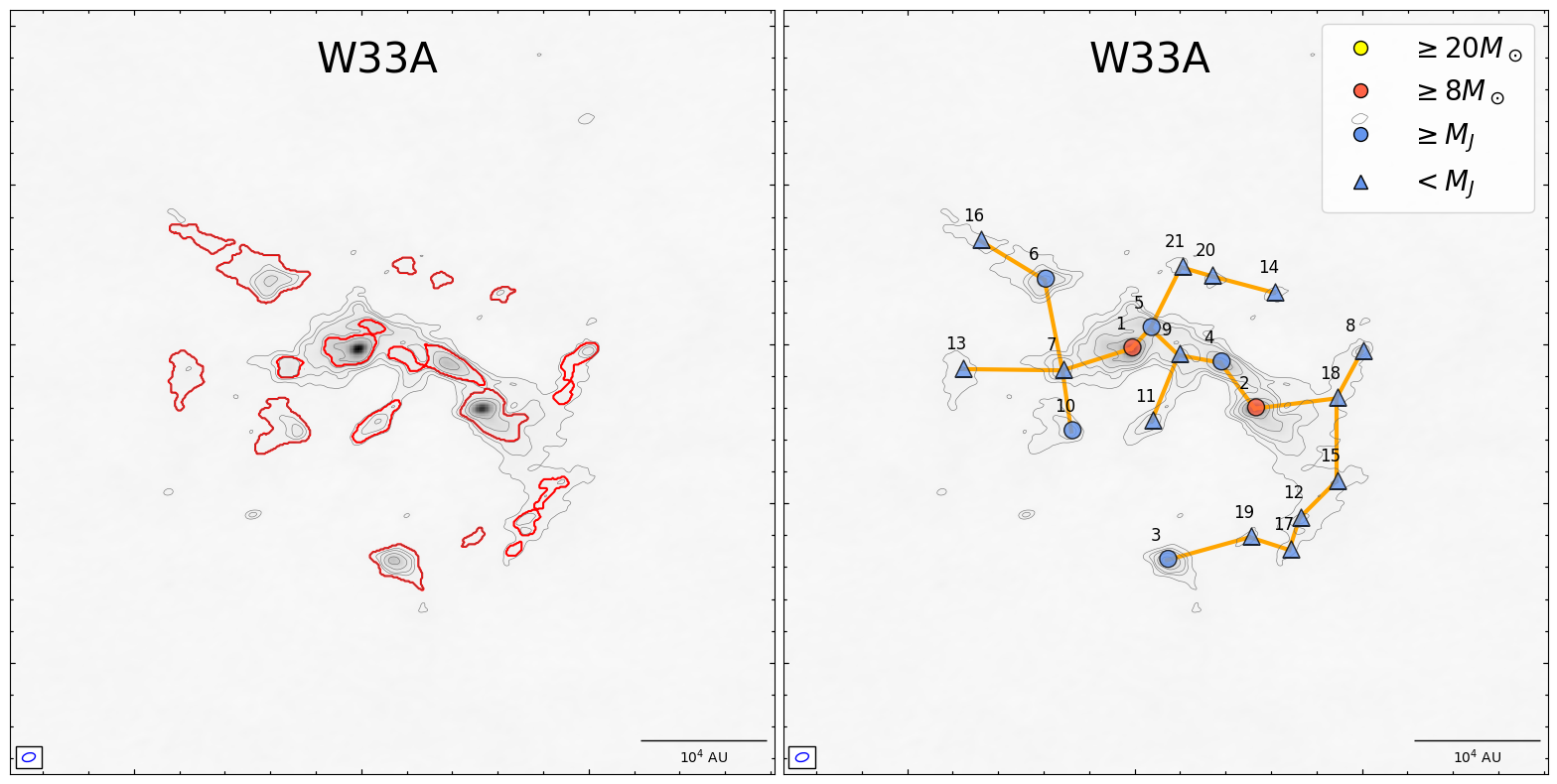} 
    \caption{
    {\it left}: Core spatial distribution with red contours identified as leaves by dendrogram. The background shows the 1.33 mm continuum image with the contour levels at [5, 10, 15, 30, 60, 90] $\times \sigma$.
    {\it right}: The left figure represented core mass and overlaid with the result of MST. 
    The number is ordered by the peak intensity.
    The orange edges are regarded as core separations for the spatial distribution.
    }
    \label{fig:mst}
\end{figure*}

In Figure \ref{fig:mst}, we show the results of the MST analysis for W33A with a distance of 2.5 kpc, as an example.  In this clump, the edges of MST appear to follow the cloud structure relatively well.
For W33A, we identified 21 cores whose masses range from 0.17 to 14.6 M$_\odot$ with a mean of 2.5 M$_\odot$. 
In this area, a significant number of cores appears to be distributed along the elongated structure running from north-east to south-west, in which 
most massive cores reside. 
The core separations derived from MST range from 1130 to 19800 au with a mean of 4840 au. 
The mean value is about 6 times larger than the beam size of 830 au.
The results from the MST analysis for other clumps are displayed in Appendix \ref{asec:targets}.

Figure \ref{fig:sep_obs_ori} shows the separation distribution and the cumulative ratio for the whole DIHCA sample. The peaks of angular and physical separation are $\sim$1\farcs48 and $\sim$4500 au, respectively, which is defined by the position at the maximum value of the logarithm probability density function (log-pdf) with a Gaussian Kernel (python module \texttt{gaussian\_kde} in SciPy)\footnote{\url{https://docs.scipy.org/doc/scipy/reference/generated/scipy.stats.gaussian_kde.html}}.
In angular scale, the peak of the separation distribution is $\sim$5 times the average angular resolution, $\sim 0\farcs3$ (Figure \ref{fig:sep_obs_ori} top panel).  
The cumulative distribution of the core separations normalized to the beam size is presented in the middle panel of Figure \ref{fig:sep_obs_ori}. A total of 80\% of core separations is resolved by more than 3 beams. If we separate the sample by distance, there are no clear difference in the distribution.

In physical scale and not yet considering projection effects (Figure \ref{fig:sep_obs_ori} bottom panel), the separation distribution seems to have a peak at $\sim$4500 au with a shoulder at around $10^3$ au and a tail at $\gtrsim 10^4$ au.

The core separation distribution appears to depend on the spatial resolution.
Once the core distribution is classified into 4 groups with different distances: $d < 2$ kpc ({\it green} histogram), $2-3$ kpc ({\it yellow}), $3-4$ kpc ({\it orange}), and $4-5$ kpc ({\it red}), we find that the shoulder at $10^3$ au comes from the targets with the closest distance ($< 2$ kpc) and the tail at larger separation is mostly explained by the  
more distant clumps ($\geq 2$ kpc).
We will discuss the effect of the spatial resolution in more detail in the following section. 
The statistics of measured separations of each region are listed in Table \ref{tab:sep_each}.

\begin{deluxetable*}{lcccccccccc}[htb]
\tabletypesize{\scriptsize}
\tablecaption{Derived Jeans parameter and measured separations of each regions.
\label{tab:sep_each}}
\tablewidth{0pt}
\tablehead{
\colhead{Source} & \colhead{$\JM$} & \colhead{$\JL$} & \multicolumn4c{$S_{3D}$ (\textit{Original})} & \multicolumn4c{$S_{3D}$ (\textit{Smoothed})}
\\\cline{4-7}\cline{8-11}
\colhead{Clump} & & & \colhead{Min} & \colhead{Max} & \colhead{Median} & \colhead{Mean}& \colhead{Min} & \colhead{Max} & \colhead{Median} & \colhead{Mean}\\
& \colhead{($M_\odot$)} & \colhead{(au)} & \colhead{(au)} & \colhead{(au)} & \colhead{(au)} & \colhead{(au)}& \colhead{(au)} & \colhead{(au)} & \colhead{(au)} & \colhead{(au)}
}
\startdata
G333.23-0.06 & 1.1 & 8100 & 3300 & 91600 & 7800 & 13300 & -- & -- & -- & -- \\
G335.579-0.272 & 1.1 & 7500 & 6200 & 47900 & 7200 & 15500 & 6200 & 9600 & 7200 & 7400 \\
IRAS16547-4247 & 1.2 & 6500 & 1300 & 13500 & 3500 & 4400 & 4300 & 20500 & 10600 & 10200 \\
IRAS16562-3959 & 3.6 & 13300 & 1500 & 38700 & 8900 & 12200 & 5800 & 20900 & 10900 & 11600 \\
NGC6334I & 0.5 & 2500 & 500 & 5500 & 1500 & 2100 & 5600 & 7300 & 7000 & 6600 \\
NGC6334I(N) & 0.5 & 3400 & 600 & 14000 & 1500 & 3000 & 3200 & 14400 & 5600 & 6800 \\
G29.96-0.02 & 2.6 & 11300 & 4300 & 89400 & 12200 & 18500 & -- & -- & -- & -- \\
G34.43+0.24MM1 & 0.8 & 5400 & 3000 & 17200 & 6000 & 8800 & 2900 & 17300 & 9600 & 9700 \\
G35.03+0.35A & 2.0 & 9800 & 2800 & 8000 & 5000 & 5000 & 3000 & 10000 & 7500 & 6800 \\
G35.20-0.74N & 1.7 & 8900 & 2400 & 33700 & 4100 & 5700 & 3200 & 14300 & 6100 & 6200 \\
IRAS18151-1208 & 2.7 & 12100 & 3200 & 41500 & 6900 & 11600 & 3200 & 50900 & 6800 & 12700 \\
IRAS18182-1433 & 1.5 & 9400 & 3500 & 20700 & 5800 & 7500 & 3500 & 20700 & 5800 & 7500 \\
IRDC18223-1243 & 0.8 & 10000 & 2300 & 26200 & 5200 & 7800 & 2800 & 26200 & 5200 & 8200 \\
G10.62-0.38 & 1.1 & 5600 & 3400 & 31300 & 9800 & 10900 & -- & -- & -- & -- \\
G11.1-0.12 & 1.4 & 13400 & 2700 & 13400 & 5100 & 6600 & 2700 & 13400 & 10400 & 9200 \\
G11.92-0.61 & 2.0 & 13800 & 4000 & 30700 & 7700 & 9100 & 4000 & 30700 & 9300 & 10400 \\
G5.89-0.37 & 1.3 & 6000 & 3200 & 12600 & 6000 & 6800 & 3700 & 14200 & 7100 & 7700 \\
IRAS18089-1732 & 1.0 & 6700 & 3300 & 27300 & 7100 & 9100 & 3300 & 27100 & 9900 & 11500 \\
IRAS18162-2048 & 1.9 & 7200 & 900 & 8700 & 1600 & 3000 & 6200 & 11200 & 8700 & 8700 \\
W33A & 1.3 & 8200 & 2800 & 10200 & 6400 & 6300 & 3500 & 12400 & 7800 & 7900 \\
G14.22-0.50S & 0.8 & 8200 & 3900 & 20000 & 9300 & 10200 & 2300 & 24700 & 13800 & 13900 \\
G351.77-0.54 & 0.5 & 2400 & 800 & 3100 & 2000 & 2000 & 11300 & 11300 & 11300 & 11300 \\
G24.60+0.08 & 1.3 & 12600 & 2600 & 36900 & 8800 & 10400 & 2600 & 36800 & 8800 & 10400 \\
IRAS18337-0743 & 1.8 & 14100 & 2900 & 45200 & 7300 & 12000 & 2900 & 45200 & 7300 & 12000 \\
G333.12-0.56 & 0.7 & 7000 & 3000 & 20200 & 5700 & 6700 & 3000 & 20200 & 5800 & 6800 \\
G335.78+0.17 & 1.3 & 8500 & 2500 & 35200 & 5600 & 9300 & 2600 & 35100 & 5600 & 9300 \\
G333.46-0.16 & 1.3 & 8000 & 2500 & 26100 & 7200 & 9300 & 3800 & 30200 & 9000 & 10900 \\
G336.01-0.82 & 0.8 & 6000 & 2600 & 25700 & 6400 & 7400 & 3000 & 26800 & 6900 & 8100 \\
G34.43+0.24MM2 & 1.2 & 9100 & 2700 & 14000 & 4900 & 5900 & 3100 & 14000 & 6200 & 6800 \\
G35.13-0.74 & 1.0 & 8000 & 1400 & 25200 & 3700 & 6200 & 5400 & 33500 & 11700 & 15600 \\
\enddata
\tablecomments{
Thermal Jeans Mass are derived as $\JM = \frac{4\pi}{3}\rho_{\rm cl} \left( \frac{\JL}{2} \right)^3$.
}
\end{deluxetable*}
%
%
%
\begin{figure}[htb!]
    \centering
    \includegraphics[width=\linewidth]{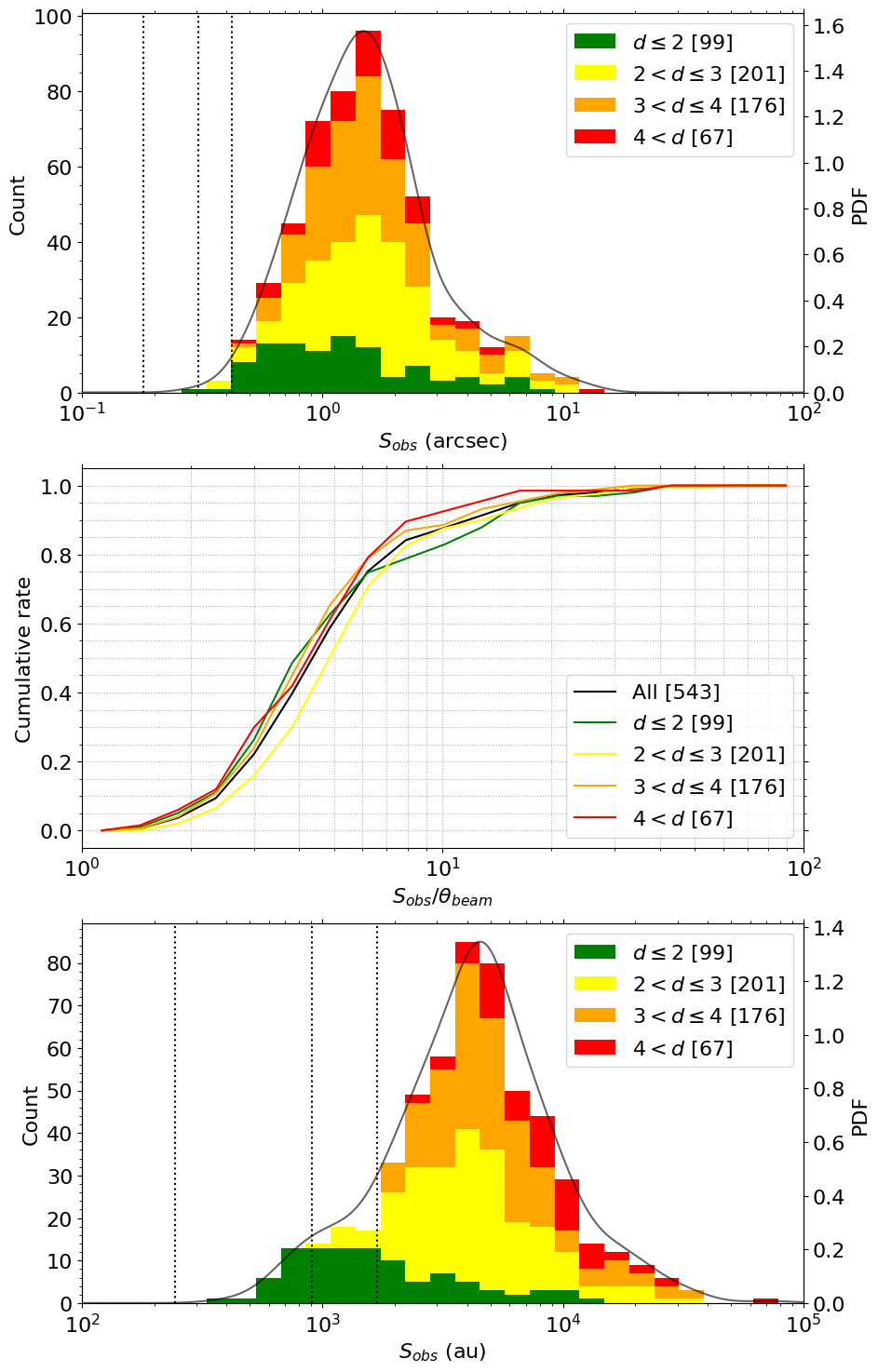}
    \caption{
    {\it top}: Distribution of the angular separation obtained from MST for the entire sample. The solid line shows the log-pdf produced by Gaussian Kernel Density Estimation. Vertical lines represent minimum, mean, and maximum angular resolutions, respectively.
    {\it middle}: Cumulative distribution of core separations divided by the angular resolution of the observations for clumps in each distance. 
    {\it bottom}: Separation distribution in au scale by converting the angular separation into a physical separation using the source distance. Vertical lines represent minimum, mean, and maximum linear resolutions, respectively.
    }
    \label{fig:sep_obs_ori}
\end{figure}
%
\begin{figure*}[htb!]
    \centering
    \includegraphics[width= \linewidth]{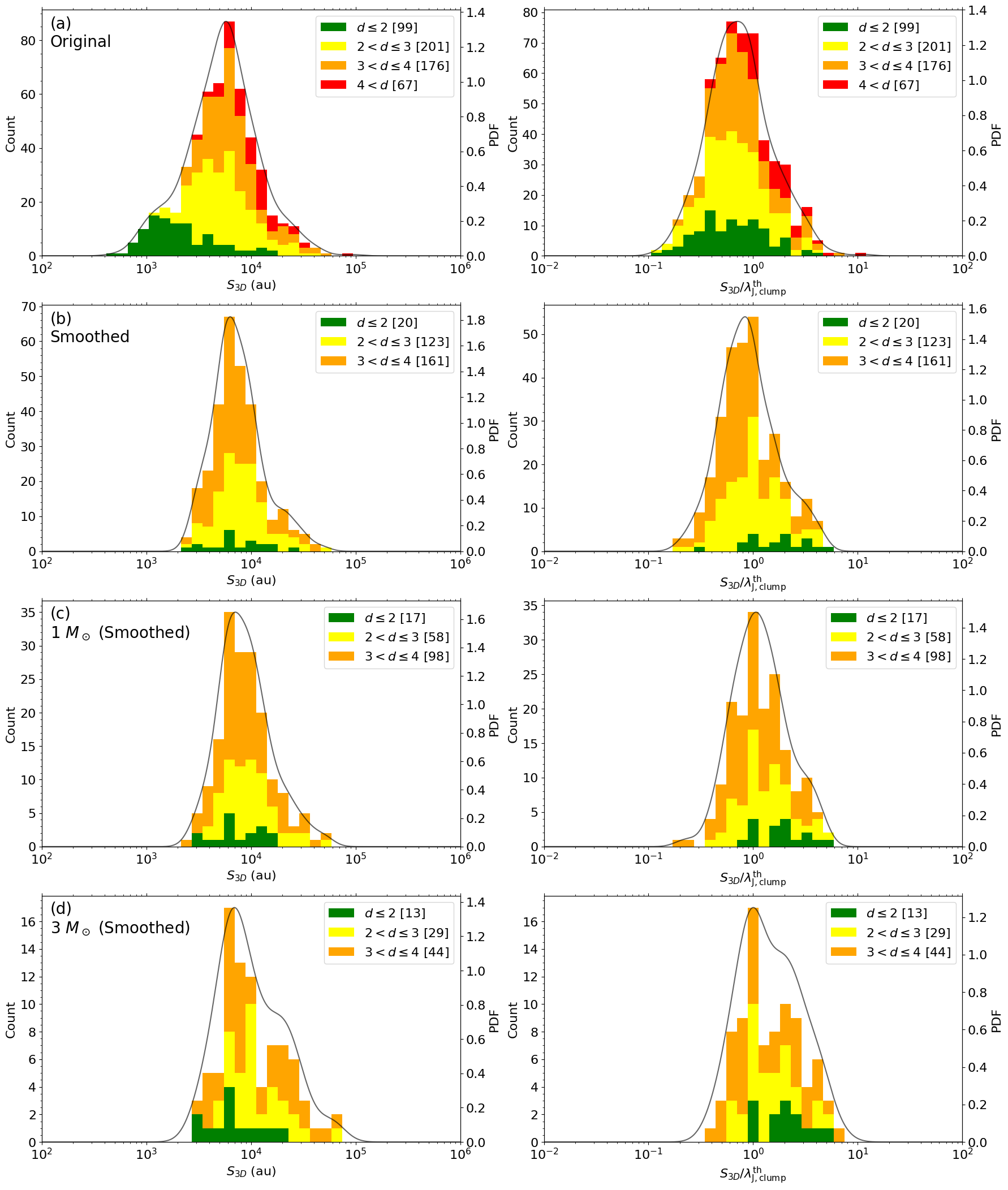}
    \caption{
    Separation distribution in various styles. Left panels display them in physical scale. Right panels display them in normalized to the clump thermal Jeans length. Each column represents as follows. Core separation measured from
    (a) original images,
    (b) smoothed images,
    (c) for cores whose mass is higher than $1~M_\odot$, and
    (d) the cores whose mass is higher than  $3~M_\odot$.
    }
    \label{fig:sep_3d}
\end{figure*}
%

\section{Discussion} \label{sec:discussion}
\subsection{Characteristic fragmentation scale}

\subsubsection{Core separation normalized to the clump Jeans length}
\label{subsec:fragmentation_scale}

We compare the core separations measured with MST to a  characteristic length scale, the clump thermal Jeans length ($\JL$).  The clump thermal Jeans lengths are calculated from the clump temperature and density listed in Table \ref{tab:sources}. The Jeans length ranges from 2000 and 14000 au. The derived values are listed in Table \ref{tab:sep_each}.

We show the core separation distribution normalized to the clump Jeans length in Figure \ref{fig:sep_3d} (a).
The left panel shows the core distribution corrected for projection effects, while the right panel is that normalized to the clump thermal Jeans length.
The projection effect on the actual 3D separation distribution has been discussed in \citet{Sanhueza_2019} and \citet{Li_2021}, for example. Assuming that the observed cores have a spherically uniform distribution, a correction factor of $4/\pi$ is applied to the projected distance $S_{obs}$. This is based on the probability density of the solid angle between core separation and the line-of-sight direction. Thus, the deprojected separation $S_{3D}$ is estimated as follows;
\begin{equation}
\label{eq:sep_3d}
S_{3D} = \frac{S_{obs}}{\frac{1}{4\pi} \int_{-\frac{\pi}{2}}^{\frac{\pi}{2}} \sin{\theta} \times 2 \pi \sin{\theta} \mathrm{d}\theta} = S_{obs} \times \frac{4}{\pi} .
\end{equation}

The peak of the separation distribution corrected by projection effects is $\sim$5800 au.  
The normalized separation appears to have a single peak at 0.71$\JL$ with no clear shoulder due to close star-forming regions. The tail at large values also becomes more subtle. The clumps with the closest distances have larger clump densities, and thus the Jeans lengths tend to be smaller.
The impact of varying the core identification parameters on the core separation distribution is discussed in Appendix \ref{asec:dedro}.

\citet{Beuther_2018} observed 20 high-mass star-forming clumps with similar properties to those in the DIHCA sample at a similar angular resolution with NOEMA. They identified 123 cores using clumpfind, finding a 3D core separation distribution with a peak at around 2500 au. To make a direct comparison with their work, we have re-analized their images and identified 129 cores using dendrograms under the same parameters we have used in the DIHCA sample. The newly derived separation distribution, corrected by projection effects, has a peak at $\sim$5600 au, consistent with the value derived in the DIHCA sample.
The reason why the difference occurred despite similar number of cores is mainly different core identification method. Despite similar parameter setting with dendrogram, clumpfind tends to identify compact and sharp sources. Thus,  separation distribution by clumpfind is distributed in smaller scale than that of dendrogram.

%
\subsubsection{The effects of different spatial resolution}
\label{sec:bias}

As mentioned in Section \ref{sec:core_sep}, the separation distribution may be affected by the spatial resolution. Because the spatial resolution is proportional to the distance of the target under similar beam-size ($\sim$0\farcs3 in this work), the broad distance range of the clumps tends to make the separation distribution more flattened. 
We minimize the effect of having different spatial resolutions in the core separation analysis by smoothing the continuum images. 
We used the CASA task \texttt{imsmooth} and smoothed the dust continuum images to the angular resolution equivalent to $\sim$1100 au. We decided to exclude three clumps that are more distant than 4 kpc (corresponding to a spatial resolution of $\gtrsim$1600 au) to keep the spatial resolution range as small as possible after smoothing. We identified the cores in the smoothed images and made the separation distribution of cores using the MST analysis with the same procedure applied in the original images (Section \ref{subsec:core identification}). As a result, the total number of identified cores is reduced from 573 in the 30 clumps to 331 in 27 clumps. Figure \ref{fig:sep_3d} (b) shows the separation distribution between cores obtained from the smoothed images. 
The separation distribution is similar to the original. 
The peak separation is about 6300 au and 0.84 $\JL$, ($\sim 8$\%) larger than the values obtained from the original images.

\begin{figure*}[htb!]
\centering
\includegraphics[width=\linewidth]{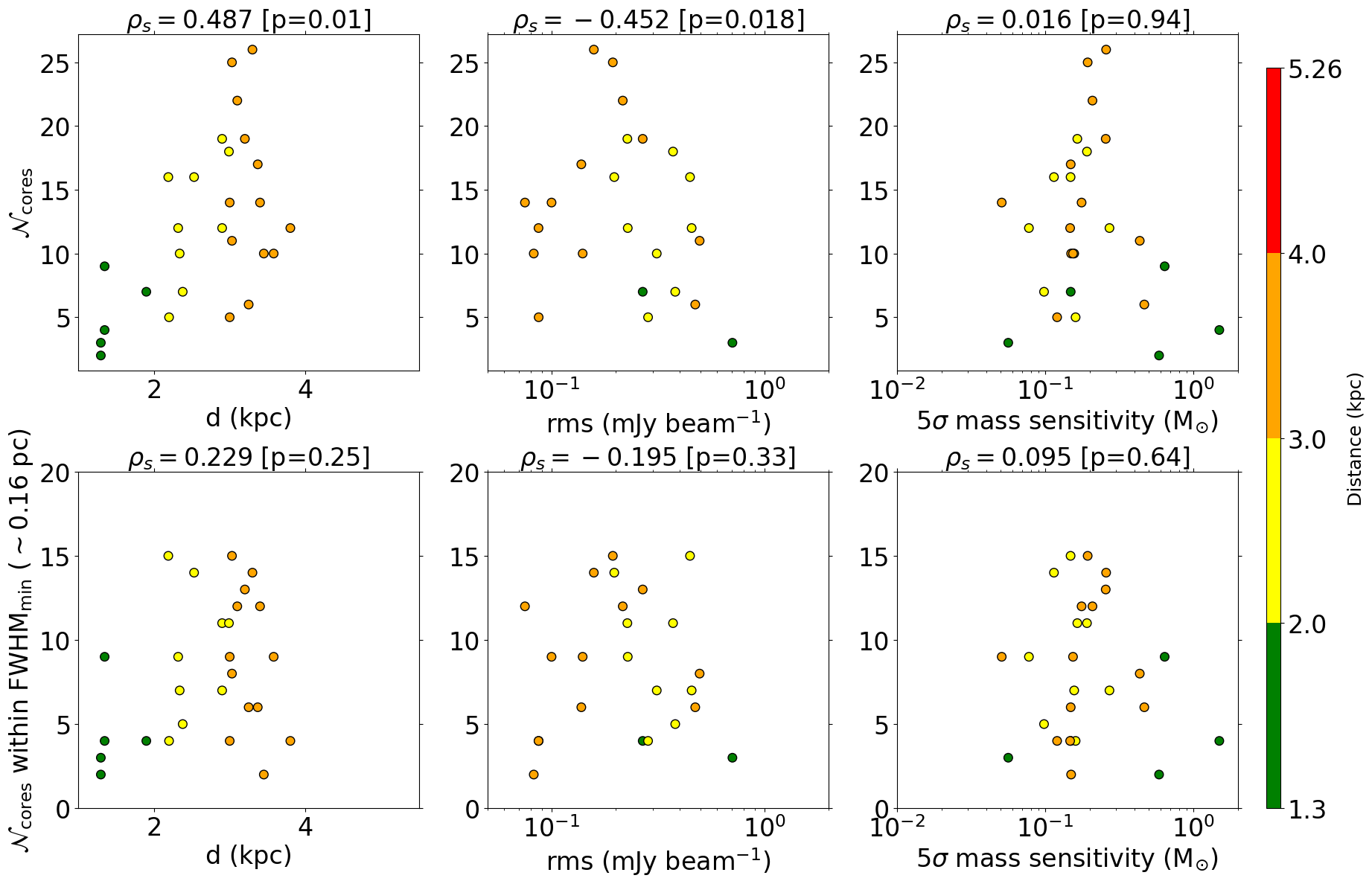}
\caption{Same as Fig. \ref{fig:correlations} except for the sample from smoothed images.
\label{fig:correlations_smooth}}
\end{figure*}

Similarly as done in Figure \ref{fig:correlations}, we check for a correlation between the number of cores from the smoothed images with observational quantities. Figure \ref{fig:correlations_smooth} shows the correlations between distance, rms noise, and mass sensitivity. For both the total number of cores and the number of cores within 0.08 pc, there is a weak or no correlation with each quantity.  

%
%
%
%

\subsubsection{The effects of completeness}
\label{sec:completeness}

\begin{figure*}[htb!]
\centering
\includegraphics[width=\linewidth]{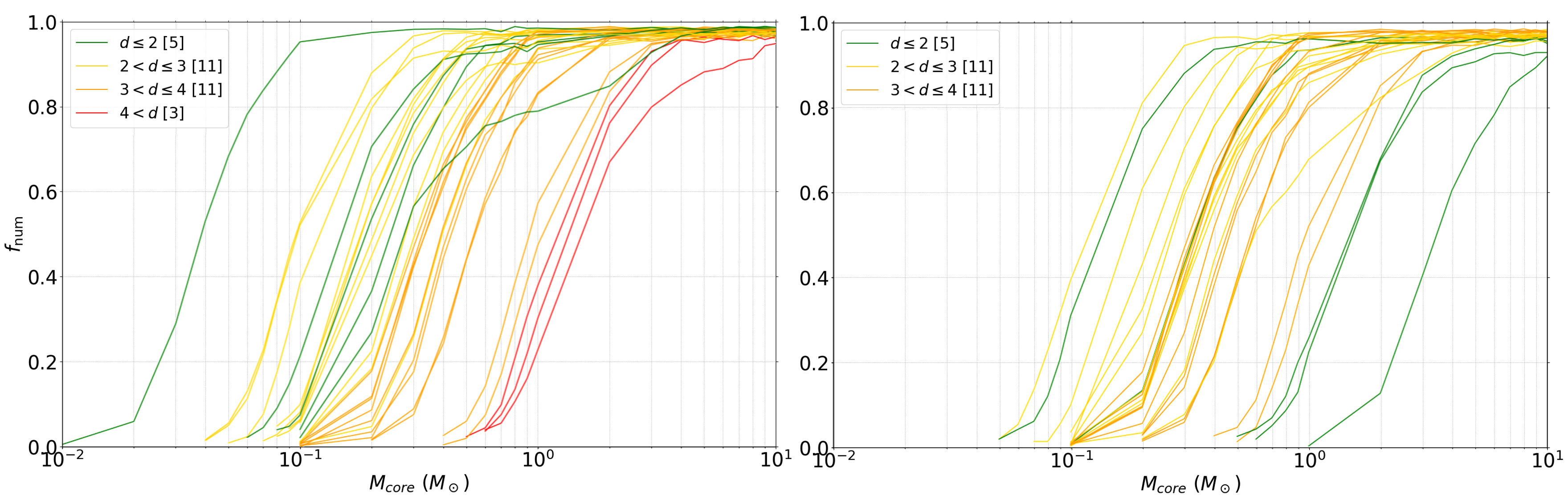}
\caption{The mass detection rate of each region for original (left) and smoothed (right) images. The 3 clumps whose distance is $\geq$4 kpc are excluded (see Section \ref{sec:bias}). The color coding is same as previous figures.
\label{fig:completeness}}
\end{figure*}

To compute the mass completeness limit, we insert a circular, beam-size, flat core in each image per trial. 
The positions of the artificial cores are determined with a uniform probability in the field of view. Then, we applied dendrogram to identify cores. We repeated this simulation 2000 times. Sometimes, the inserted core overlaps and merges with an existing core. In such a case, the core is counted as one detection.
Figure \ref{fig:completeness} shows the completeness analysis for a given mass.
The core detection rate increases with increasing core mass. For most of the clumps, the detection rate shows a similar dependence on the core mass. In a few regions, the morphology of the continuum emission affects the actual detection rate. For example, the detection rate of NGC6334I has a bump at around 0.4 M$_\odot$. The clump looks centrally condensed and the overlap of the inserted cores may somewhat reduce the detection rate. 
The G10.62-0.38 region shows the poorest mass completeness since it is one of the most distant clump in our sample ($d = 4.95$ kpc).
However, the detection rate reaches more than 90 \% by 3 M$_\odot$ for all the clumps, except for three targets that have the worst mass sensitivity (G333.23-0.06, G29.96-0.02, and G10.62-0.38, which correspond to the three clumps located at distances larger than 4 kpc).   

\begin{figure*}[htb!]
\centering
\includegraphics[width=\linewidth]{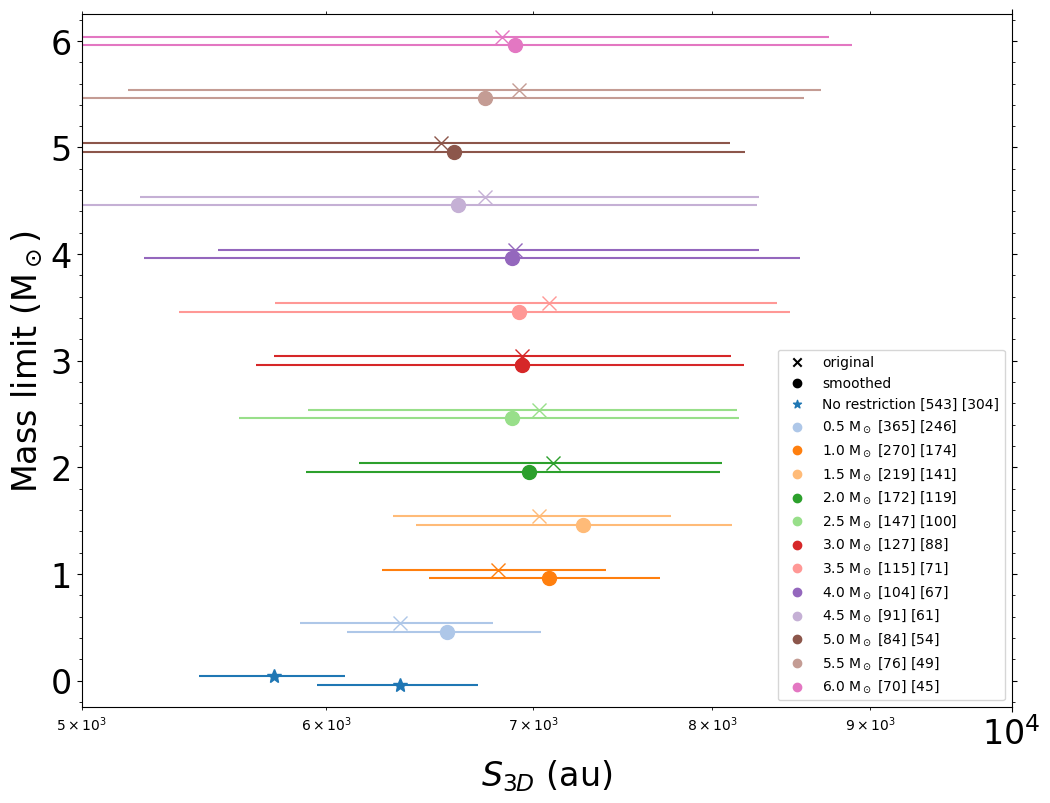}
\caption{The peak positions of probability density function (pdf) for each separation distribution which core identification was restricted by the core mass. The errorbar represents the dispersion of the separation distribution. To avoid the duplication of errorbars, the y locations are slightly shifted from actual values.
For the square brackets in the legend, the left and right values of each marker display the total sample of MST from original images and smoothed images, respectively.
\label{fig:peak_by_mass_limit}}
\end{figure*}

To explore the effect of having different mass sensitivities, we set the minimum mass for the smoothed images as $M_{\rm min}=$ 1 M$_\odot$ and 3 M$_\odot$, and apply the MST to measure the core separations.  
For $M_{\rm min} = 1$ M$_\odot$ and 3 M$_\odot$, the core separation peaked at 7200 au and 7100 au, respectively (see also Figure \ref{fig:sep_3d} (c), (d)).  
Figure \ref{fig:peak_by_mass_limit} summarizes the dependence of the core separation on the minimum core mass. The marker and errorbar illustrate the peak position and the dispersion of the separation distribution.
The higher mass restriction, the smaller sample size and the larger dispersion of the separation distribution. Except for no restriction case, all of the peak positions are within the margin of error and the peak separation appears not to be sensitive to the minimum core mass.
Thus, we conclude that the core separation peaks at around 7000 au, irrespective of the observational bias such as the different angular resolution and mass sensitivity.
The peak values of each separation distribution defined by different mass limits, as shown in Figure~\ref{fig:peak_by_mass_limit}, are listed in Table \ref{atab:sep_peak}.

%
%
%
%
%
%

\subsection{Characteristic fragmentation scale in high-mass star-forming clumps}
\label{sec:fragmentation_characteristics}
In the previous section, we concluded that the peak separation is comparable to  the thermal Jeans length of the clumps and there is no need to invoke turbulent Jeans fragmentation. 
However, numerical simulations on cluster formation suggest that 
a number of density structures are created by diverse physical processes, e.g., fragmentation, merging,  accretion, and turbulence, and such structure formation proceeds hierarchically
\citep[e.g.,][]{Klessen_2000, Bate_2012, Myers_2013, Federrath_2015, Wu_2017, Hennebelle_2019, Vazquez-Semadeni_2019}. Local turbulence tends to compress and form elongated denser substructures, such as filaments and sheets. Within these denser regions, which have weaker turbulence, fragmentation into cores is expected to take place. Filaments and sheets 
can also be generated as a result of anisotropicgravitational contraction
\citep[e.g.,][]{Burkert_2004, Heitsch_2009, Gomez_2014}. Then, protostars (or sink particles) created by fragmentation can gain additional mass from the surroundings so that their final stellar masses are determined \citep{Bate_2019, Pelkonen_2021}. Here, we consider such a case in which the turbulent clumps create hierarchical structure during their global contraction, forming sheets or filaments as substructures from which dense cores are formed by gravitational fragmentation.

Below, we compare the clump thermal Jeans length with the characteristic fragmentation scale of the filaments and sheets. 
For simplicity, we first consider an isothermal axi-symmetric cylindrical cloud in dynamical equilibrium.  The equilibrium solution was derived by \citet{Stodolkiewicz_1963}. 
According to this work, the cloud radius can be expressed as
\begin{equation}
    R_{\rm cyl} = c_s \sqrt{\frac{2}{\pi G \rho_{\rm cyl}}} \ ,
\end{equation}
where $\rho_{\rm cyl}$ is the density of the cylinder.
The fragmentation wavelength of this cylindrical cloud has been obtained by several authors \citep[e.g.,][]{Larson_1985, Inutsuka_1992}.
\citet{Stodolkiewicz_1963} obtained the critical wavelength above which the perturbation grows in time. The wavelength of the maximum linear growth rate is about twice the critical wavelength, which is about 8 times longer than the cylinder radius.
\begin{equation}
    \lambda_{\rm cyl} \simeq 8 R_{\rm cyl}
\end{equation}
If the cylinder density is $f$ times higher than the clump density, $\rho_{\rm cyl} = f \rho _{\rm cl}$, the fragmentation scale of the cylinder can be expressed as
\begin{equation}
    \lambda_{\rm cyl} \simeq 8 c_s \sqrt{\frac{2}{\pi G \rho_{\rm cyl}}} \sim 1.14 \left( \frac{f}{10} \right)^{-1/2} \JL.
\end{equation}

For an isothermal sheet, the thickness of the equilibrium sheet \citep{Spitzer_1978} can be given as
\begin{equation}
    H_{\rm sh} = \frac{c_s}{\sqrt{2 \pi G \rho_{\rm sh}}} .
\end{equation}
The fragmentation scale for the equilibrium sheet \citep[e.g.,][]{Larson_1985} can be estimated as
\begin{equation}
    \lambda_{\rm sh} \simeq 4\pi H_{\rm sh} \sim 0.89 \left( \frac{f}{10} \right)^{-1/2} \JL ,
\end{equation}
where $\rho_{\rm sh} = f \rho _{\rm cl}$. 

Numerical simulations of cluster formation show that the densities of substructures created by either local turbulent compression or anisotropic gravitational contraction are about $\gtrsim 10$ times higher, that is, $f \gtrsim 10$
\citep[e.g.,][]{Klessen_2000, Bate_2012, Bate_2019, Myers_2013, Wu_2017, Hennebelle_2019, Pelkonen_2021}.
We also estimate the average density from our ALMA data to roughly evaluate $f$. Here, we consider that the structures identified as trunks\footnote{see here for a visual explanation of the differences between leaves, branches, and trunks as defined in dendrograms: https://dendrograms.readthedocs.io/en/stable/} are the direct precursors of filaments and sheets. We then derive the average density ($\rho_{\rm trunk}$) as the total trunk mass divided by the trunk volume for each target. We calculate the volume of a trunk assuming a spherical symmetry with a radius of $\sqrt{{\rm Area}/ \pi}$, excluding trunks that have no substructures (leaves). 
For all the cases, the trunks having hierarchical structures reside near the center of the FoV. 
The ratio between the trunk radius and the Maximum Recoverable scale ranges between 0.09 and 1.08, with about 90\% of the values smaller than 1. This means that for some trunks, the determined density is a lower limit.
The values of $f = \rho_{\rm cl}/\rho_{\rm trunk}$ are estimated to be several tens with a mean value of $f\simeq 40$ and a median value of $f\simeq 27$. Therefore,  considering that for some trunks the density is a lower limit, we conclude that $f \sim 10$ is a reasonable estimation.

In such a case, the fragmentation scale of either a cylinder or a sheet is almost equal 
to the clump thermal Jeans length (within 13\%), $\lambda_{\rm cyl}, \lambda_{\rm sh} \approx 0.89-1.14 \JL$.
This analysis indicates that our peak separation is consistent with both clump thermal Jeans length and the fragmentation scale of a dense cylinder or sheet. However, it is worth noting that if the fragmentation scale is primarily determined by filament fragmentation, the filament width should be about 
7000 au / 4 = 1750 au ($\simeq 8.5\times 10^{-3}$ pc), which is about 10 times smaller than that of the proposed constant filament width ($\approx 0.1$ pc) model \citep{Andre_2014}. This implies that  filament fragmentation is not a dominant process to create dense cores or the filament width is not universal and it is much more smaller than those of nearby star-forming regions.
In reality, recent several observations found filaments with small width in Orion A
\citep[e. g.,][]{Hacar_2018, Monsch_2018, Suri_2019, Teng_2020} and Perseus \citep[e. g.,][]{Schmiedeke_2021, Chen_2022}.
At the average clump density of $10^6$ cm$^{-3}$, the Jeans length is 5600 au or 0.027 pc, and in the central denser parts of the clumps, it is much smaller.  Therefore, the Jeans length in the clumps, thus the core separation, is likely to be much shorter than that predicted by fragmentation of the 0.1 pc universal width filaments \citep{Andre_2014}. These might indicate that the filament width is unlikely to be universal in high-mass star-forming clumps.

\subsection{Comparison with previous studies}
\label{sec:comparison_previous}

We compare our results with those of similar previous studies targeting high-mass star-forming regions. The comparative studies are the following three surveys: CORE/20 clumps \citep{Beuther_2018}, SQUALO/13 clumps \citep{Traficante_2023}, and ASHES/12 clumps \citep{Sanhueza_2019}. Our sample is referred to as DIHCA. ASHES is a mosaic observation covering $\sim1$ arcmin$^2$, while the others are single-pointing observations. For CORE, we show the re-analyzed result by astrodendro (see Sect. \ref{subsec:fragmentation_scale}).

\begin{figure}[htb!]
\centering
\includegraphics[width=\linewidth]{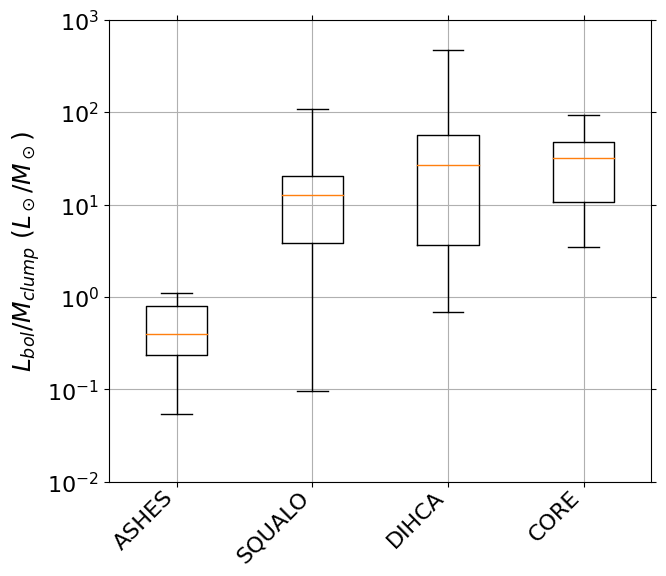}
\caption{The box plot of clump luminosity-to-mass ratio ($L_{\mathrm{bol}}/M_{\mathrm{clump}}$) for each survey. 
\label{fig:LM_plot}}
\end{figure}

\begin{figure}[htb!]
\centering
\includegraphics[width=\linewidth]{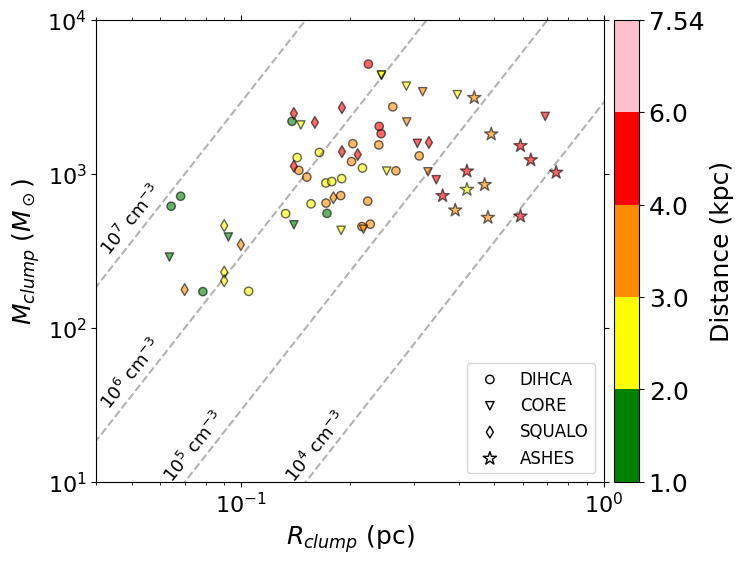}
\caption{Clump mass against radius including other surveys. The dashed lines represent corresponding average volume density assuming spherical symmetry.
The color coding shows target distance.
\label{fig:clump_M-R_comp}}
\end{figure}

Figure \ref{fig:LM_plot} shows a box plot of the clump luminosity-to-mass ratio ($L_{\mathrm{bol}}/M_{\mathrm{clump}}$) for each survey. In the ASHES survey they selected 70 $\mu$m-dark objects, representing the earliest evolutionary stage among the samples to be compared. On the other hand, 
In the CORE sample, luminous objects with $>10^4~L_\odot$ were selected, making it the most evolved sample. In SQUALO and DIHCA samples, no thresholds were considered in the luminosity or $L_{\mathrm{bol}}/M_{\mathrm{clump}}$ parameter,
and include objects at various evolutionary stages. Figure \ref{fig:clump_M-R_comp} plots the mass-size relationship of the clumps, indicating that the density of ASHES targets is $\sim10^4-10^5$ cm$^{-3}$, lower than the targets in the other surveys. 

%
%
%
%
%
%

\begin{figure}[htb!]
\centering
\includegraphics[width=\linewidth]{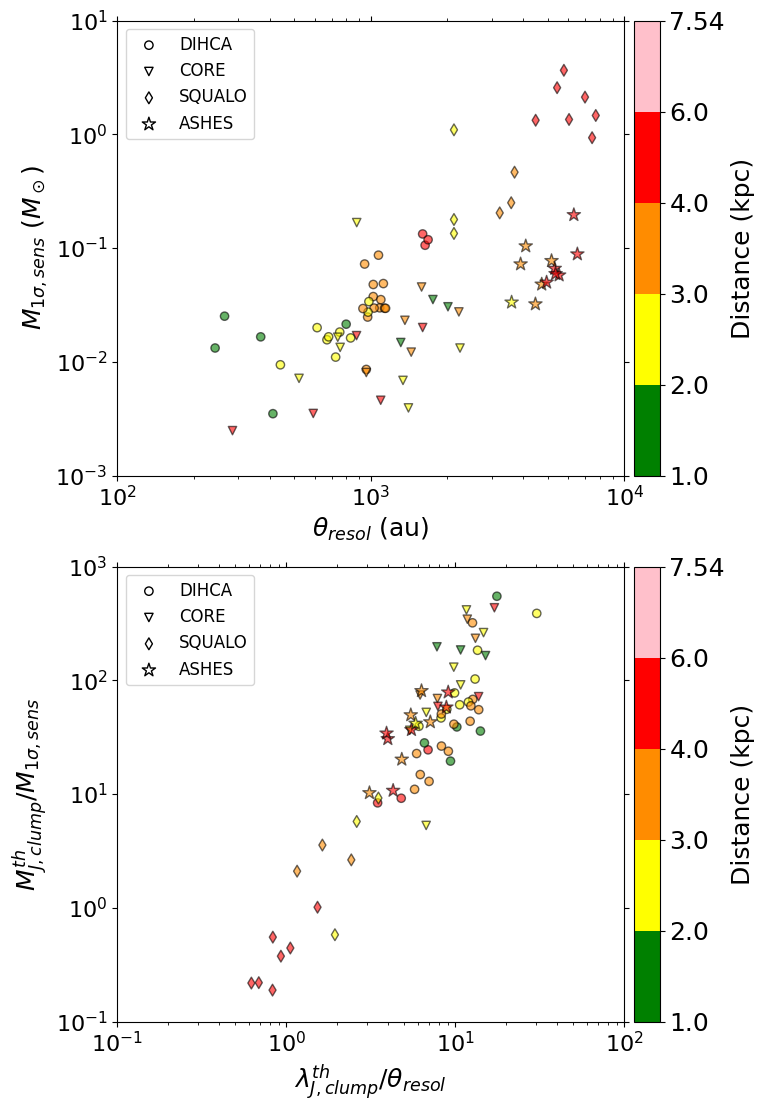}
\caption{
{\it top}: Relationship between $1\sigma$ mass sensitivities and spatial resolutions of each survey.
{\it bottom}: Relationship between thermal Jeans mass divided by $1\sigma$ mass sensitivity ($\JM / M_{\rm 1\sigma,sens}$) and thermal Jeans length divided by spatial resolution ($\JL / \theta_{\rm resol}$) of each survey.
The color coding shows target distance.
\label{fig:obs_prop_comp}}
\end{figure}

\begin{figure*}[htb!]
\centering
\includegraphics[width=\linewidth]{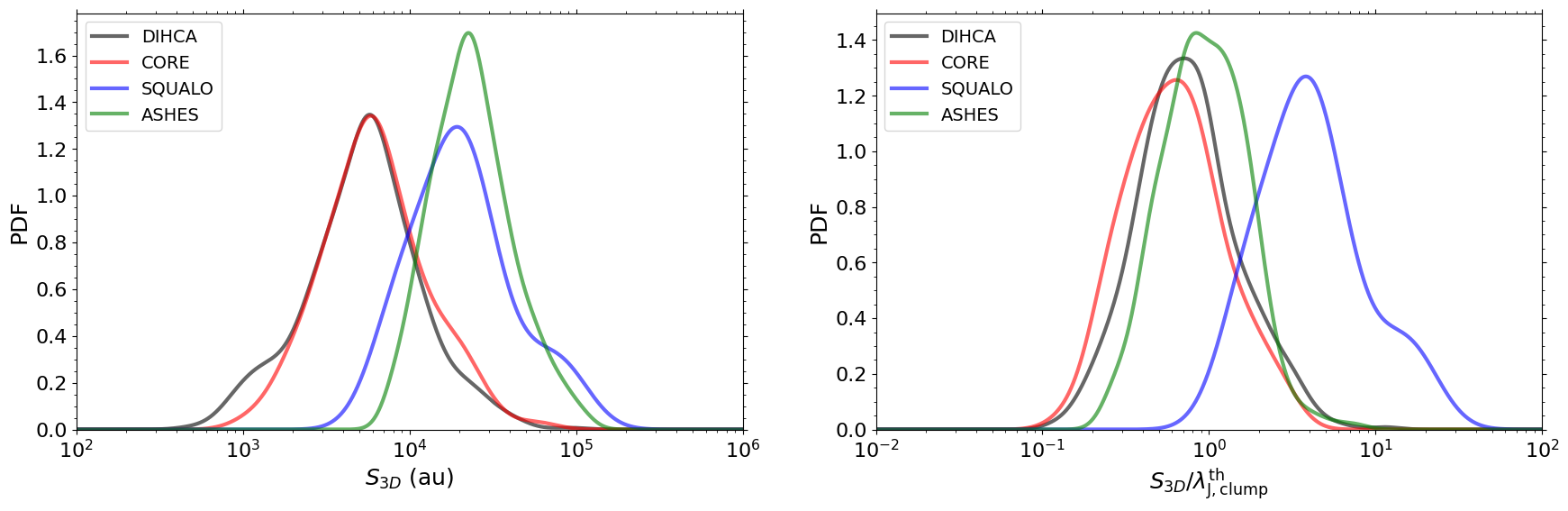}
\caption{Separation distribution in physical scale (left) and in normalized to clump thermal Jeans length (right).
\label{fig:sep_3d_comp}}
\end{figure*}

The upper panel of Figure \ref{fig:obs_prop_comp} plots the relationship between the $1\sigma$ mass sensitivity and spatial resolution for each survey. DIHCA and CORE have spatial resolutions up to $\sim2000$ au, while SQUALO and ASHES have resolutions of $\sim2000-8000$ au. The mass sensitivity of SQUALO is $\gtrsim 0.1 M_\odot$, while the others have $\lesssim 0.1 M_\odot$. The lower panel plots the ratio of the estimated thermal Jeans mass to the $1\sigma$ mass sensitivity and the ratio of the thermal Jeans length to the spatial resolution. This also shows that SQUALO can only resolve the thermal Jeans mass by $\lesssim 10$ and the thermal Jeans length by $\lesssim 3$, which is nearly an order of magnitude smaller than the other surveys.
Our tests in Section \ref{sec:bias} and \ref{sec:completeness} showed that the separation distributions of the nearby clumps (green histograms) in Figure \ref{fig:sep_3d_comp} differ significantly between (a) original and (b), (c), (d) smoothed, suggesting that the effects of insufficient resolution and mass sensitivity are reflected. It is highly likely that a similar phenomenon is occurring in the separation distribution of SQUALO. Thus, we do not seek a physical meaning in the separation distribution of SQUALO.
For the calculation of thermal Jeans mass and thermal Jeans length in each survey, we used the mass, radius, and temperature from Table 1 of \citet{Sanhueza_2019} for ASHES, and from Table 3 of \citet{Traficante_2023} for SQUALO. For CORE, we estimated the mass and radius using the same method as ours towards SCUBA 850 $\mu$m data of 17 clumps. Although the rotational temperature of H$_2$CO emission lines has been estimated as gas kinetic temperature by \citet{Beuther_2018}, the temperature range is high at $\sim40-160$ K except for one region, which deviates from the clump temperatures of other surveys. Therefore, we assumed a temperature of 23.2 K, which is the median temperature of DIHCA clumps. Indeed, \cite{Izumi_2024} find evidence that the H$_2$CO kinetic temperature is not tracing ambient gas, but it is rather tracing the molecular outflows. 

%
%
%
%
%
%

Figure \ref{fig:sep_3d_comp} shows the log-pdf of the separation distribution for each survey (same as Figure \ref{fig:sep_3d}). Comparing the distribution of physical scales in the left panel, it is clear that the distributions of CORE and DIHCA are distinctly different from those of ASHES and SQUALO. These differences in distribution are thought to be mainly due to differences in spatial resolution and evolutionary stage. Spatial resolution determines the lower limit of resolvable separation and inevitably affects the separation distribution. Regarding the difference in evolutionary stage, the effect of tighter separation due to gravity as evolution progresses may also be reflected \citep[e.g.,][]{Traficante_2023, Xu_2024}. Interestingly, the peak position of DIHCA and CORE is $\sim$6000 au. This value has also been found in previous studies such as \citet{Xu_2024} (ASSEMBLE survey), \citet{Tang_2019} (W51 North), \citet{Palau_2018} (OMC-1S), and \citet{Lu_2020} (CMZ), suggesting the existence of a universal fragmentation scale.

Looking at the distribution of scales normalized by the thermal Jeans length in the right panel of Figure \ref{fig:sep_3d_comp}, CORE, DIHCA and ASHES are around $\sim$0.8. 
ASHES clumps are at an earlier evolutionary stage compared to CORE and DIHCA, but it shows a separation distribution that is similarly consistent with the thermal Jeans length. In this way, previous studies also show consistent results, robustly supporting the hypothesis that thermal Jeans fragmentation is dominant in $\lesssim1$ pc scale.

This result is supported by other previous studies up to $\sim1$ pc scale that focus on aspects other than the fragmentation scale. \citet{Gutermuth_2011} proposed a simple picture of thermal fragmentation of high-density gas in an isothermal self-gravitating layer to explain the correlation between YSO surface density and gas column density they found in eight molecular clouds within 1 kpc. \citet{Palau_2015} compared the ``fragmentation level" with the Jeans mass and Jeans number of 19 massive dense core with a size of $\sim0.1$ pc. They found that it shows a good correlation when thermally supported rather than turbulent supported. 
Thus, the overall results of this work along with the results of previous works seem to consistently support a scenario where thermal Jeans fragmentation is at work.

\section{Summary} \label{sec:summary}
We have presented the core properties of 30 high-mass star-forming clumps obtained by the Digging into the Interior of Hot Cores with ALMA (DIHCA) survey, at a typical angular resolution of $\sim$0\farcs3. 

We have searched for dense cores using the 1.3 mm continuum emission maps and identified 579 cores in total using dendrograms. We have derived the separation between cores using the Minimum Spanning Tree (MST) technique and the obtained separation distribution that has a peak at $\sim$5800 au. 

In order to remove possible observational biases produced by the variation of spatial resolution and sensitivity in the observed sample, we smoothed the data and run completeness tests. 
We have concluded that the characteristic fragmentation scale of the observed high-mass star-forming regions, which are representative of the Galactic population, is $\sim$7000 au. 
This characteristic scale is comparable to the thermal Jeans length of the observed clumps. 

Numerical simulations of the dense cores suggest that circumstellar disks tend to fragment to form binaries or multiple systems. In that case, the separation is likely to be $<$1000 au. In a few cases, we indeed observed short separations among cores, especially in the closest  regions ($d  \lesssim 2$ kpc). Disk fragmentation is difficult to resolve from the present data used in this work, but it is possible with the extended configuration observations taken as part of the DIHCA survey. Disk fragmentation will be studied in detail in forthcoming works.


\begin{acknowledgments}
This work was supported in part by The Graduate University for Advanced Studies, SOKENDAI. PS was partially supported by a Grant-in-Aid for Scientific Research (KAKENHI Number JP22H01271 and JP23H01221) of JSPS. 
This paper makes use of the following ALMA data: ADS/JAO.ALMA\#2016.1.01036.S, ADS/JAO.ALMA\#2017.1.00237.S. 
ALMA is a partnership of ESO (representing its member states), NSF (USA) and NINS (Japan), together with NRC (Canada), MOST and ASIAA (Taiwan), and KASI (Republic of Korea), in cooperation with the Republic of Chile. The Joint ALMA Observatory is operated by ESO, AUI/NRAO and NAOJ.
Data analysis was in part carried out on the Multi-wavelength Data Analysis System operated by the Astronomy Data Center (ADC), National Astronomical Observatory of Japan.
\end{acknowledgments}

\facilities{ALMA, IPAC, Spitzer, APEX, JCMT}.
\software{CASA \citep{McMullin_2007}, astrodendro \citep{Rosolowsky_2008}, Astropy \citep{Price_2018_astropy}, SciPy \citep{Virtanen_2020_SciPy}, Matplotlib \citep{Hunter_2007_Matplotlib}}.

\clearpage
%
\appendix
\restartappendixnumbering
\section{The effect of core identification parameters} \label{asec:dedro}

Here we confirm the impact toward separation distributions in case of adopting the different core identification parameters (Section \ref{subsec:core identification}).
We tested several cases changing the minimum intensity of the core $S_{\mathrm{min}}$ and the minimum step to distinguish the neighboring structures $\delta_{\mathrm{min}}$. 
The ranges of $S_{\mathrm{min}}$ and $\delta_{\mathrm{min}}$ are between $5-7\sigma$ and $1-2\sigma$ with the step of $1\sigma$, respectively.

\begin{figure*}[htb!]
\centering
\includegraphics[width=\linewidth]{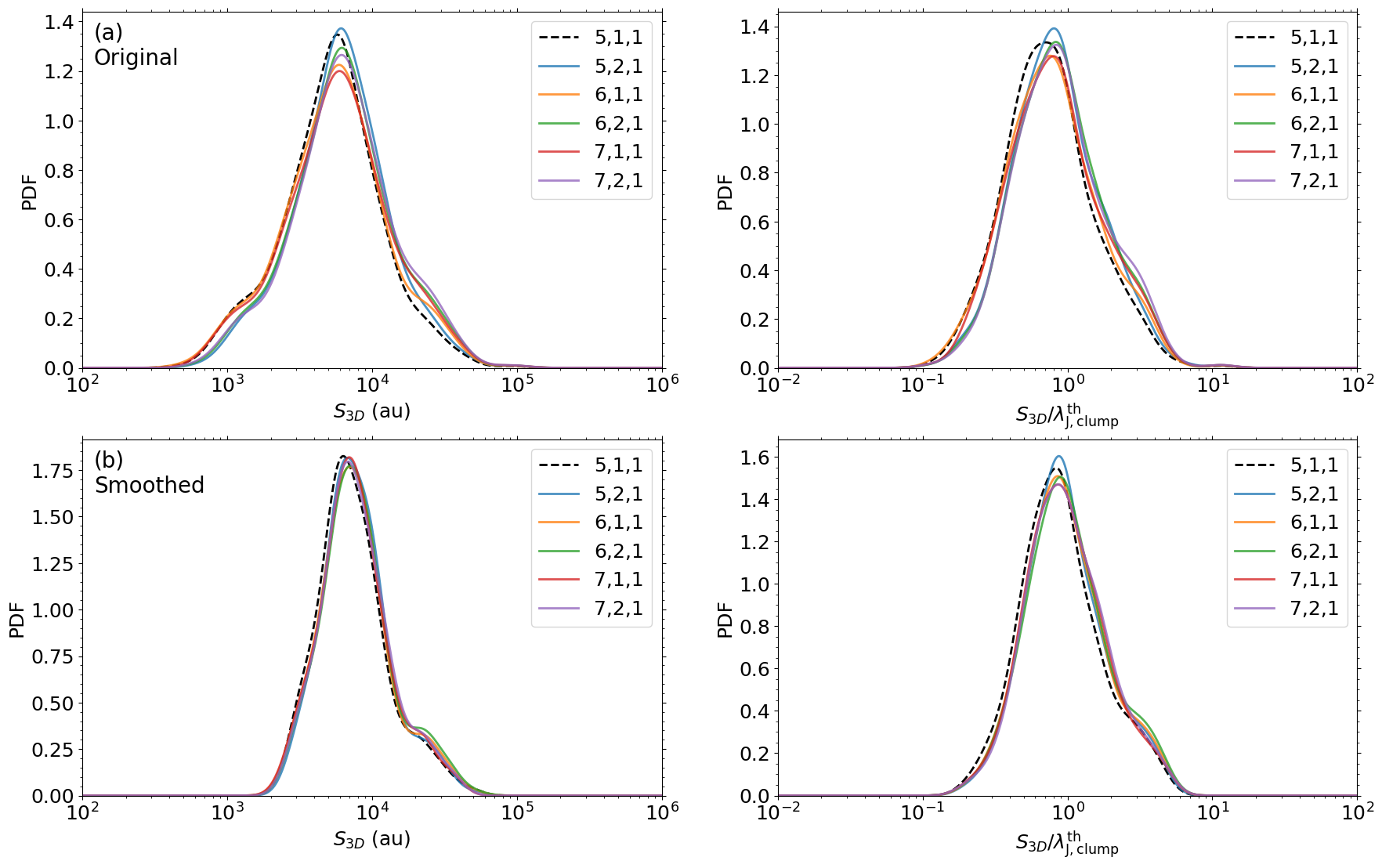}
\caption{The probability density functions (pdf) of each separation distribution for different dendrogram parameters in physical scale (left) and normalized scale (right). Core separation measured from (a) original images and (b) smoothed images, respectively.
\label{afig:sep_other_params}}
\end{figure*}

\begin{figure*}[htb!]
\centering
\includegraphics[width=\linewidth]{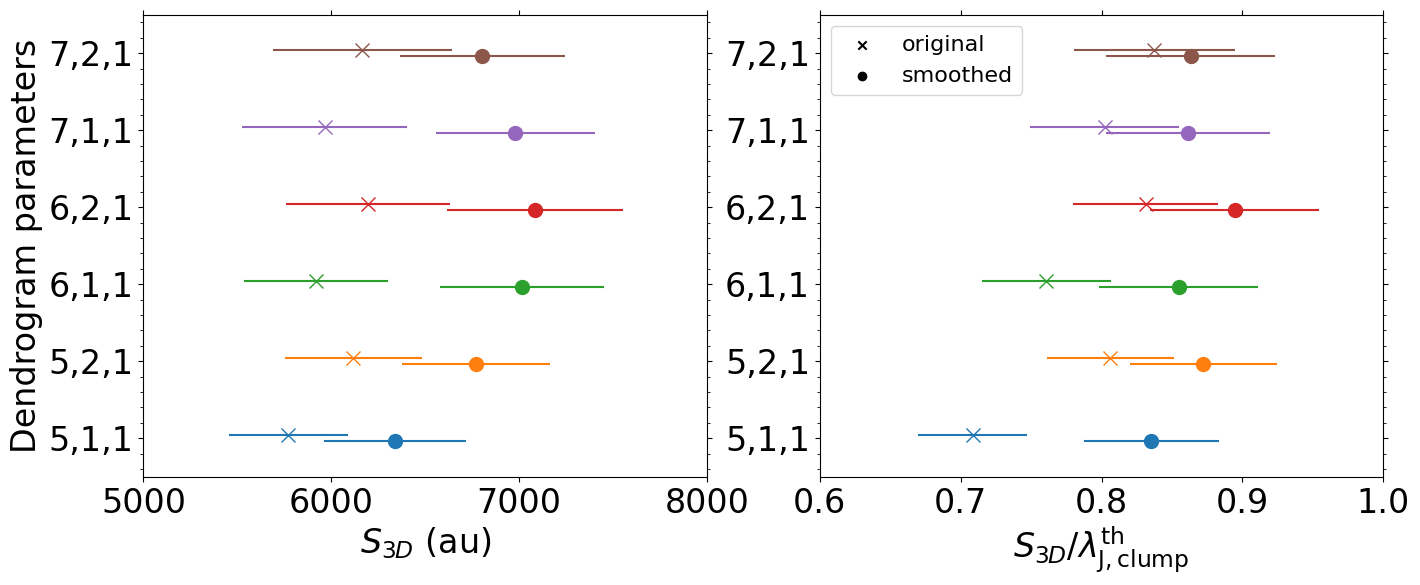}
\caption{The peak positions of probability density function (pdf) for each separation distribution \Add{whose} core identification was restricted by the core mass. The errorbar represents the dispersion of the separation distribution. To avoid the duplication of errorbars, the y locations are slightly shifted from actual values.
For the square brackets in the legend, the left and right values of each marker display the total sample of MST from original images and smoothed images, respectively.
\label{afig:peak_by_other_params}}
\end{figure*}

Table \ref{atab:sep_peak_other_params} lists the numbers of identified cores, peak positions in angular, physical, and normalized scales of each case.
Figure \ref{afig:sep_other_params} shows each separation distributions in physical scale and normalized scale. The peak positions for different dendrogram parameters is shown in Figure \ref{afig:peak_by_other_params}. In both scales, we can find that the separation distributions are almost indistinguishable, mostly keeping the same shape and peak position, for different input dendrogram parameters. 
Thus, we concluded that the difference of dendrogram parameters has no large impact on peak positions and shape of separation distributions.

\begin{deluxetable}{lcccc}[htb!]
\tabletypesize{\scriptsize}
\tablecaption{Peak values of separation distribution for each dendrogram parameters.
\label{atab:sep_peak_other_params}}
\tablewidth{0pt}
\tablehead{
\colhead{Identification} & \colhead{$\mathcal{N}_{\mathrm{core}}$} & \multicolumn3c{$S_{3D}^{peak}$}
\\\cline{3-5}
\colhead{Condition} && \colhead{(arcsecond)} & \colhead{(au)} & \colhead{($\JL$)}
}
\startdata
\textit{Original}\\\hline
(5,1,1) & 573 & 1.48 & 5800 & 0.71 \\
(5,2,1) & 486 & 1.63 & 6100 & 0.81 \\
(6,1,1) & 477 & 1.50 & 5900 & 0.76 \\
(6,2,1) & 420 & 1.58 & 6200 & 0.83 \\
(7,1,1) & 406 & 1.52 & 6000 & 0.8 \\
(7,2,1) & 374 & 1.61 & 6200 & 0.84 \\
\hline\textit{Smoothed}\\\hline
(5,1,1) & 331 & 1.67 & 6300 & 0.84 \\
(5,2,1) & 296 & 1.92 & 6800 & 0.87 \\
(6,1,1) & 274 & 1.88 & 7000 & 0.85 \\
(6,2,1) & 257 & 1.91 & 7100 & 0.89 \\
(7,1,1) & 242 & 1.83 & 7000 & 0.86 \\
(7,2,1) & 235 & 1.84 & 6800 & 0.86 \\
\enddata
\tablecomments{
For the first column, the three numbers in parentheses represent the coefficients of  $S_{\mathrm{min}}$ ($\times \sigma$, \textit{left}), $\delta_{\mathrm{min}}$ ($\times \sigma$, \textit{middle}), and $\theta_{\mathrm{min}}$ ($\times \theta_{\mathrm{beam}}$, where $\theta_{\mathrm{beam}}$ means the corresponding number of pixels contained in a synthesized beam \textit{right}), respectively.}
The peak value is defined by the position at the maximum value of the logarithm probability density function (log-pdf) with a Gaussian Kernel.
\end{deluxetable}


%
%
%
%
%
%
%
\restartappendixnumbering
\section{The effect of evolutionary processes} \label{asec:dynamical}

Here we confirm the impact toward separation distributions by evolutionary processes.
Our analysis depends on the assumption that fragmentation scale reflect on the core separations. However, in the more evolved clumps, the assumption can break because of the stellar feedback or dynamical process.
Actually G5.89-0.37 has a shell structure in the central region created by explosive event \citep{Zapata_2020,Fernandez-Lopez_2021}. The central shell is identified as 4 cores by dendrogram.
The separation distributions in such regions may not keep fragmentation scale.
To consider the effect, we derived the separation distribution without the most evolved 12 regions which are more likely to be affected by stellar feedback or dynamical process. 
The most evolved regions are selected based on the existence of extended 8$\mu$m emission which means characteristic of PDR emission produced mostly by PAHs  (Polycyclic Aromatic Hydrocarbons) emission. 
The selected 12 regions are IRAS16562-3959, NGC6334I, C35.20-0.74(N), G29.96-0.02, IRAS1812-1433, G10.62-0.38, G5.89-0.38, IRAS18162-2048, W33A, G351.77-0.54, G333.46-0.16, and G336.01-0.82. 

\begin{figure*}[htb!]
\centering
\includegraphics[width=\linewidth]{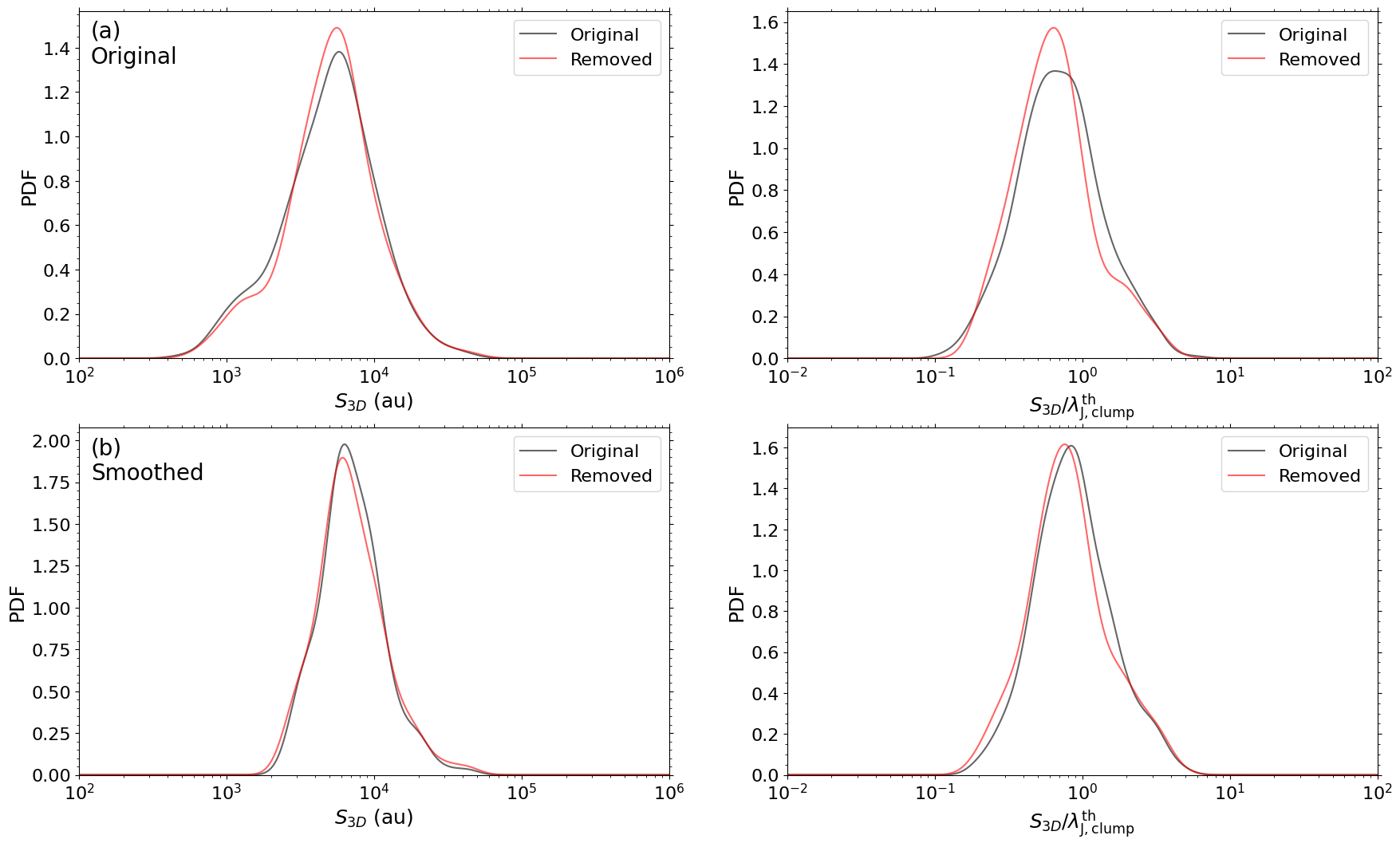}
\caption{The probability density functions (pdf) of each separation distribution for different evolutionary stages in physical scale (left) and normalized scale (right). Core separation measured from (a) original images and (b) smoothed images, respectively.
\label{afig:sep_evolution}}
\end{figure*}

Figure \ref{afig:sep_evolution} shows each separation distributions. The black line and red line represent the separation distribution of all regions and remaining 18 regions which excluded most evolved 12 regions, respectively.
We concluded that the effect of dynamical effect have no large impact on peak positions and shape of separation distributions.

%
%
%
%
\restartappendixnumbering
\section{Statistics of separations}
\label{asec:separation}
Here, Table \ref{atab:sep_peak} lists sample number and peak separation of each mass restriction for core identification \textit{Original} and \textit{Smoothed}.

\begin{deluxetable}{lcccc}[htb!]
\tabletypesize{\scriptsize}
\tablecaption{Peak values of separation distribution for each condition.
\label{atab:sep_peak}}
\tablewidth{0pt}
\tablehead{
\colhead{Identification} & \colhead{$\mathcal{N}_{\mathrm{core}}$} & \multicolumn3c{$S_{3D}^{peak}$}
\\\cline{3-5}
\colhead{Condition} && \colhead{(arcsecond)} & \colhead{(au)} & \colhead{($\JL$)}
}
\startdata
\textit{Original}\\\hline
no restriction & 573 & 1.48 & 5800 & 0.71 \\
$\geq$ 0.5 M$_\odot$ & 395 & 1.59 & 6300 & 0.87 \\
$\geq$ 1.0 M$_\odot$ & 299 & 1.69 & 6800 & 0.94 \\
$\geq$ 1.5 M$_\odot$ & 248 & 1.75 & 7000 & 1.02 \\
$\geq$ 2.0 M$_\odot$ & 201 & 1.72 & 7100 & 0.97 \\
$\geq$ 2.5 M$_\odot$ & 175 & 1.67 & 7000 & 0.94 \\
$\geq$ 3.0 M$_\odot$ & 152 & 1.65 & 6900 & 0.92 \\
$\geq$ 3.5 M$_\odot$ & 140 & 1.63 & 7100 & 0.97 \\
$\geq$ 4.0 M$_\odot$ & 129 & 1.67 & 6900 & 0.96 \\
$\geq$ 4.5 M$_\odot$ & 115 & 1.75 & 6800 & 0.94 \\
$\geq$ 5.0 M$_\odot$ & 104 & 1.71 & 6500 & 0.93 \\
$\geq$ 5.5 M$_\odot$ & 94 & 1.87 & 6900 & 0.99 \\
$\geq$ 6.0 M$_\odot$ & 87 & 1.94 & 6800 & 1.77 \\
\hline\textit{Smoothed}\\\hline
no restriction & 331 & 1.67 & 6300 & 0.84 \\
$\geq$ 0.5 M$_\odot$ & 273 & 1.8 & 6600 & 0.84 \\
$\geq$ 1.0 M$_\odot$ & 200 & 1.95 & 7100 & 1.07 \\
$\geq$ 1.5 M$_\odot$ & 167 & 2.12 & 7300 & 1.08 \\
$\geq$ 2.0 M$_\odot$ & 145 & 1.93 & 7000 & 1.04 \\
$\geq$ 2.5 M$_\odot$ & 126 & 1.9 & 6900 & 1.06 \\
$\geq$ 3.0 M$_\odot$ & 113 & 1.87 & 6900 & 1.01 \\
$\geq$ 3.5 M$_\odot$ & 96 & 1.95 & 6900 & 1.07 \\
$\geq$ 4.0 M$_\odot$ & 92 & 1.89 & 6900 & 1.05 \\
$\geq$ 4.5 M$_\odot$ & 83 & 1.82 & 6600 & 0.99 \\
$\geq$ 5.0 M$_\odot$ & 72 & 1.8 & 6600 & 1.01 \\
$\geq$ 5.5 M$_\odot$ & 66 & 1.75 & 6800 & 1.03 \\
$\geq$ 6.0 M$_\odot$ & 62 & 1.75 & 6900 & 1.02 \\
\enddata
\tablecomments{
The peak value is defined by the position at the maximum value of the logarithm probability density function (log-pdf) with a Gaussian Kernel.
}
\end{deluxetable}

\restartappendixnumbering
\section{The Individual Targets}
\label{asec:targets}
Here we illustrate clump scale single-dish images composed of submillimeter continuum data and infrared data. In addition, the results of dendrogram analysis and MST analysis are also presented. 
The complete figure set (30 images) is available in the online journal.

%
%
\figsetstart
\figsetnum{1}
\figsettitle{Individual targets}
\figsetgrpstart

\figsetgrpstart
\figsetgrpnum{D.1}
\figsetgrptitle{G333.23-0.06}
\figsetplot{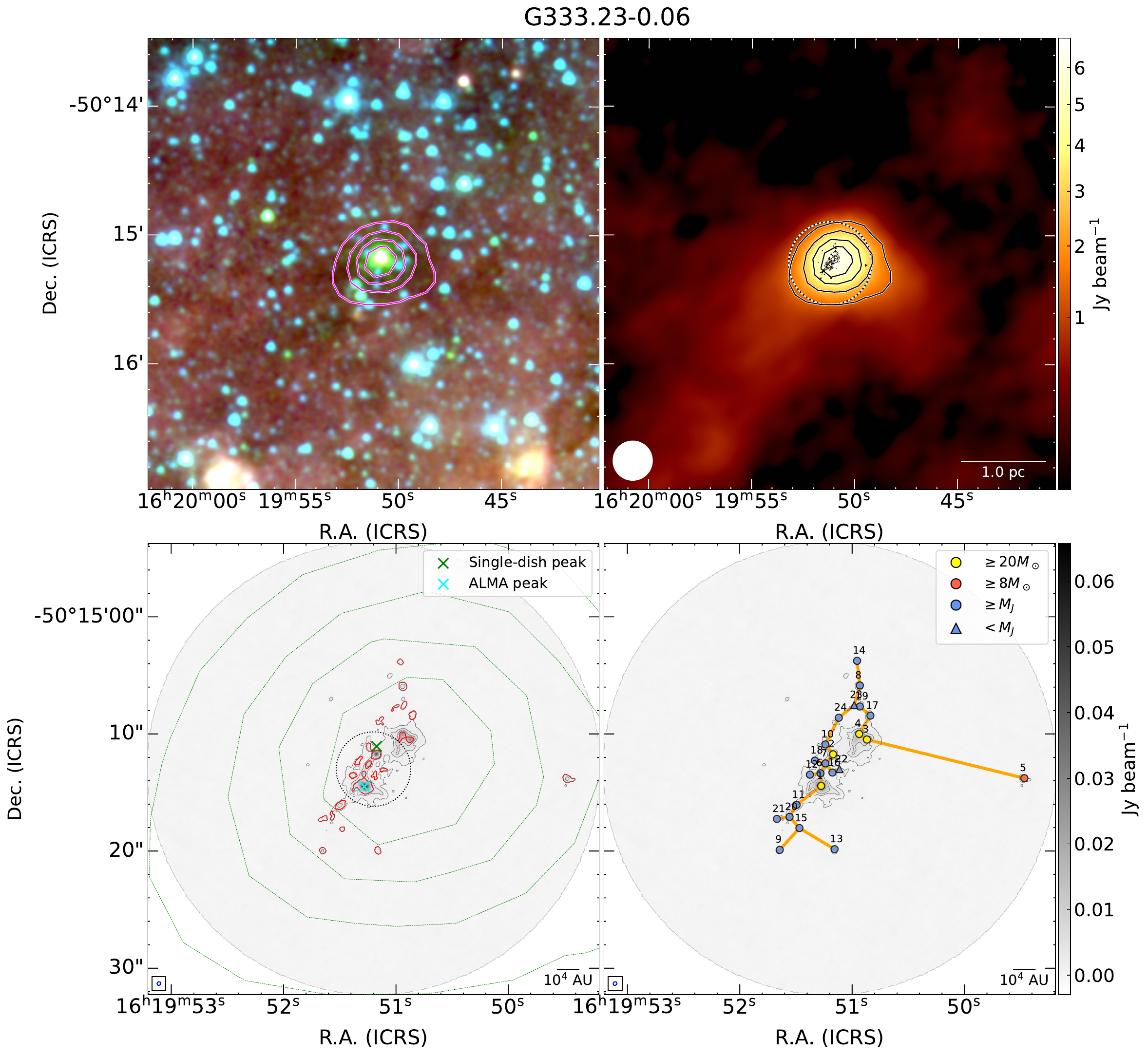}
\figsetgrpnote{
        Top left: Large-scale overview images of \textit{Spitzer}/IRAC 3.6, 4.5 and 8.0 $\mu$m images overlaid ATLASGAL 870 $\mu$m dust continuum emission contours. 
        The contour levels for the ATLASGAL 870 $\mu$m continuum are [0.2, 0.4, 0.6, 0.8] $\times$ peak flux, with peak flux $=6.9$ Jy beam$^{-1}$.
        Top right: ATLASGAL 870 $\mu$m continuum colormap and the contour same as top left.
        In addition, ALMA 1.33 mm continuum contours are plotted internal white dotted circle which represents field of view for ALMA.
        The contour levels are \Add{[5, 10, 15, 30, 60, 90]} $\times \sigma$ , with $\sigma=0.206$ mJy beam$^{-1}$.
        Bottom left: ALMA 1.33 mm continuum gray-scale map and the red contours as the results of core identification by \texttt{astrodendro}.
        Green dashed contour shows ATLASGAL 870 $\mu$m continuum same as the top panels. Green cross shows the peak position of ATLASGAL 870 $\mu$m continuum emission.
        Cyan cross shows that of ALMA 1.33 mm continuum emission.
        Bottom right: The result of core separation analysis by MST.
        }
\figsetgrpend

\figsetgrpstart
\figsetgrpnum{D.2}
\figsetgrptitle{G335.579-0.272}
\figsetplot{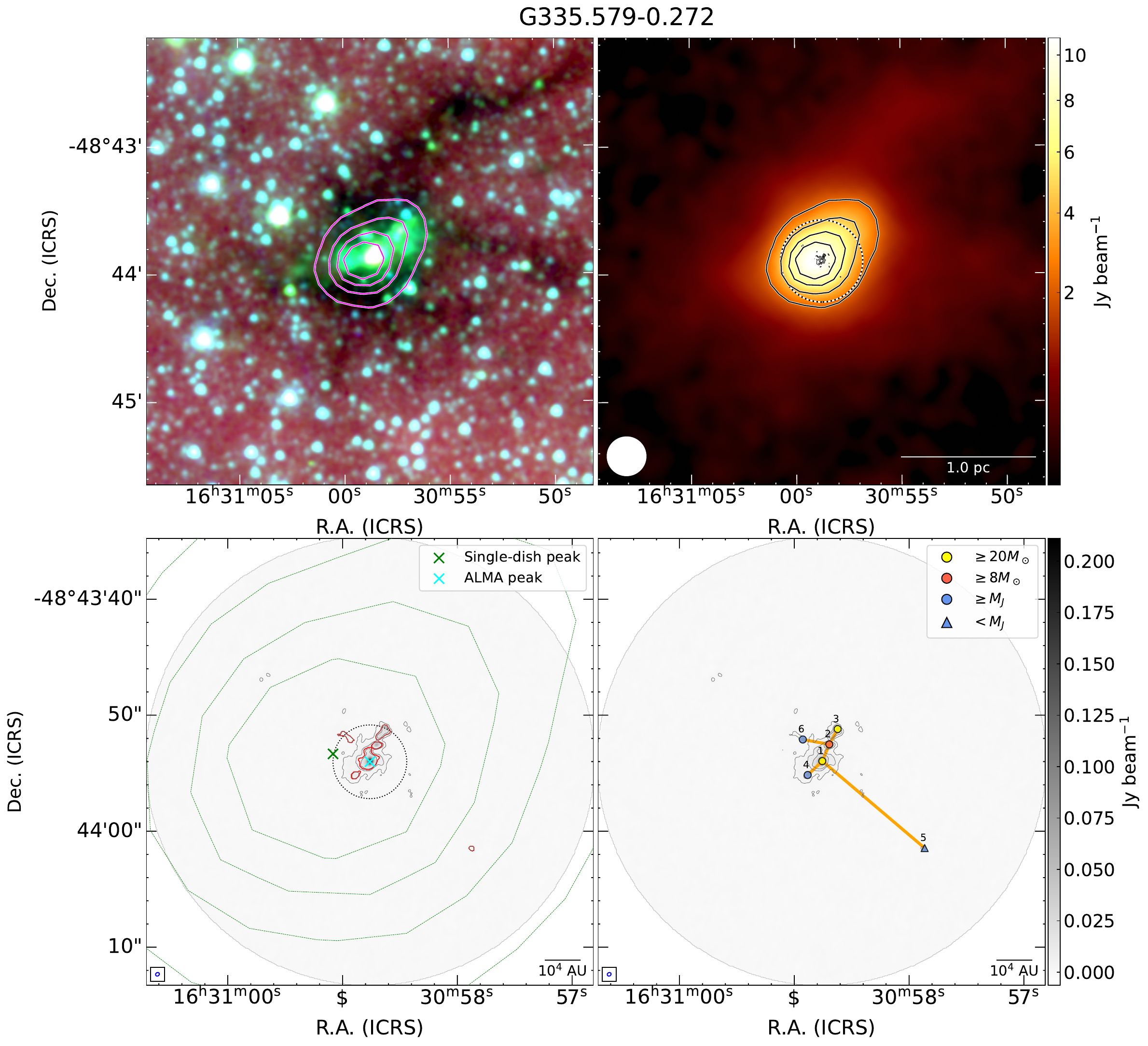}
\figsetgrpnote{Same as figure D.1, except for peak flux $=10.8$ Jy beam$^{-1}$ of ATLASGAL 870 $\mu$m continuum and 
        $\sigma=0.438$ mJy beam$^{-1}$ of ALMA 1.33 mm continuum.}
\figsetgrpend

\figsetgrpstart
\figsetgrpnum{D.3}
\figsetgrptitle{IRAS16547-4247}
\figsetplot{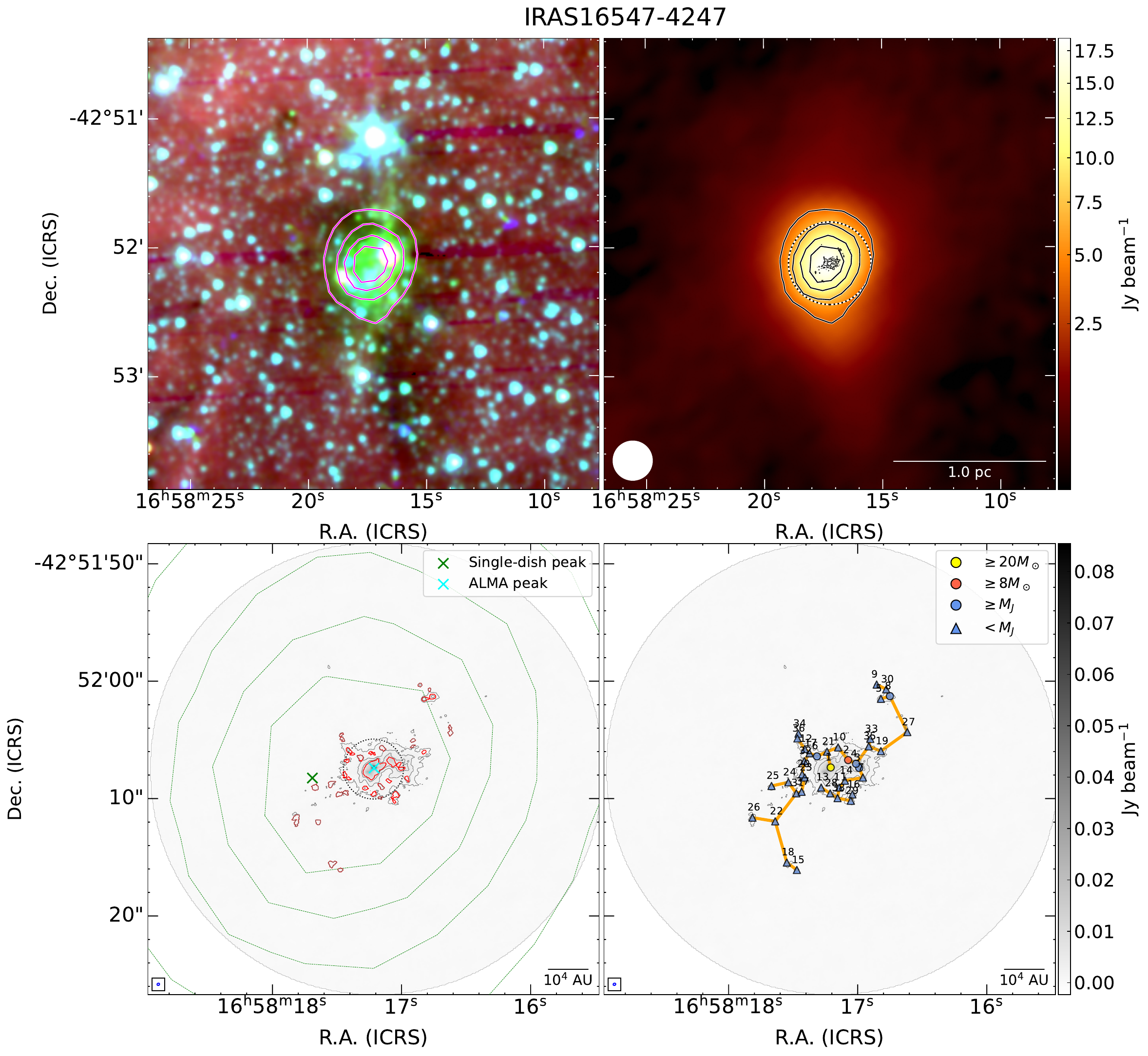}
\figsetgrpnote{Same as figure D.1, except for peak flux $=18.5$ Jy beam$^{-1}$ of ATLASGAL 870 $\mu$m continuum and 
        $\sigma=0.167$ mJy beam$^{-1}$ of ALMA 1.33 mm continuum.}
\figsetgrpend

\figsetgrpstart
\figsetgrpnum{D.4}
\figsetgrptitle{IRAS16562-3959}
\figsetplot{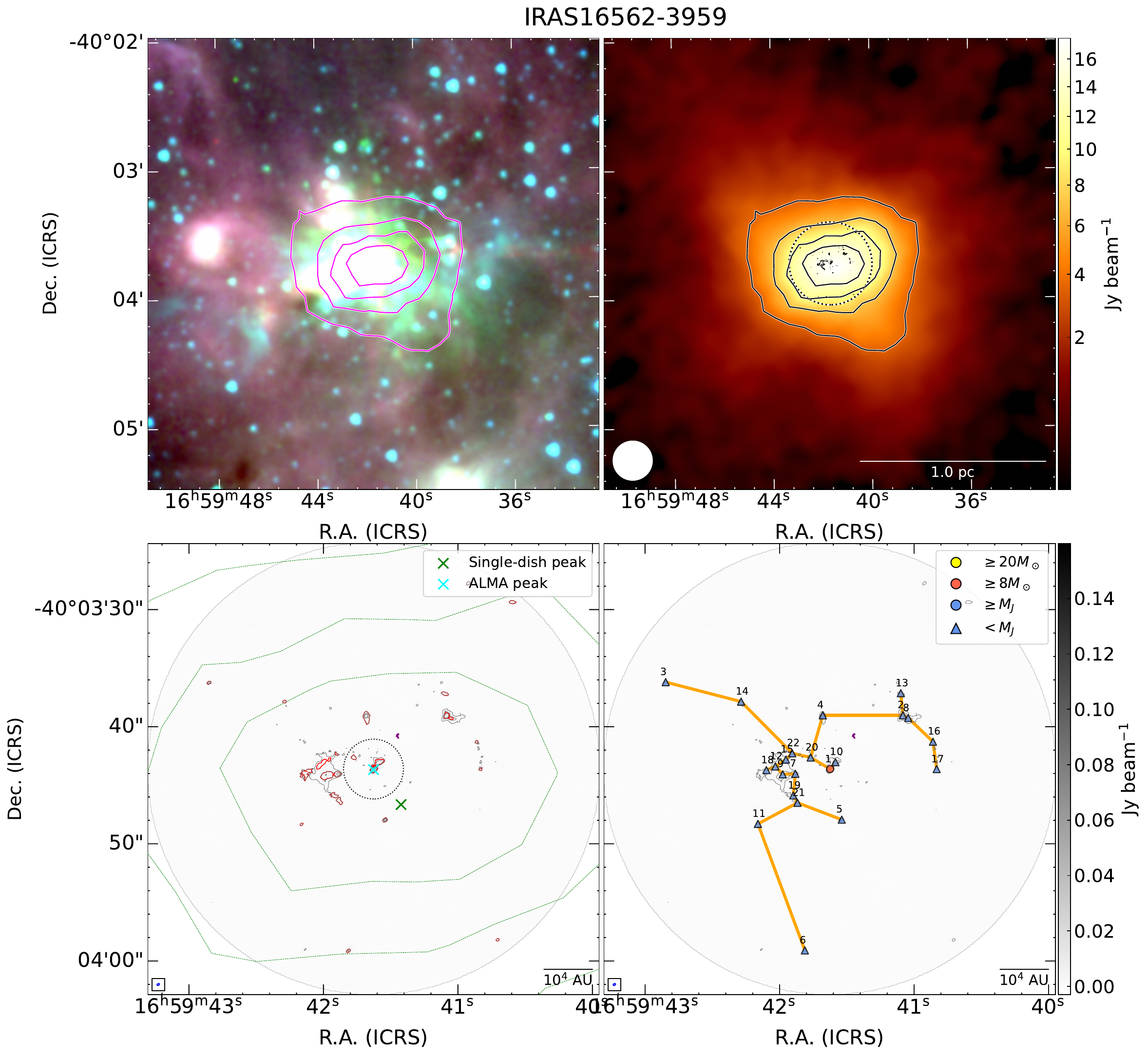}
\figsetgrpnote{Same as figure D.1, except for peak flux $=17.6$ Jy beam$^{-1}$ of ATLASGAL 870 $\mu$m continuum and 
        $\sigma=0.182$ mJy beam$^{-1}$ of ALMA 1.33 mm continuum.}
\figsetgrpend

\figsetgrpstart
\figsetgrpnum{D.5}
\figsetgrptitle{NGC6334I}
\figsetplot{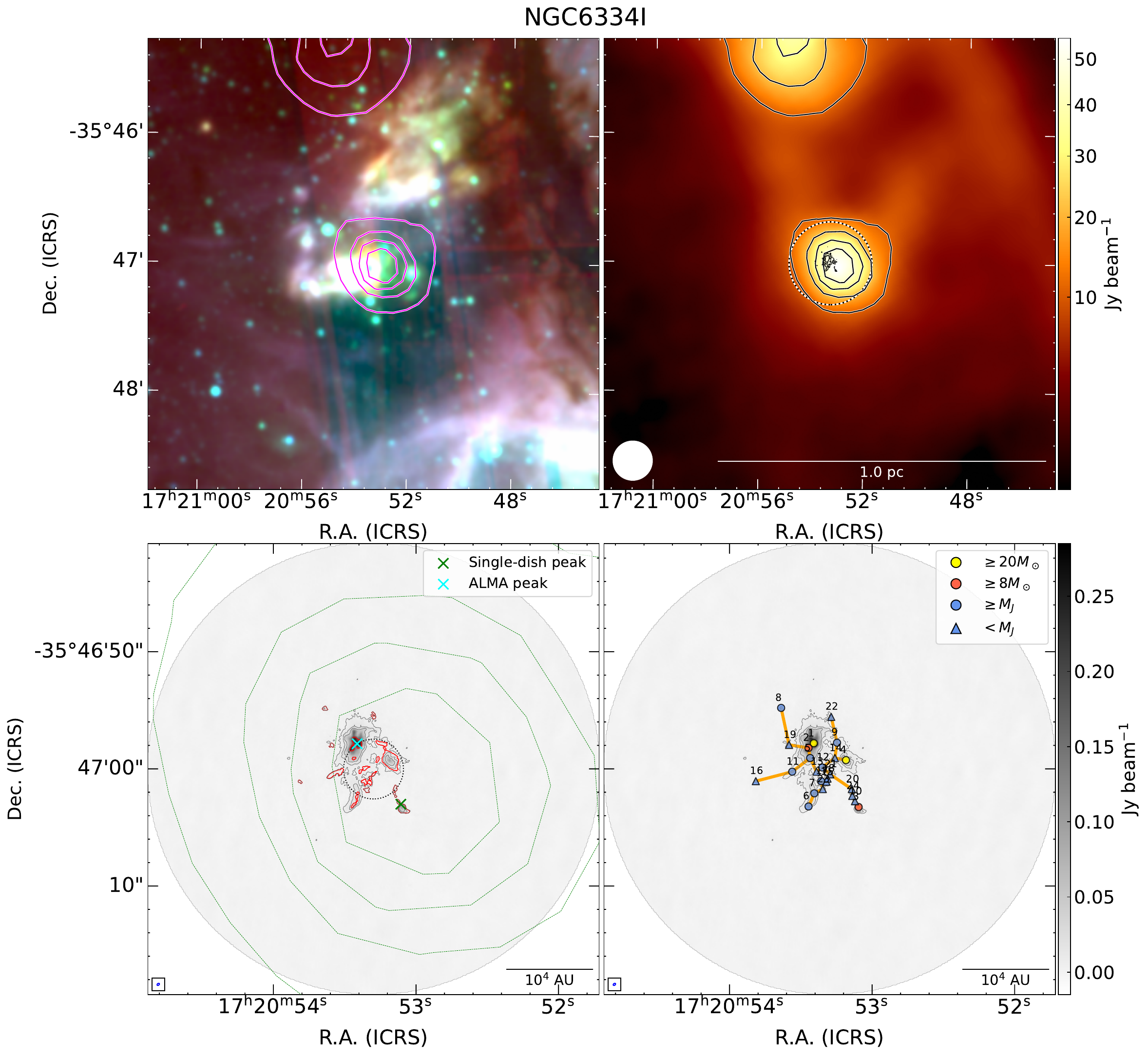}
\figsetgrpnote{Same as figure D.1, except for peak flux $=55.0$ Jy beam$^{-1}$ of ATLASGAL 870 $\mu$m continuum and 
        $\sigma=1.040$ mJy beam$^{-1}$ of ALMA 1.33 mm continuum.}
\figsetgrpend

\figsetgrpstart
\figsetgrpnum{D.6}
\figsetgrptitle{NGC6334I(N)}
\figsetplot{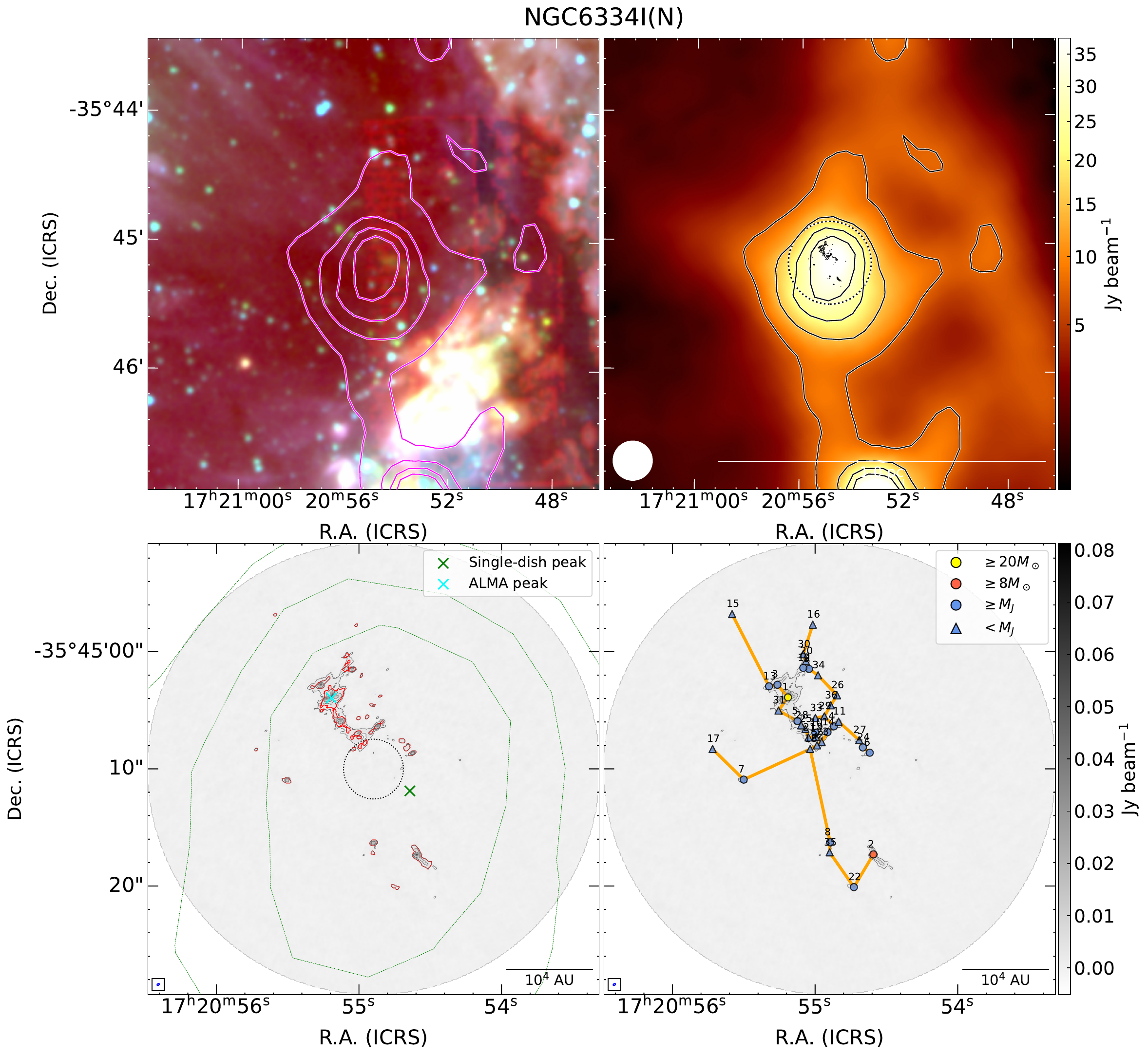}
\figsetgrpnote{Same as figure D.1, except for peak flux $=37.6$ Jy beam$^{-1}$ of ATLASGAL 870 $\mu$m continuum and 
        $\sigma=0.352$ mJy beam$^{-1}$ of ALMA 1.33 mm continuum.}
\figsetgrpend

\figsetgrpstart
\figsetgrpnum{D.7}
\figsetgrptitle{G29.96-0.02}
\figsetplot{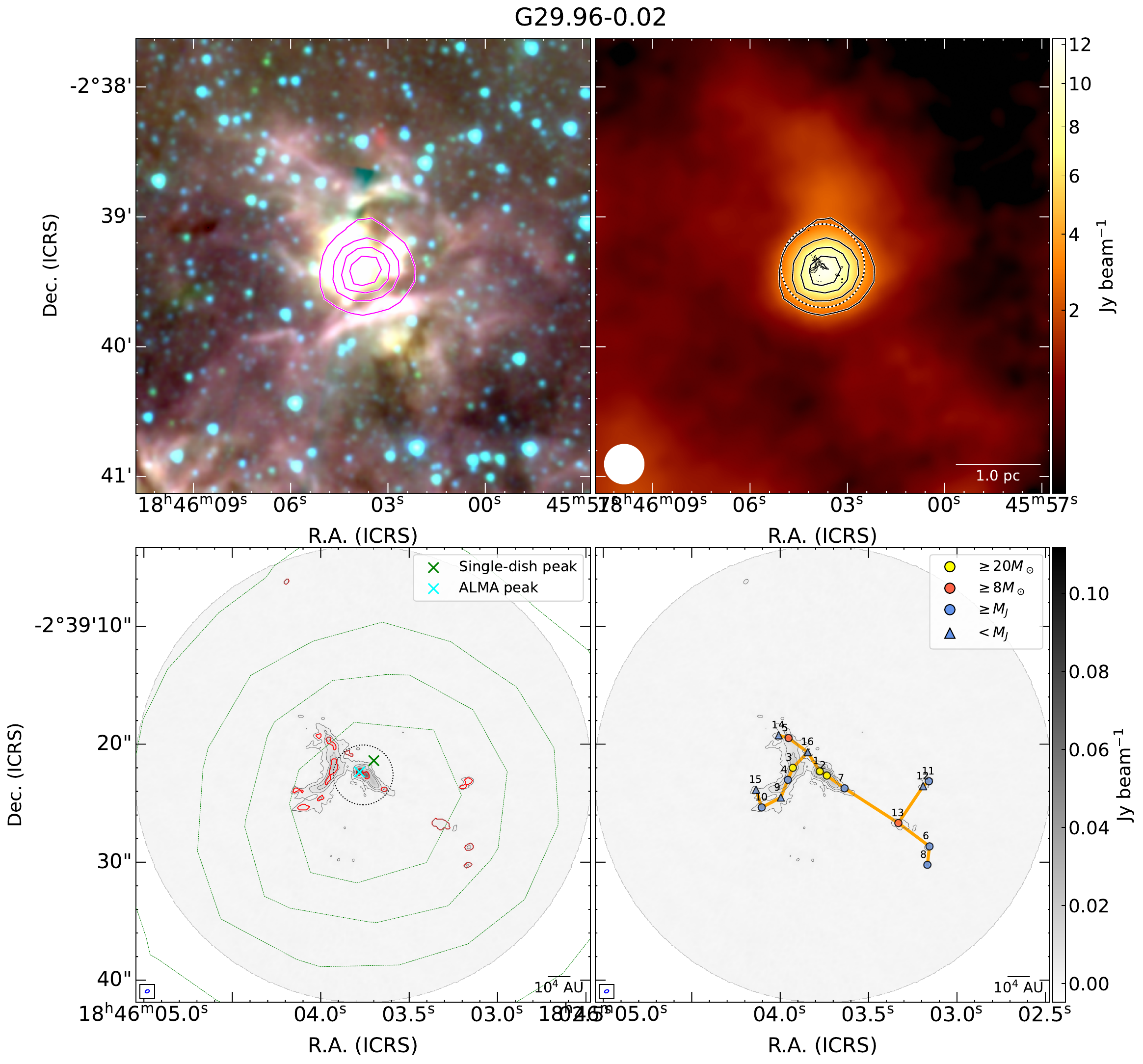}
\figsetgrpnote{Same as figure D.1, except for peak flux $=12.3$ Jy beam$^{-1}$ of ATLASGAL 870 $\mu$m continuum and 
        $\sigma=0.343$ mJy beam$^{-1}$ of ALMA 1.33 mm continuum.}
\figsetgrpend

\figsetgrpstart
\figsetgrpnum{D.8}
\figsetgrptitle{G34.43+0.24MM1}
\figsetplot{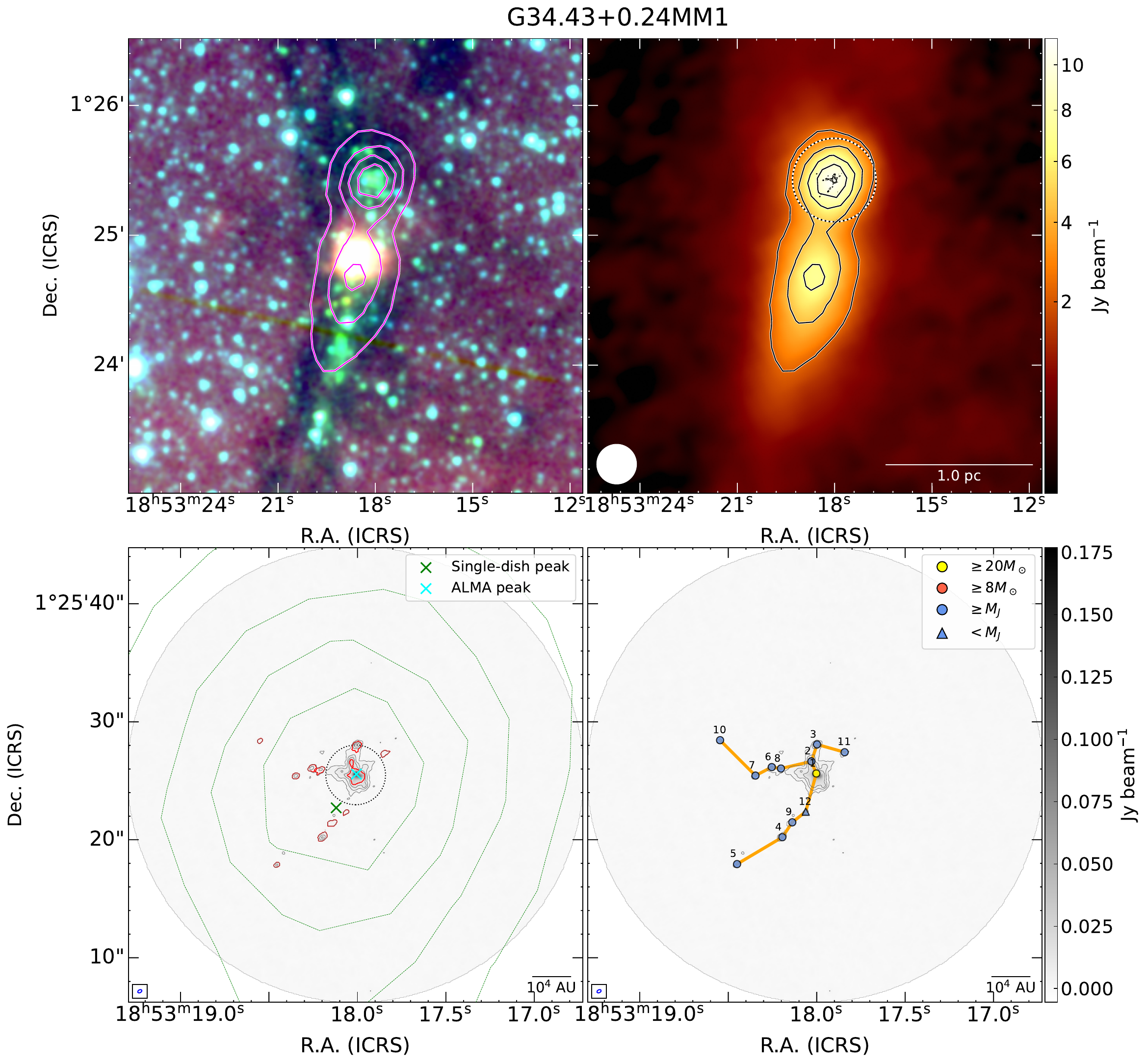}
\figsetgrpnote{Same as figure D.1, except for peak flux $=11.3$ Jy beam$^{-1}$ of ATLASGAL 870 $\mu$m continuum and 
        $\sigma=0.412$ mJy beam$^{-1}$ of ALMA 1.33 mm continuum.}
\figsetgrpend

\figsetgrpstart
\figsetgrpnum{D.9}
\figsetgrptitle{G35.03+0.35A}
\figsetplot{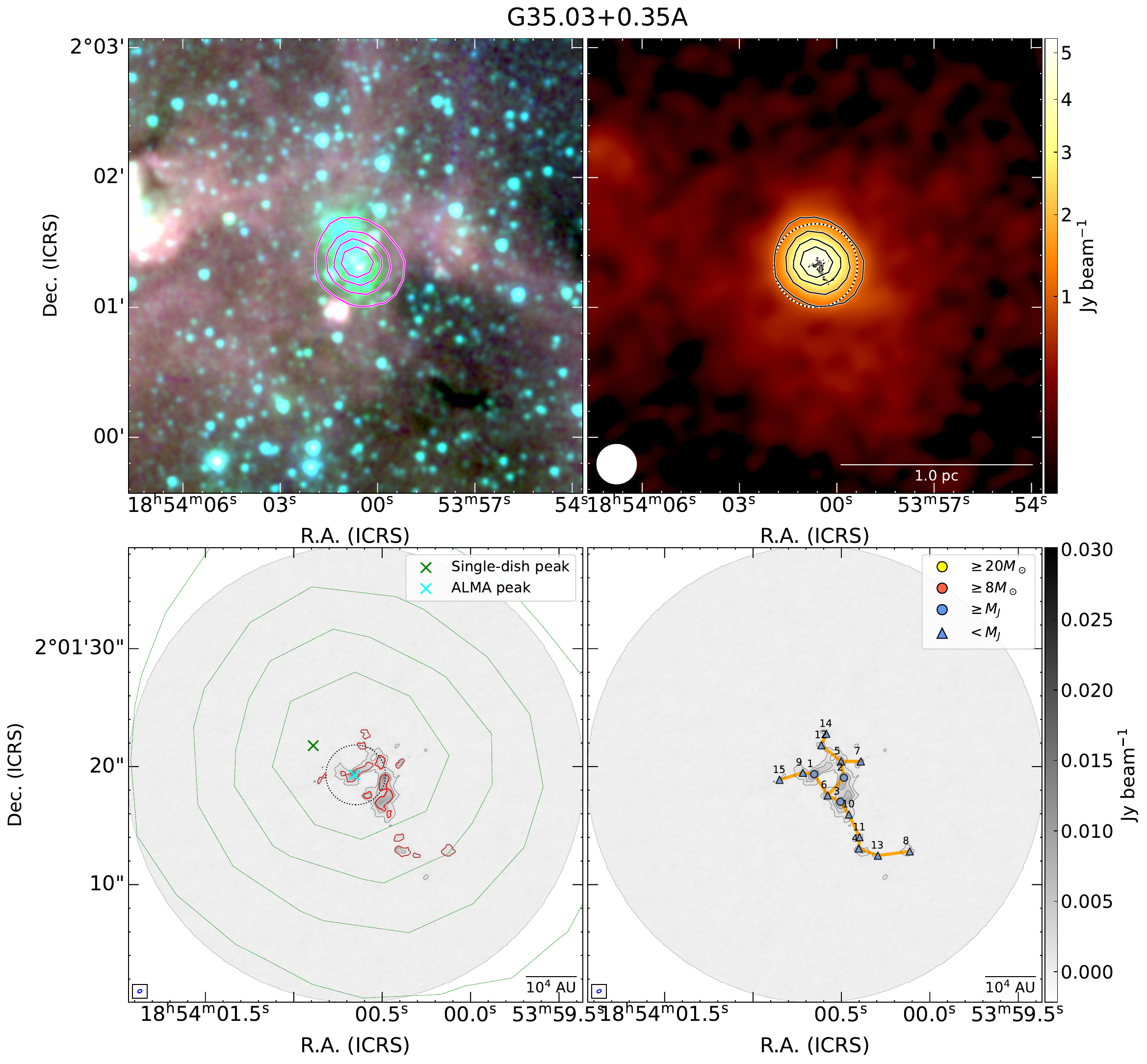}
\figsetgrpnote{Same as figure D.1, except for peak flux $=5.3$ Jy beam$^{-1}$ of ATLASGAL 870 $\mu$m continuum and 
        $\sigma=0.161$ mJy beam$^{-1}$ of ALMA 1.33 mm continuum.}
\figsetgrpend

\figsetgrpstart
\figsetgrpnum{D.10}
\figsetgrptitle{G35.20-0.74N}
\figsetplot{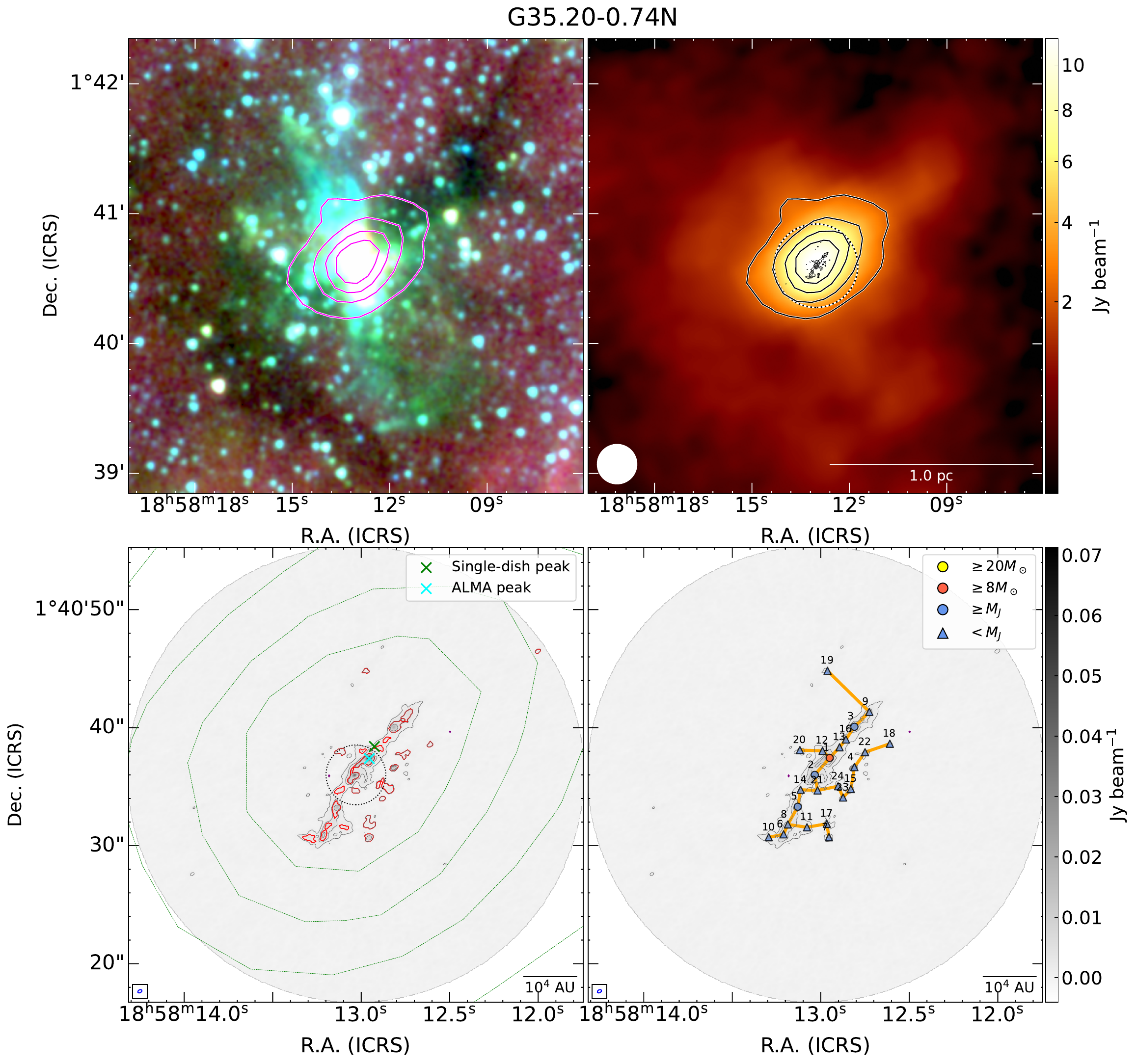}
\figsetgrpnote{Same as figure D.1, except for peak flux $=11.3$ Jy beam$^{-1}$ of ATLASGAL 870 $\mu$m continuum and 
        $\sigma=0.249$ mJy beam$^{-1}$ of ALMA 1.33 mm continuum.}
\figsetgrpend

\figsetgrpstart
\figsetgrpnum{D.11}
\figsetgrptitle{IRAS18151-1208}
\figsetplot{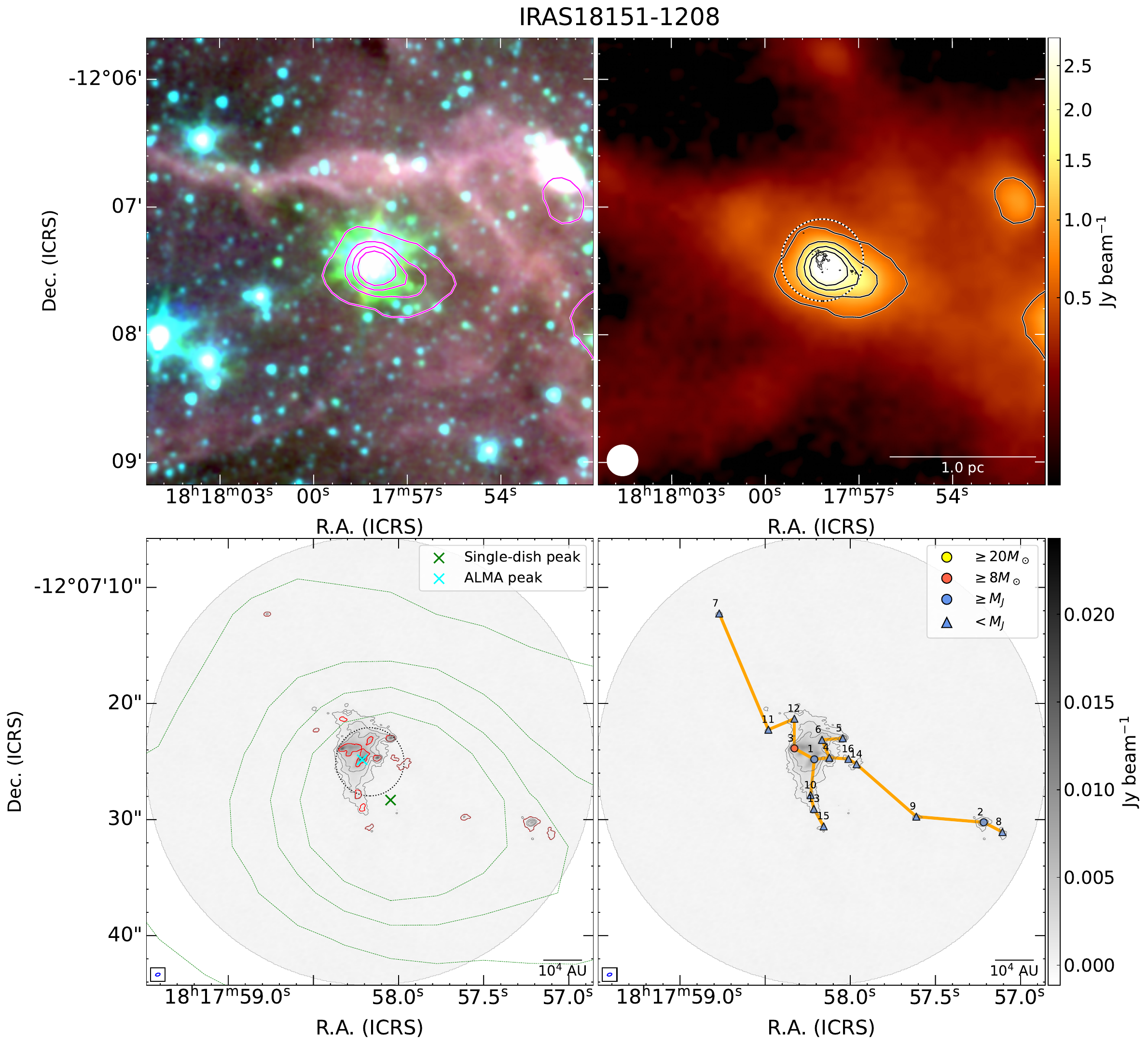}
\figsetgrpnote{Same as figure D.1, except for peak flux $=2.8$ Jy beam$^{-1}$ of ATLASGAL 870 $\mu$m continuum and 
        $\sigma=0.084$ mJy beam$^{-1}$ of ALMA 1.33 mm continuum.}
\figsetgrpend

\figsetgrpstart
\figsetgrpnum{D.12}
\figsetgrptitle{IRAS18182-1433}
\figsetplot{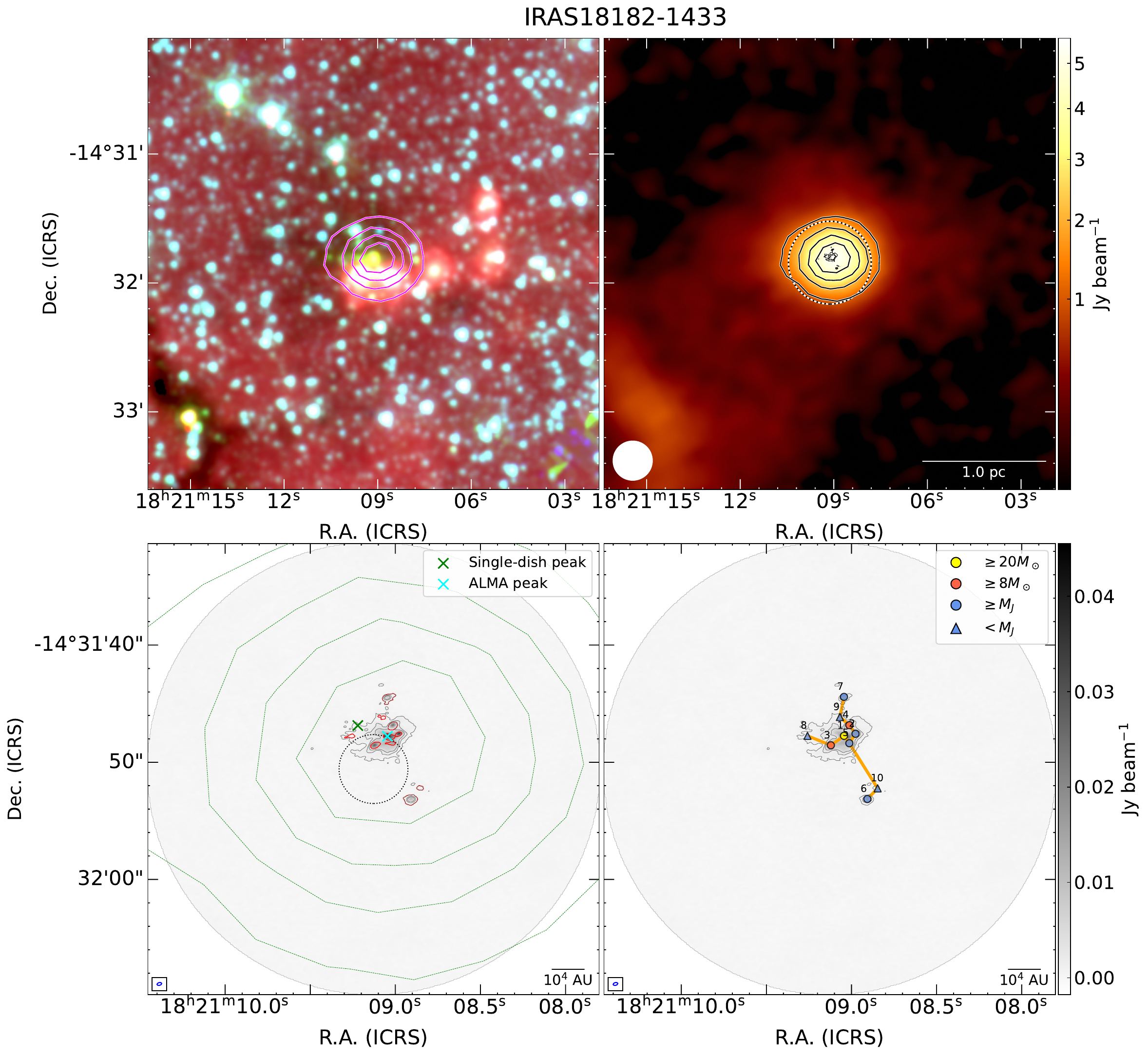}
\figsetgrpnote{Same as figure D.1, except for peak flux $=5.6$ Jy beam$^{-1}$ of ATLASGAL 870 $\mu$m continuum and 
        $\sigma=0.134$ mJy beam$^{-1}$ of ALMA 1.33 mm continuum.}
\figsetgrpend

\figsetgrpstart
\figsetgrpnum{D.13}
\figsetgrptitle{IRDC18223-1243}
\figsetplot{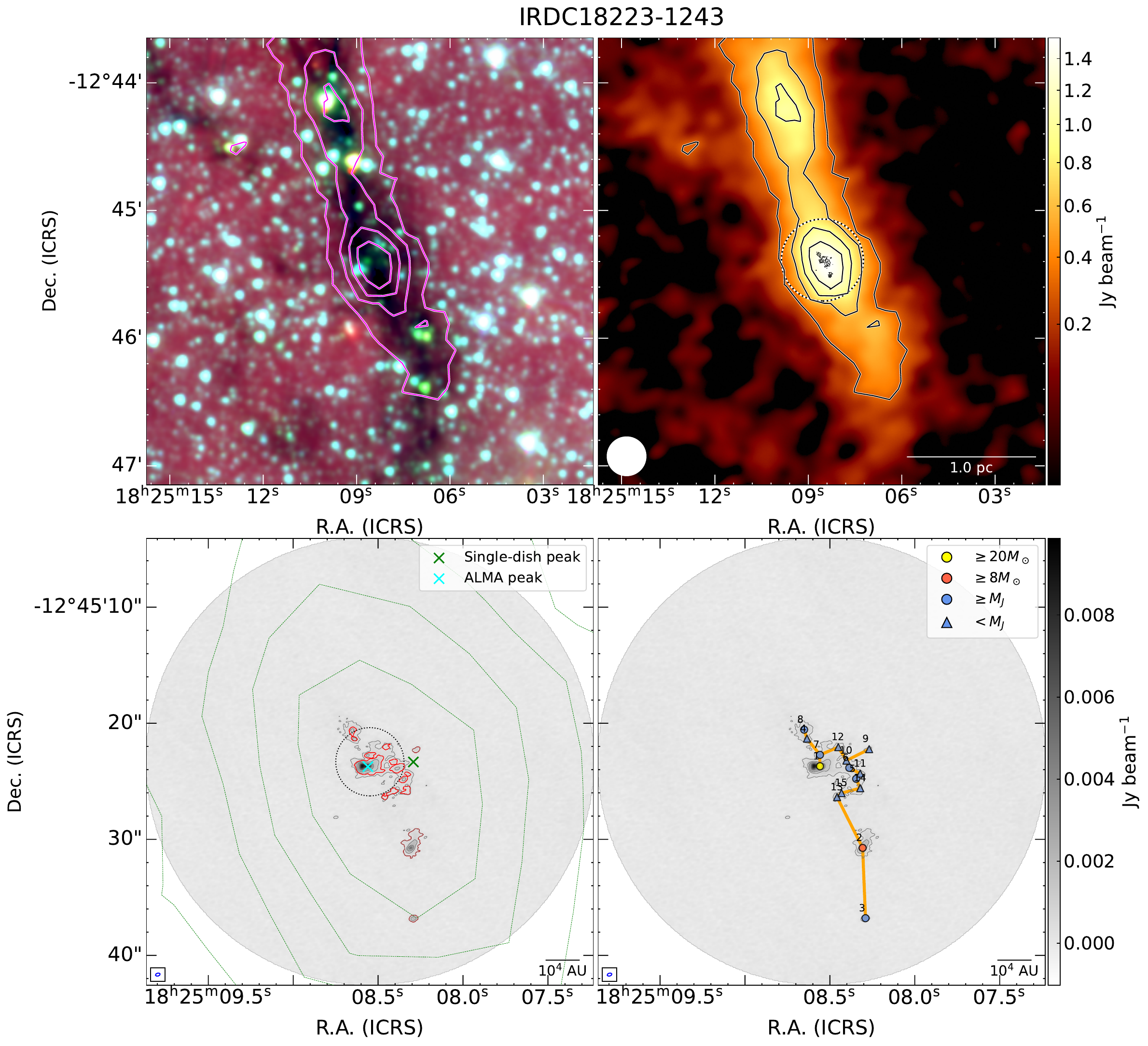}
\figsetgrpnote{Same as figure D.1, except for peak flux $=1.5$ Jy beam$^{-1}$ of ATLASGAL 870 $\mu$m continuum and 
        $\sigma=0.075$ mJy beam$^{-1}$ of ALMA 1.33 mm continuum.}
\figsetgrpend

\figsetgrpstart
\figsetgrpnum{D.14}
\figsetgrptitle{G10.62-0.38}
\figsetplot{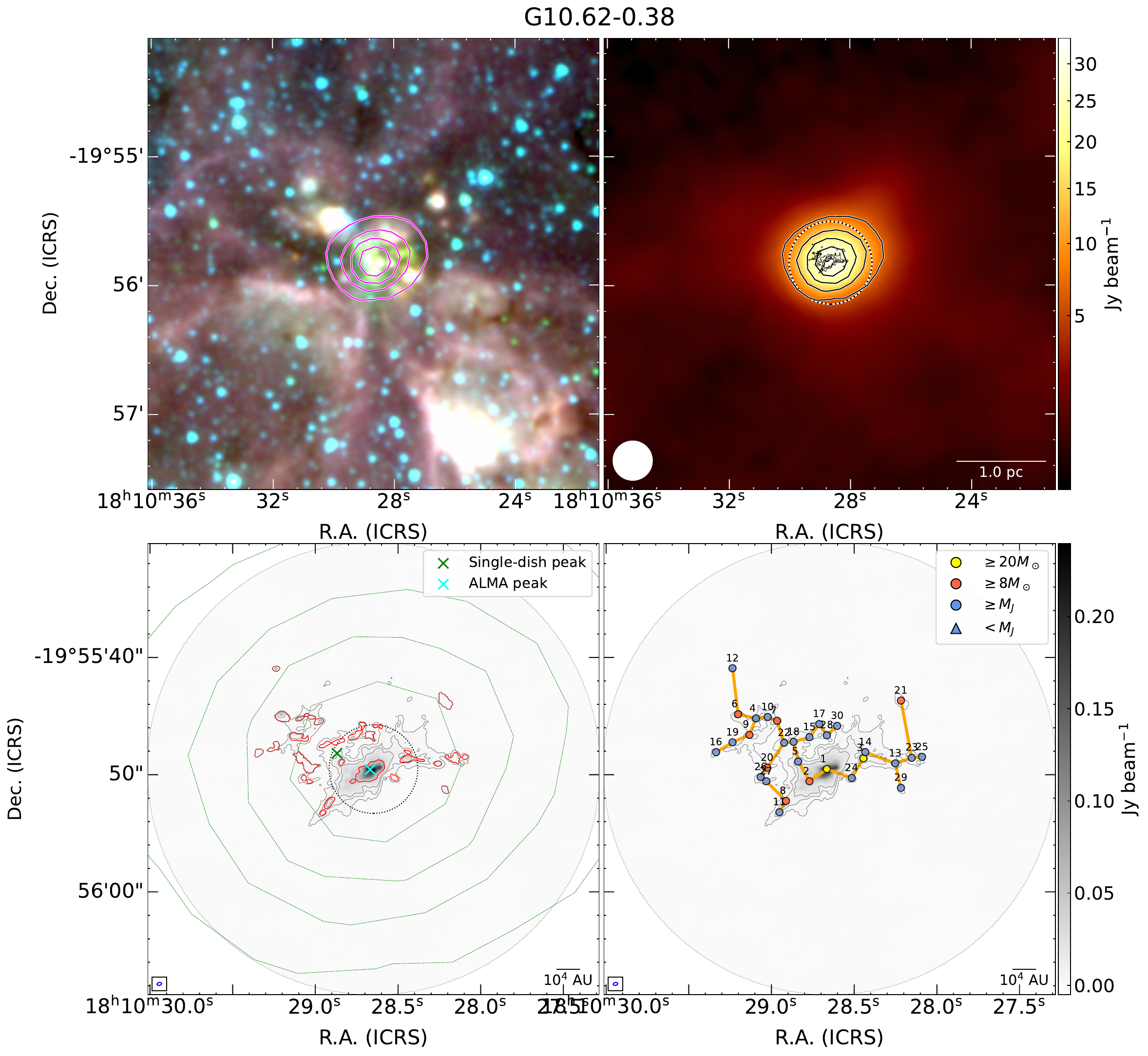}
\figsetgrpnote{Same as figure D.1, except for peak flux $=33.7$ Jy beam$^{-1}$ of ATLASGAL 870 $\mu$m continuum and 
        $\sigma=0.415$ mJy beam$^{-1}$ of ALMA 1.33 mm continuum.}
\figsetgrpend

\figsetgrpstart
\figsetgrpnum{D.15}
\figsetgrptitle{G11.1-0.12}
\figsetplot{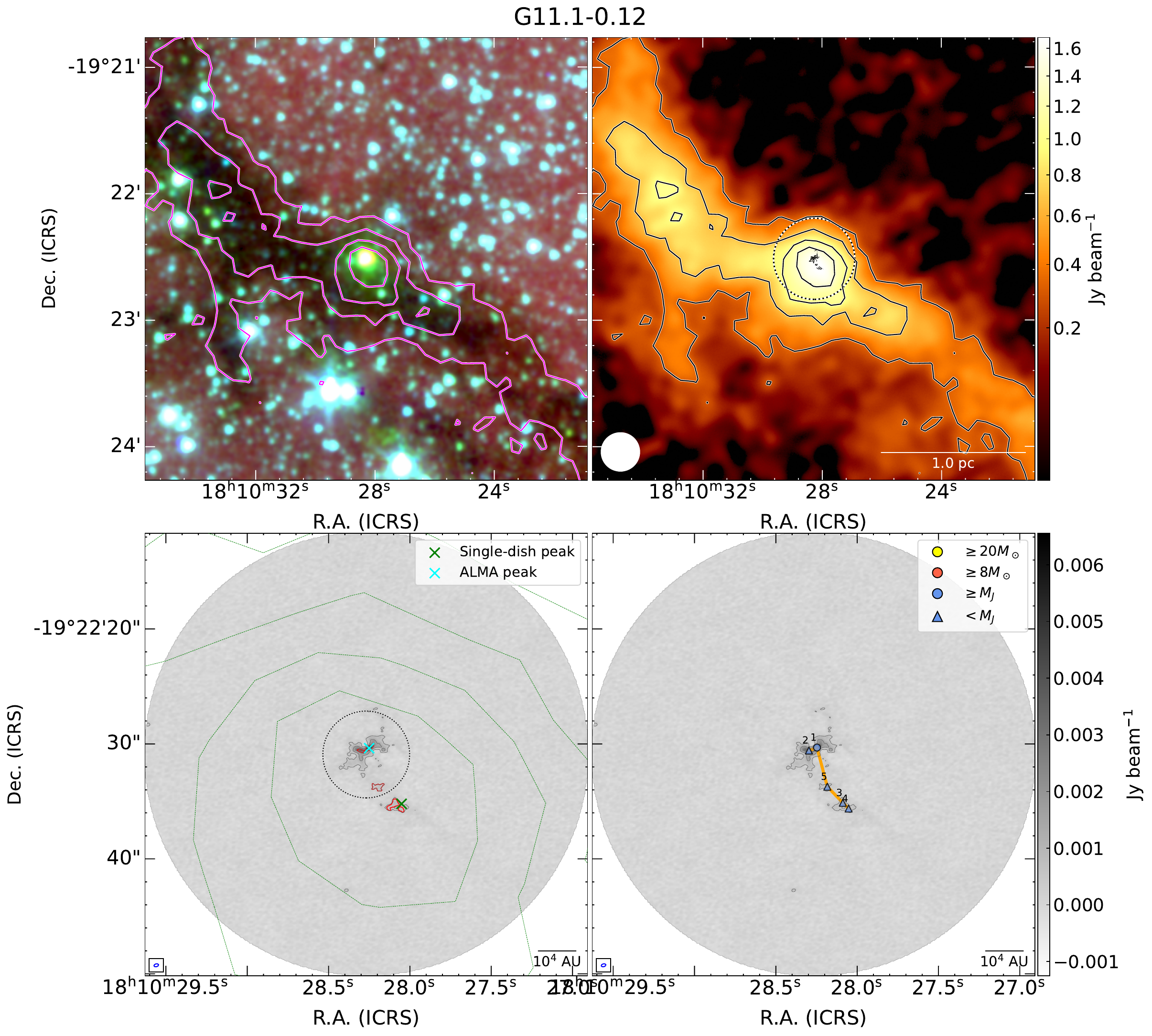}
\figsetgrpnote{Same as figure D.1, except for peak flux $=1.7$ Jy beam$^{-1}$ of ATLASGAL 870 $\mu$m continuum and 
        $\sigma=0.089$ mJy beam$^{-1}$ of ALMA 1.33 mm continuum.}
\figsetgrpend

\figsetgrpstart
\figsetgrpnum{D.16}
\figsetgrptitle{G11.92-0.61}
\figsetplot{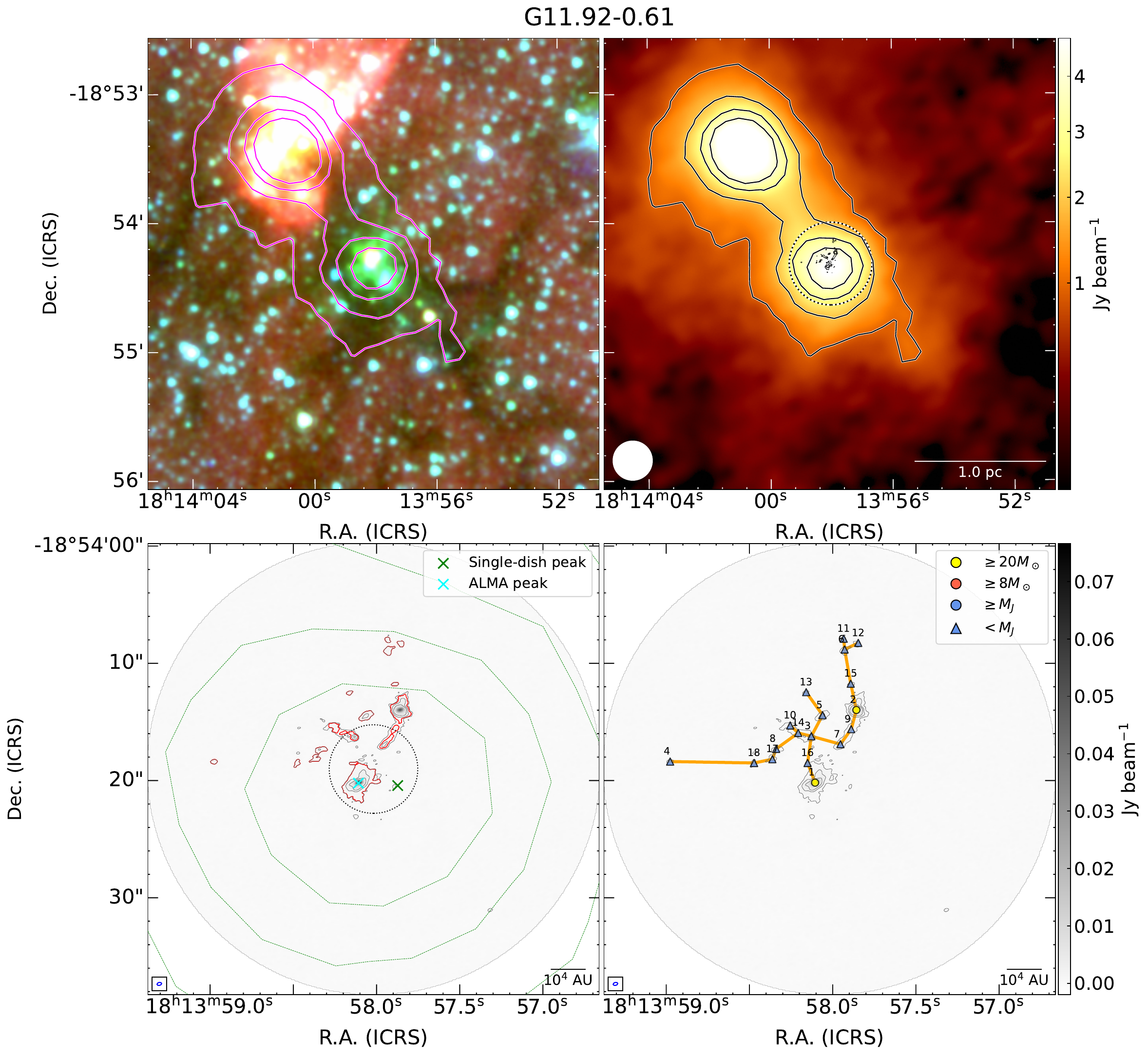}
\figsetgrpnote{Same as figure D.1, except for peak flux $=4.8$ Jy beam$^{-1}$ of ATLASGAL 870 $\mu$m continuum and 
        $\sigma=0.138$ mJy beam$^{-1}$ of ALMA 1.33 mm continuum.}
\figsetgrpend

\figsetgrpstart
\figsetgrpnum{D.17}
\figsetgrptitle{G5.89-0.37}
\figsetplot{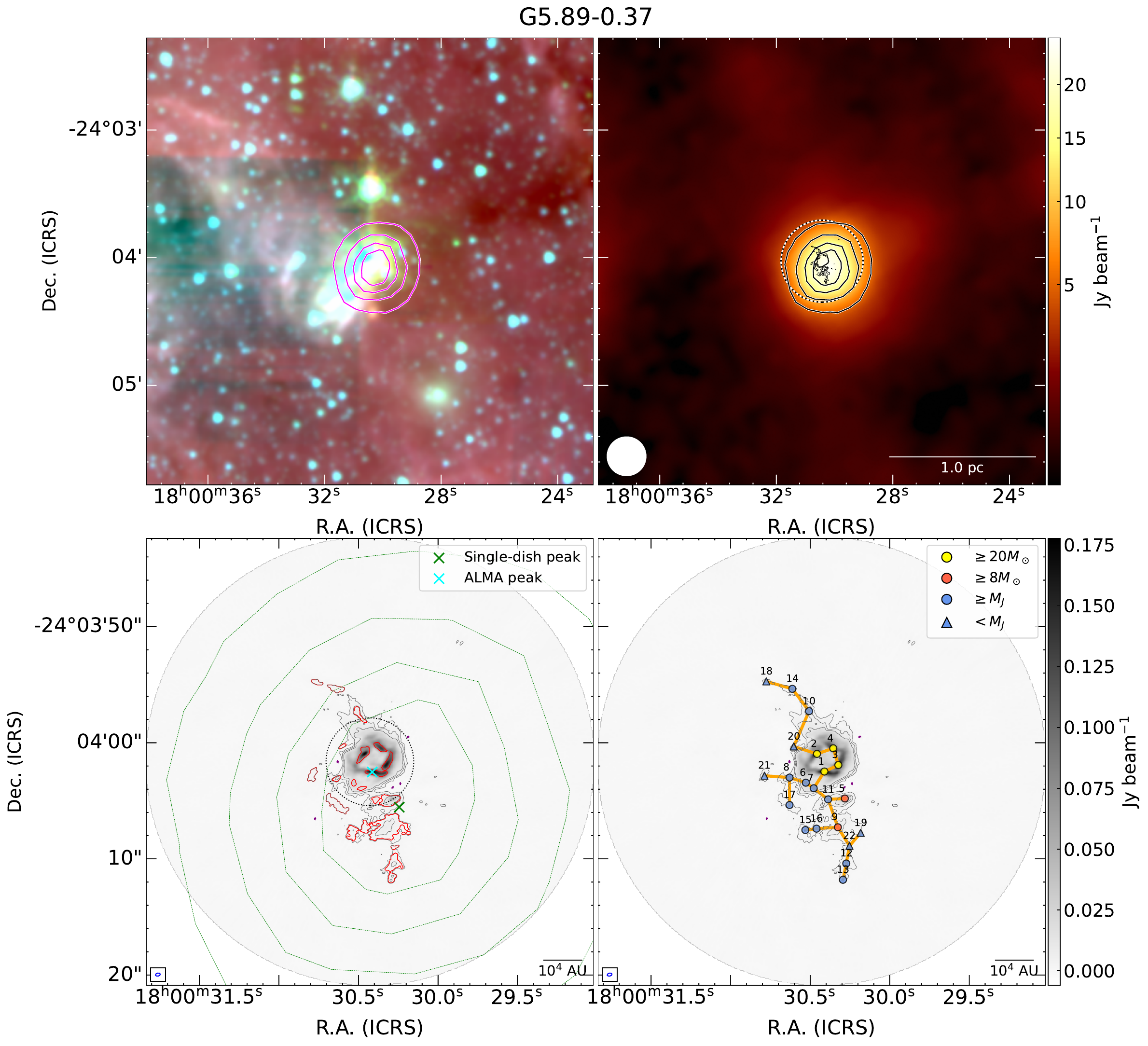}
\figsetgrpnote{Same as figure D.1, except for peak flux $=24.9$ Jy beam$^{-1}$ of ATLASGAL 870 $\mu$m continuum and 
        $\sigma=0.328$ mJy beam$^{-1}$ of ALMA 1.33 mm continuum.}
\figsetgrpend

\figsetgrpstart
\figsetgrpnum{D.18}
\figsetgrptitle{IRAS18089-1732}
\figsetplot{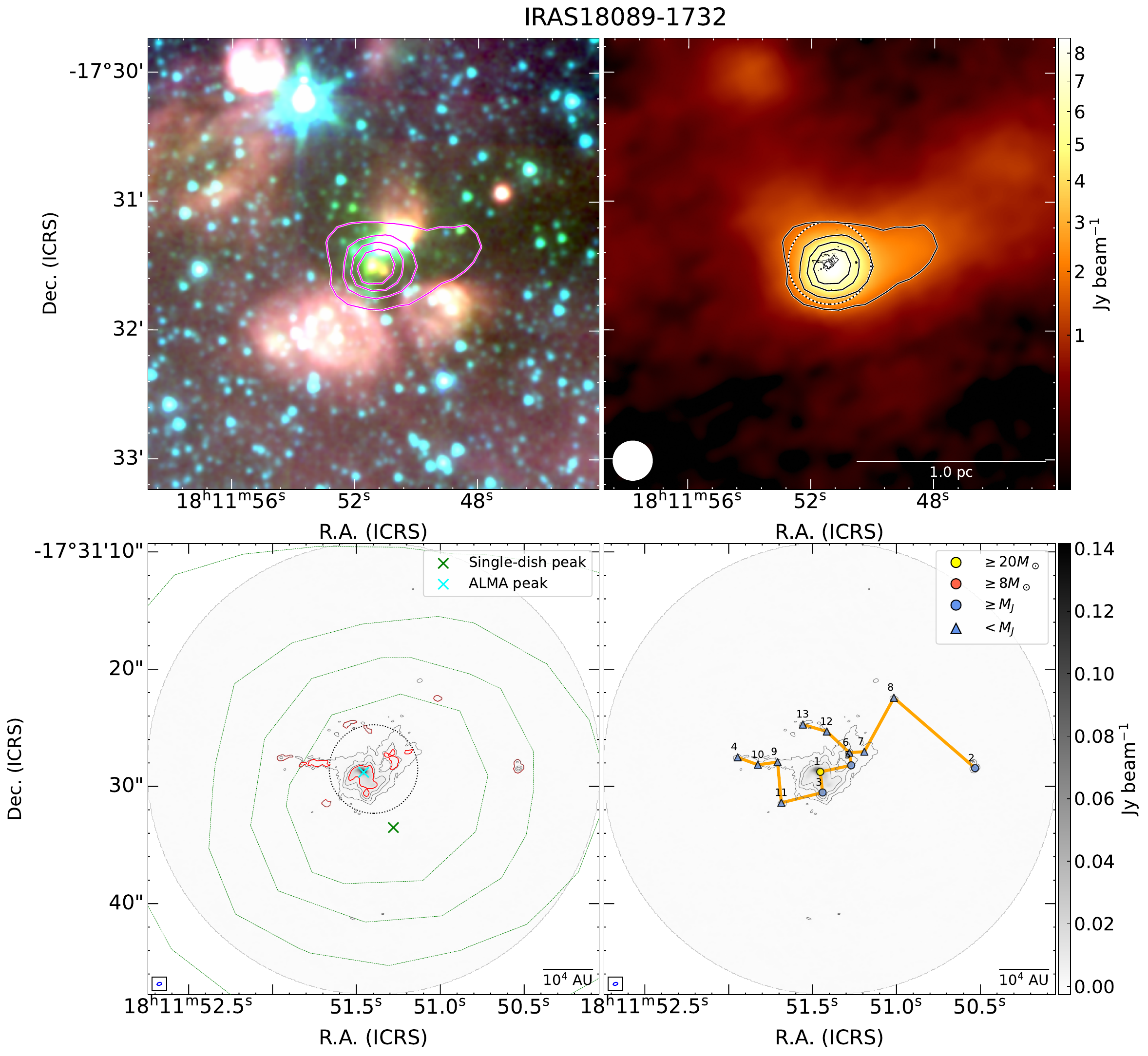}
\figsetgrpnote{Same as figure D.1, except for peak flux $=8.6$ Jy beam$^{-1}$ of ATLASGAL 870 $\mu$m continuum and 
        $\sigma=0.181$ mJy beam$^{-1}$ of ALMA 1.33 mm continuum.}
\figsetgrpend

\figsetgrpstart
\figsetgrpnum{D.19}
\figsetgrptitle{IRAS18162-2048}
\figsetplot{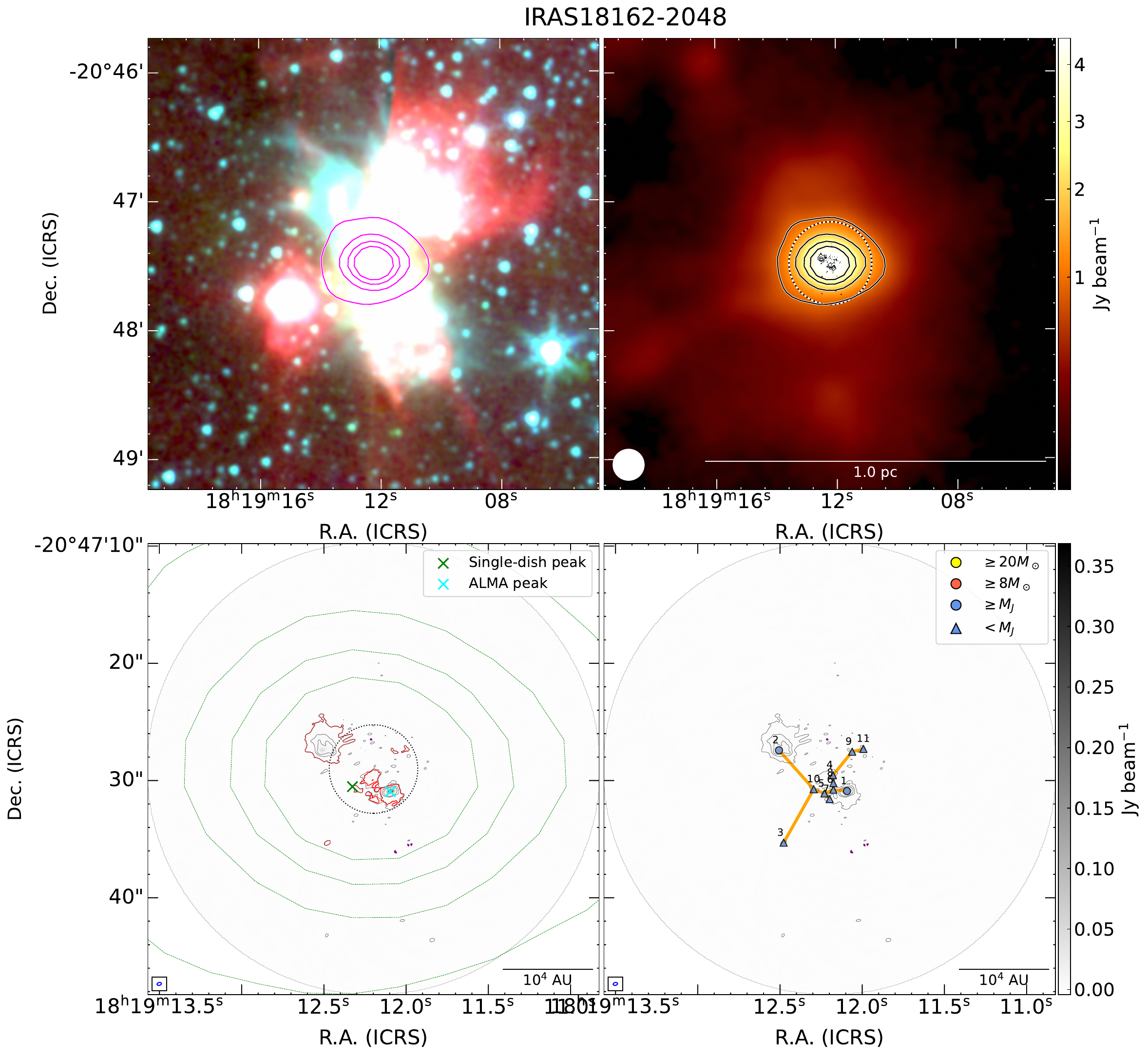}
\figsetgrpnote{Same as figure D.1, except for peak flux $=4.5$ Jy beam$^{-1}$ of ATLASGAL 870 $\mu$m continuum and 
        $\sigma=0.219$ mJy beam$^{-1}$ of ALMA 1.33 mm continuum.}
\figsetgrpend

\figsetgrpstart
\figsetgrpnum{D.20}
\figsetgrptitle{W33A}
\figsetplot{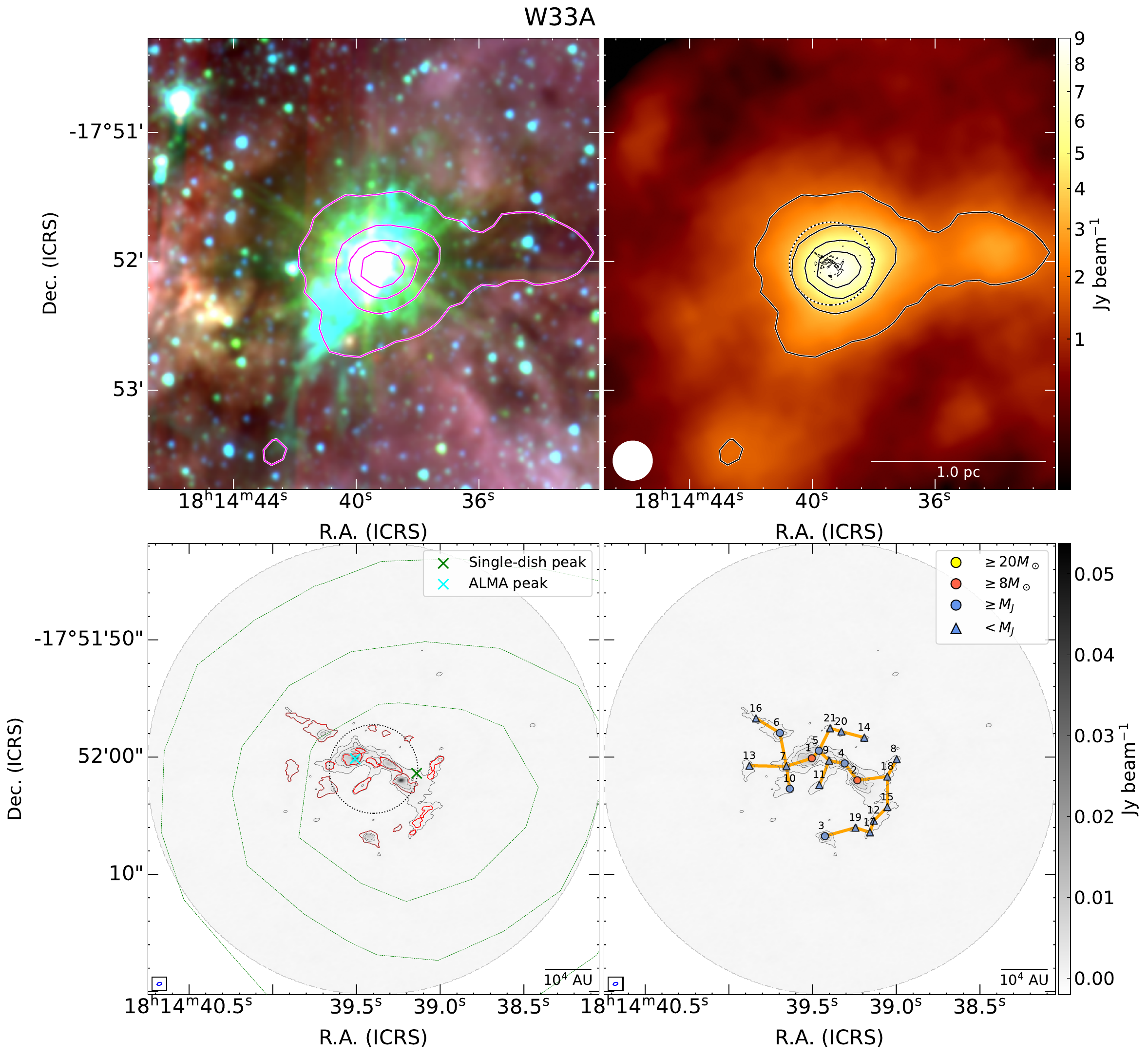}
\figsetgrpnote{Same as figure D.1, except for peak flux $=9.0$ Jy beam$^{-1}$ of ATLASGAL 870 $\mu$m continuum and 
        $\sigma=0.139$ mJy beam$^{-1}$ of ALMA 1.33 mm continuum.}
\figsetgrpend

\figsetgrpstart
\figsetgrpnum{D.21}
\figsetgrptitle{G14.22-0.50S}
\figsetplot{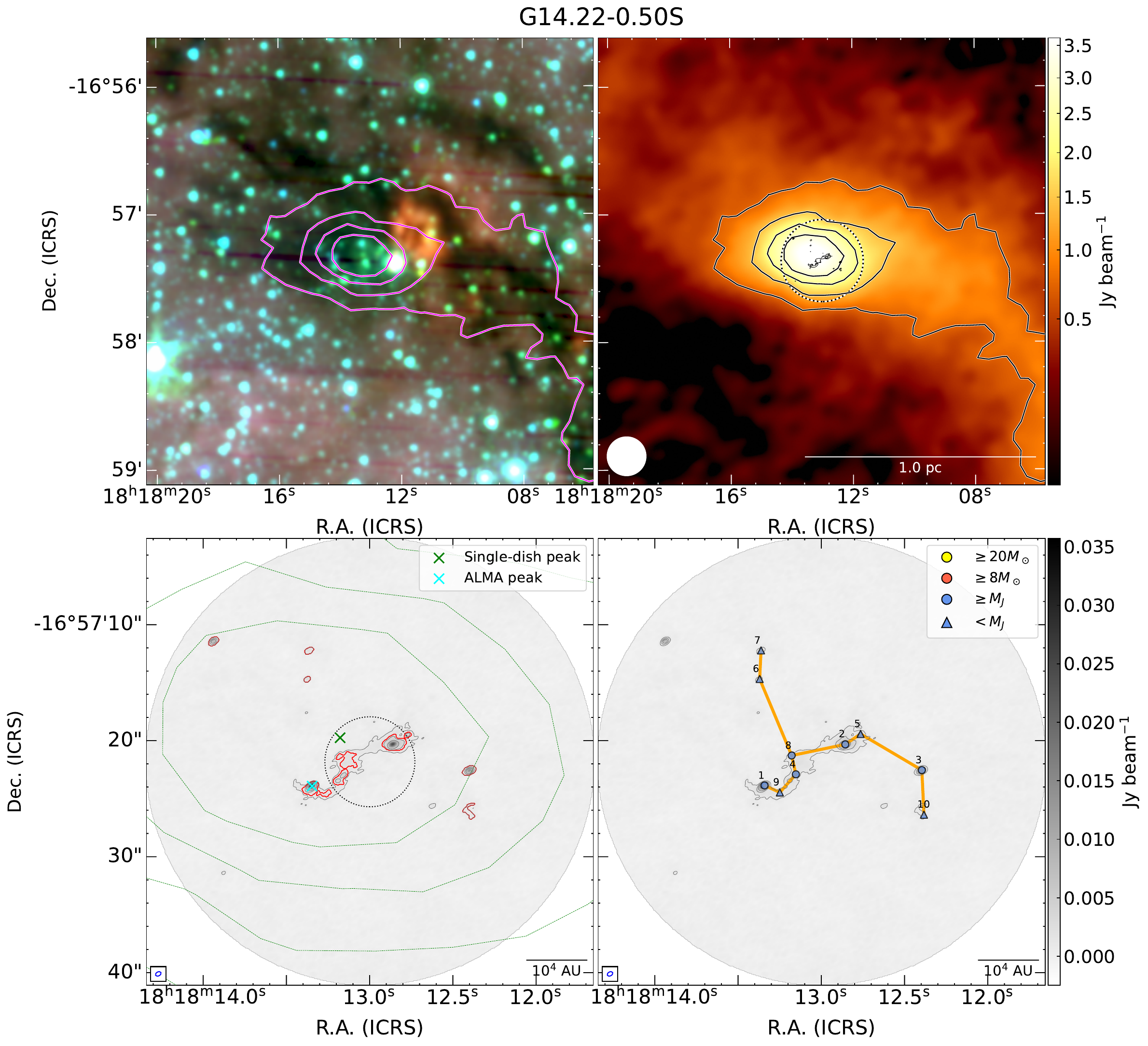}
\figsetgrpnote{Same as figure D.1, except for peak flux $=3.6$ Jy beam$^{-1}$ of ATLASGAL 870 $\mu$m continuum and 
        $\sigma=0.193$ mJy beam$^{-1}$ of ALMA 1.33 mm continuum.}
\figsetgrpend

\figsetgrpstart
\figsetgrpnum{D.22}
\figsetgrptitle{G351.77-0.54}
\figsetplot{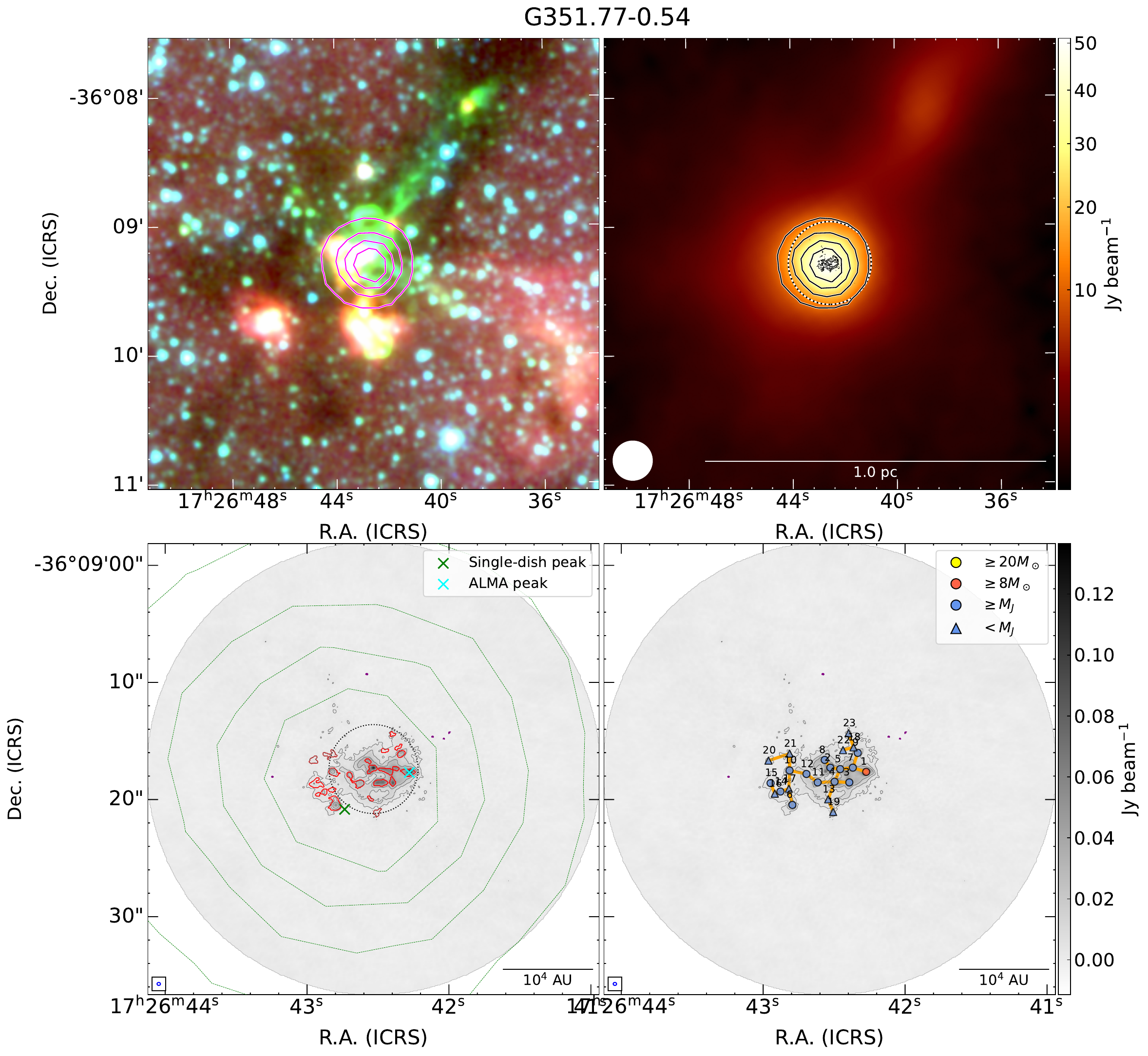}
\figsetgrpnote{Same as figure D.1, except for peak flux $=51.1$ Jy beam$^{-1}$ of ATLASGAL 870 $\mu$m continuum and 
        $\sigma=0.718$ mJy beam$^{-1}$ of ALMA 1.33 mm continuum.}
\figsetgrpend

\figsetgrpstart
\figsetgrpnum{D.23}
\figsetgrptitle{G24.60+0.08}
\figsetplot{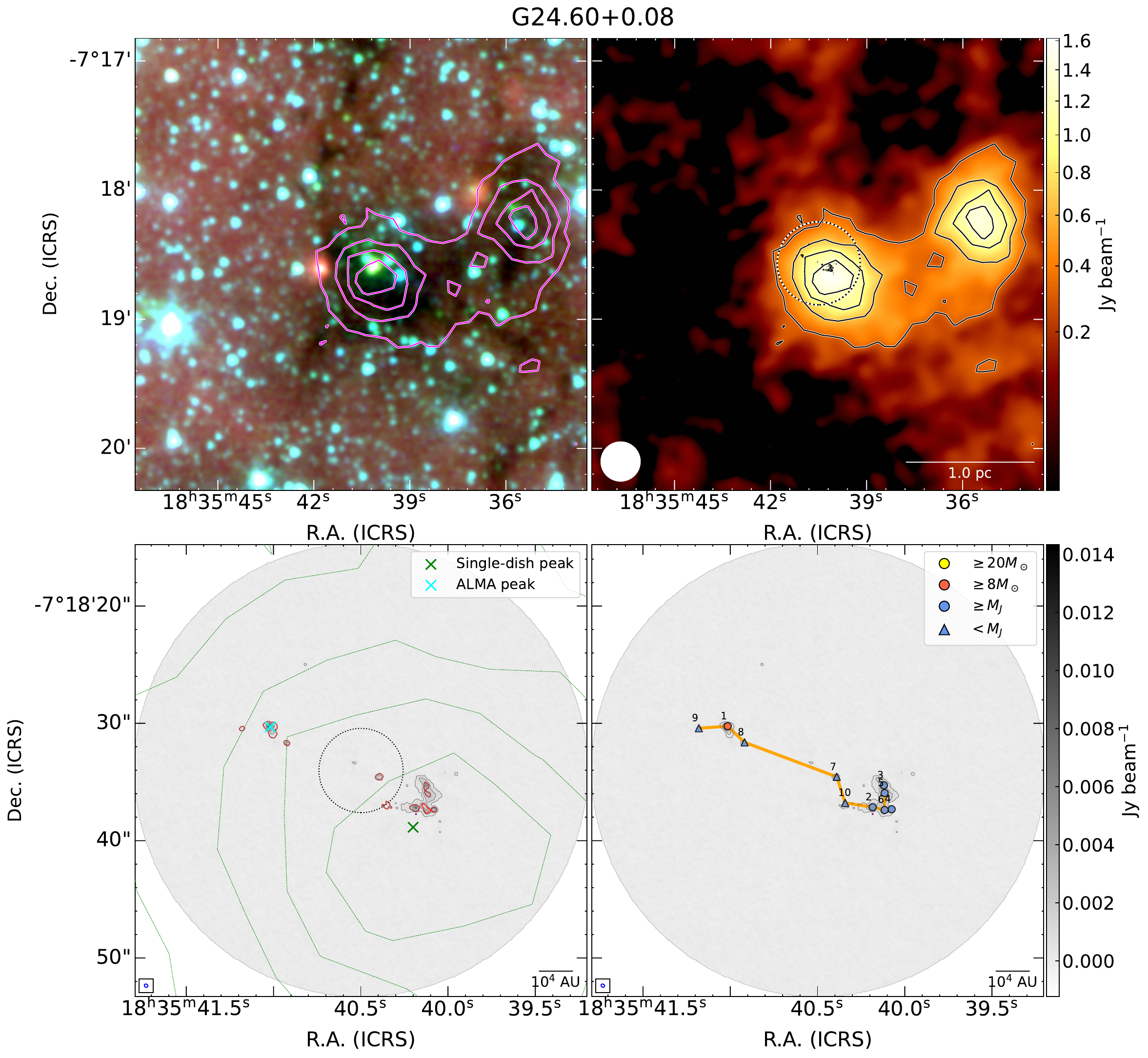}
\figsetgrpnote{Same as figure D.1, except for peak flux $=1.6$ Jy beam$^{-1}$ of ATLASGAL 870 $\mu$m continuum and 
        $\sigma=0.082$ mJy beam$^{-1}$ of ALMA 1.33 mm continuum.}
\figsetgrpend

\figsetgrpstart
\figsetgrpnum{D.24}
\figsetgrptitle{IRAS18337-0743}
\figsetplot{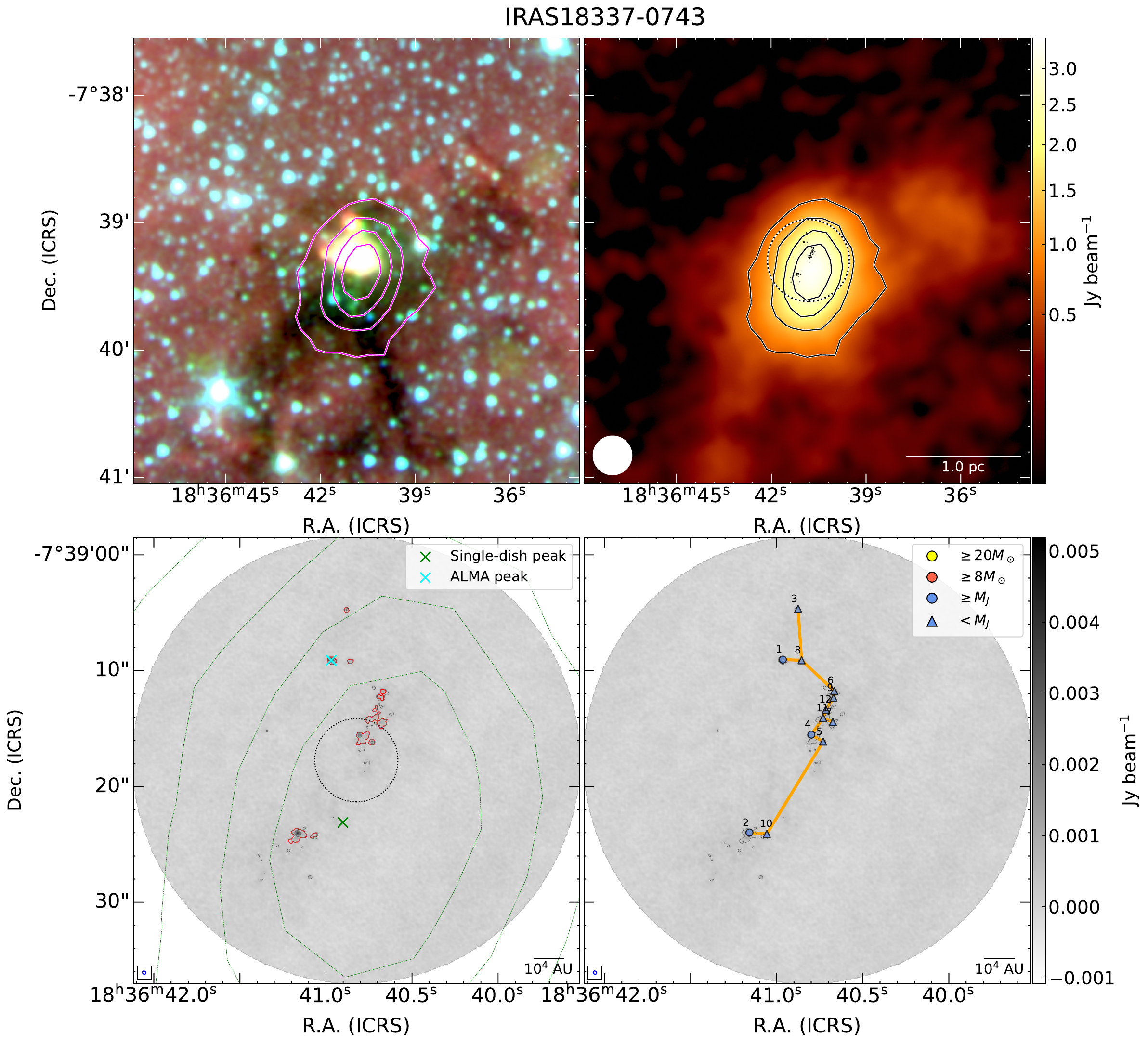}
\figsetgrpnote{Same as figure D.1, except for peak flux $=3.5$ Jy beam$^{-1}$ of ATLASGAL 870 $\mu$m continuum and 
        $\sigma=0.087$ mJy beam$^{-1}$ of ALMA 1.33 mm continuum.}
\figsetgrpend

\figsetgrpstart
\figsetgrpnum{D.25}
\figsetgrptitle{G333.12-0.56}
\figsetplot{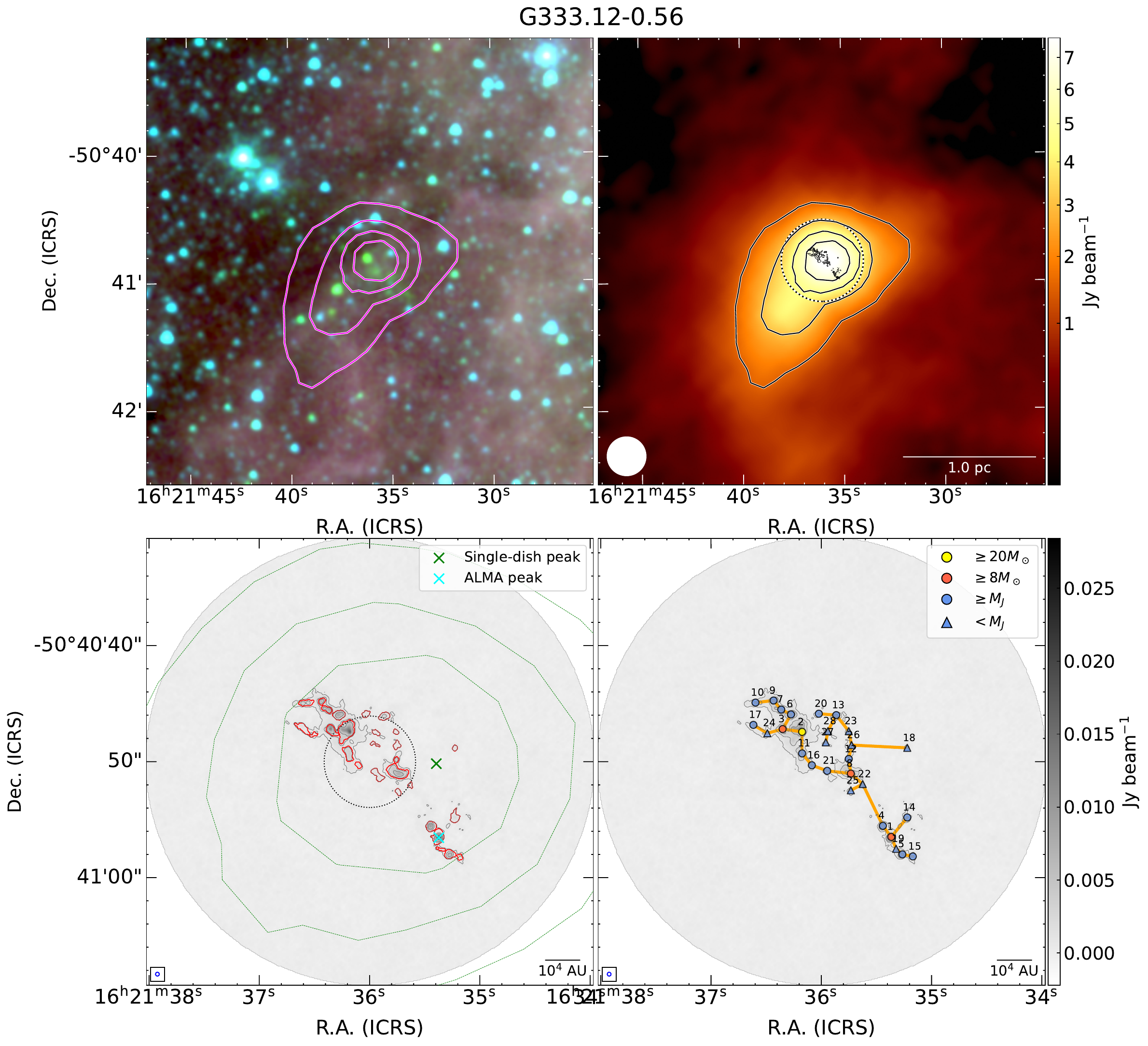}
\figsetgrpnote{Same as figure D.1, except for peak flux $=7.7$ Jy beam$^{-1}$ of ATLASGAL 870 $\mu$m continuum and 
        $\sigma=0.149$ mJy beam$^{-1}$ of ALMA 1.33 mm continuum.}
\figsetgrpend

\figsetgrpstart
\figsetgrpnum{D.26}
\figsetgrptitle{G335.78+0.17}
\figsetplot{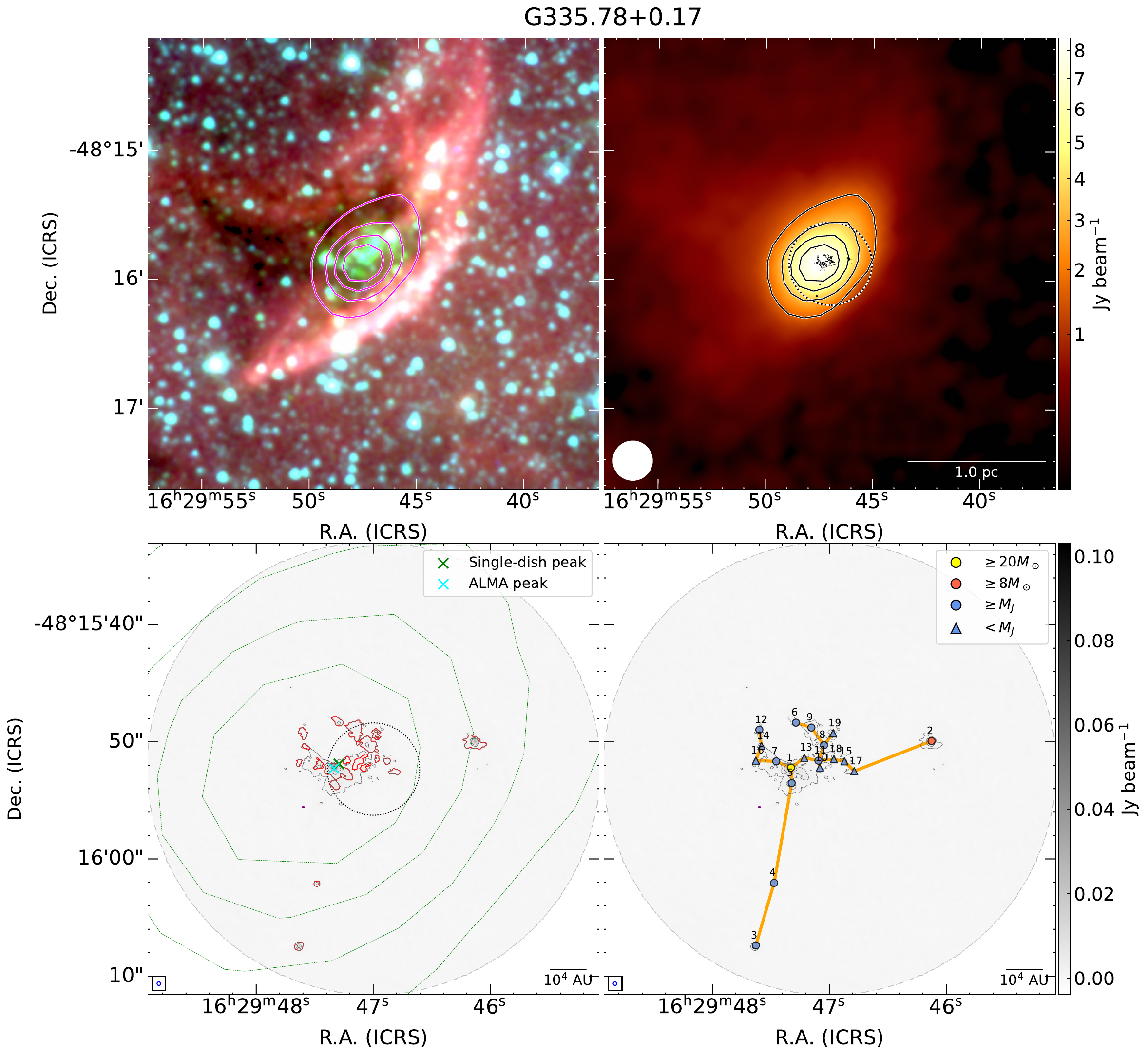}
\figsetgrpnote{Same as figure D.1, except for peak flux $=8.4$ Jy beam$^{-1}$ of ATLASGAL 870 $\mu$m continuum and 
        $\sigma=0.250$ mJy beam$^{-1}$ of ALMA 1.33 mm continuum.}
\figsetgrpend

\figsetgrpstart
\figsetgrpnum{D.27}
\figsetgrptitle{G333.46-0.16}
\figsetplot{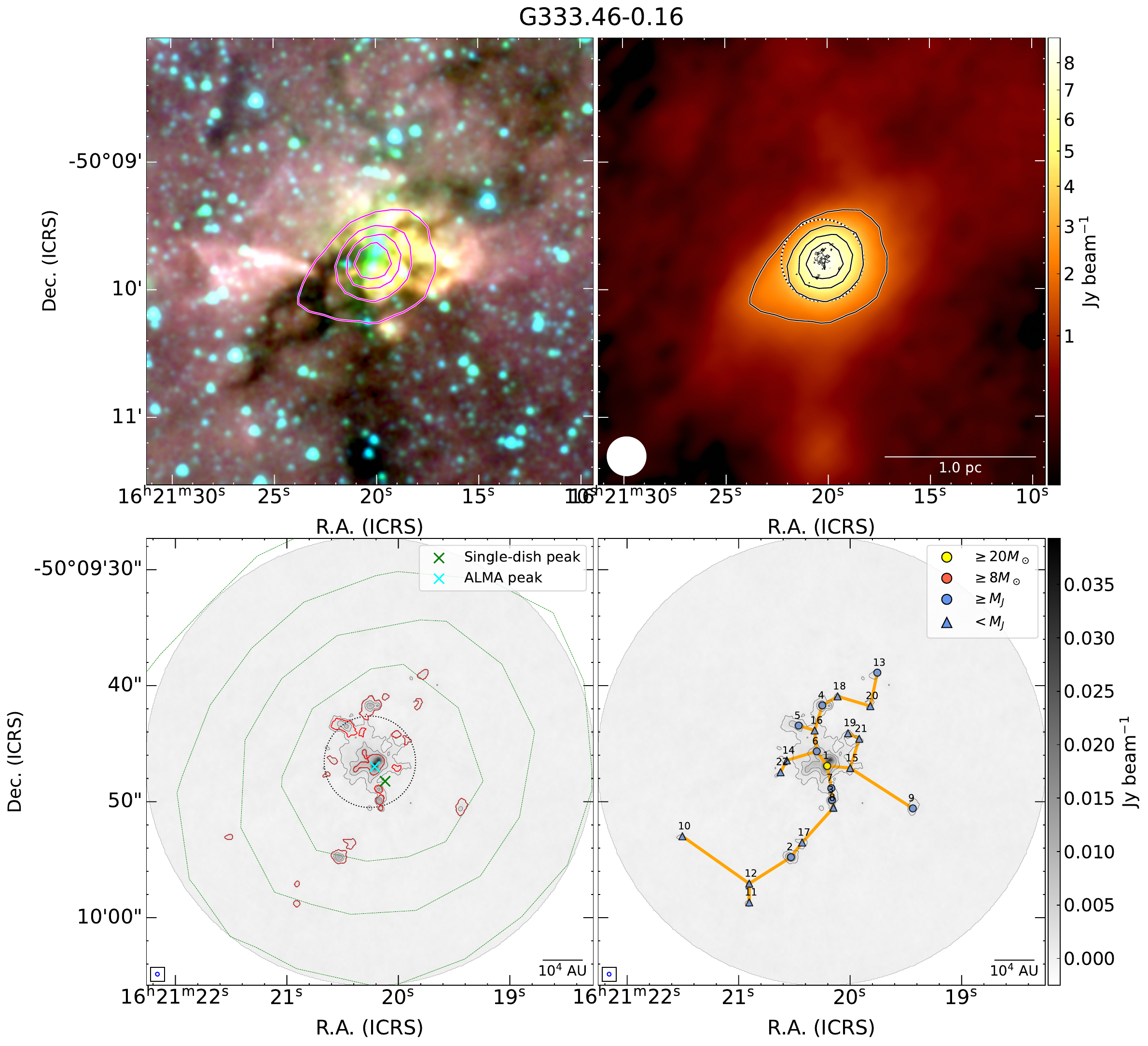}
\figsetgrpnote{Same as figure D.1, except for peak flux $=9.0$ Jy beam$^{-1}$ of ATLASGAL 870 $\mu$m continuum and 
        $\sigma=0.188$ mJy beam$^{-1}$ of ALMA 1.33 mm continuum.}
\figsetgrpend

\figsetgrpstart
\figsetgrpnum{D.28}
\figsetgrptitle{G336.01-0.82}
\figsetplot{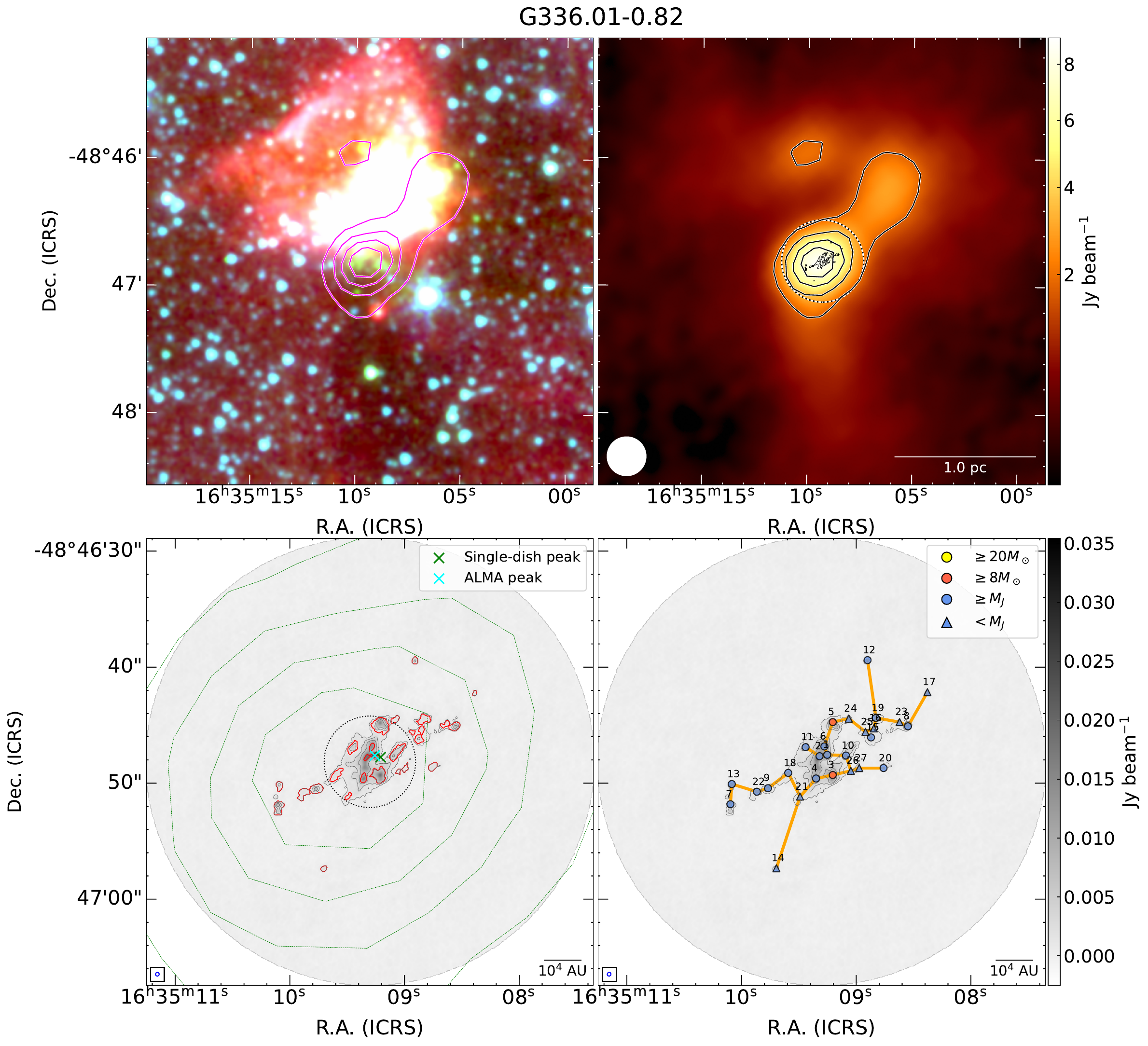}
\figsetgrpnote{Same as figure D.1, except for peak flux $=9.0$ Jy beam$^{-1}$ of ATLASGAL 870 $\mu$m continuum and 
        $\sigma=0.194$ mJy beam$^{-1}$ of ALMA 1.33 mm continuum.}
\figsetgrpend

\figsetgrpstart
\figsetgrpnum{D.29}
\figsetgrptitle{G34.43+0.24MM2}
\figsetplot{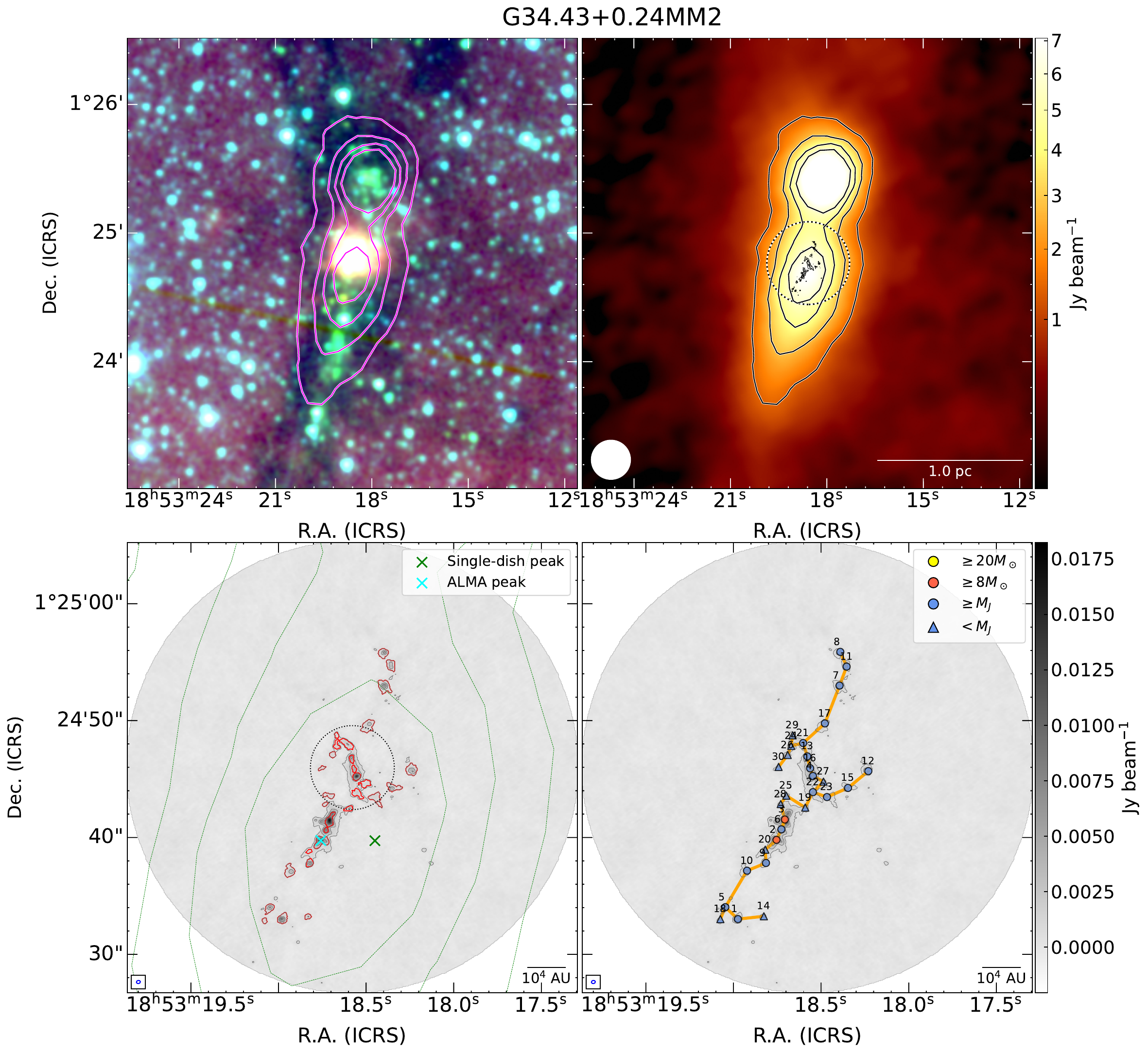}
\figsetgrpnote{Same as figure D.1, except for peak flux $=7.1$ Jy beam$^{-1}$ of ATLASGAL 870 $\mu$m continuum and 
        $\sigma=0.147$ mJy beam$^{-1}$ of ALMA 1.33 mm continuum.}
\figsetgrpend

\figsetgrpstart
\figsetgrpnum{D.30}
\figsetgrptitle{G35.13-0.74}
\figsetplot{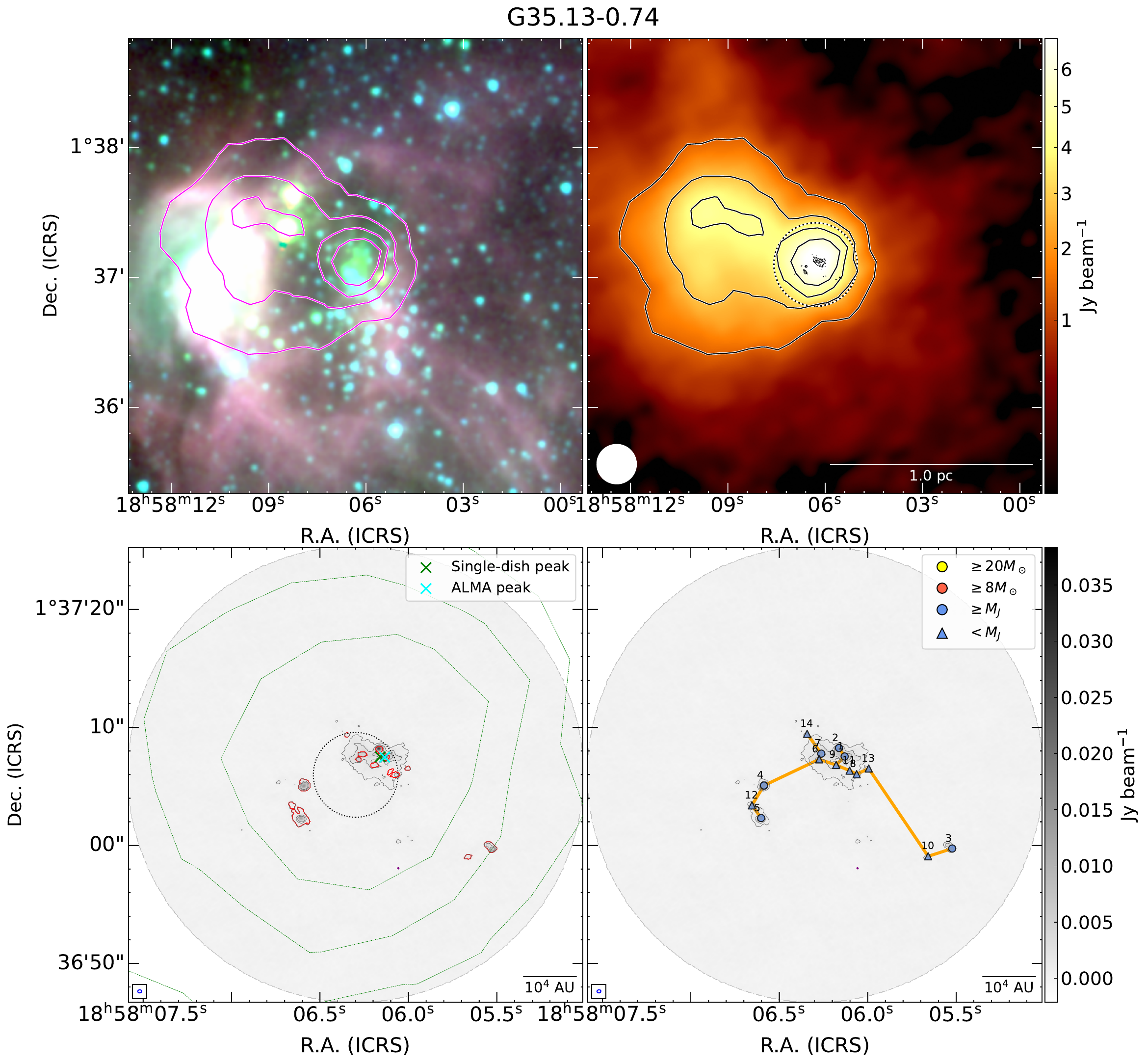}
\figsetgrpnote{Same as figure D.1, except for peak flux $=6.9$ Jy beam$^{-1}$ of ATLASGAL 870 $\mu$m continuum and 
        $\sigma=0.138$ mJy beam$^{-1}$ of ALMA 1.33 mm continuum.}
\figsetgrpend

\figsetend
\clearpage
\onecolumngrid

\bibliography{main}{}
\bibliographystyle{aasjournal}



\end{document}